\documentclass[useAMS,usenatbib]{mn2e}
\newcommand \kms{km s$^{-1}$}
\newcommand \zabs{z$_{\rm abs}$}
\newcommand \zem{z$_{\rm em}$}
\newcommand \url{}
\newcommand{\ts}{\textsuperscript}

\usepackage{longtable}
\usepackage{graphicx}
\usepackage{lscape}
\usepackage{rotating}
\usepackage{amssymb}

\def \nodata{. . .}

\def \alphafe{[$\alpha$/Fe]}
\def \HI{H\textsc{i}}
\def \sion{\textsc{ii}}

\def \npdla{46}
\def \nlpdla{41}

\def \nmdla{27}
\def \nlmdla{21}
\def \nxmdla{6}
\def \vninety{v$_{90}$}
\begin{document}

\title[Chemistry of XQ-100 DLA sample]{Chemical abundances of the Damped Lyman Alpha systems in the XQ-100 survey}


\author[Berg et al.] {
\parbox[t]{\textwidth}{
T. A. M. Berg$^1$,  S. L. Ellison$^1$, R. S\'anchez-Ram\'irez$^{2,3,4}$, J. X. Prochaska$^5$, S. Lopez$^{6}$, V. D'Odorico$^{7}$, G. Becker$^{8,9}$, L. Christensen$^{10}$, G. Cupani$^{7}$, K. Denney$^{11}$, G. Worseck$^{12}$}\\\\
$^1$ Department of Physics and Astronomy, University of Victoria, Victoria, British Columbia, V8P 1A1, Canada.\\
$^2$Unidad Asociada Grupo Ciencias Planetarias (UPV/EHU, IAA-CSIC), Departamento de F\'isica Aplicada I,\\ E.T.S. Ingenier\'ia, Universidad del Pa\'is Vasco (UPV/EHU), Alameda de Urquijo s/n, E-48013 Bilbao, Spain.\\
$^3$Ikerbasque, Basque Foundation for Science, Alameda de Urquijo 36-5, E-48008 Bilbao, Spain.\\
$^4$Instituto de Astrof\'isica de Andaluc\'ia (IAA-CSIC), Glorieta de la Astronom\'ia s/n, E-18008, Granada, Spain.\\
$^5$ Department of Astronomy and Astrophysics, University of California, Santa Cruz, Santa Cruz, CA, 95064, USA.
$^{6}$Departamento de Astronom\'{\i}a, Universidad de Chile, Casilla 36-D, Santiago, Chile.\\
$^{7}$INAF-Osservatorio Astronomico di Trieste, Via Tiepolo 11, I-34143 Trieste, Italy.\\
$^{8}$Space Telescope Science Institute, 3700 San Martin Drive, Baltimore, MD 21218, USA.\\
$^{9}$Institute of Astronomy and Kavli Institute of Cosmology, Madingley Road, Cambridge CB3 0HA, UK.\\
$^{10}$Dark Cosmology Centre, Niels Bohr Institute, University of Copenhagen, Juliane Maries Vej 30, DK-2100 Copenhagen, Denmark.\\
$^{11}$Department of Astronomy, The Ohio State University, 140 West 18th Avenue, Columbus, OH 43210, USA.\\
$^{12}$Max-Planck-Institut f\"{u}r Astronomie, K\"{o}nigstuhl 17, D-69117 Heidelberg, Germany.\\
}

\maketitle
\begin{abstract}
The XQ-100 survey has provided high signal-noise spectra of 100 redshift 3--4.5 quasars with the X-Shooter spectrograph. The metal abundances for 13 elements in the 41 damped Ly$\alpha$ systems (DLAs) identified in the XQ-100 sample are presented, and an investigation into abundances of a variety of DLA classes is conducted.  The XQ-100 DLA sample contains five DLAs within 5000 \kms{} of their host quasar (proximate DLAs; PDLAs) as well as three sightlines which contain two DLAs within 10,000 \kms{} of each other along the same line-of-sight (multiple DLAs; MDLAs). Combined with previous observations in the literature, we demonstrate that PDLAs with logN(HI)$<21.0$ show lower [S/H] and [Fe/H] (relative to intervening systems with similar redshift and N(\HI{})), whilst higher [S/H] and [Si/H] are seen in PDLAs with logN(HI)$>21.0$. These abundance discrepancies are independent of their line-of-sight velocity separation from the host quasar, and the velocity width of the metal lines (\vninety{}). Contrary to previous studies, MDLAs show no difference in \alphafe{} relative to single DLAs matched in metallicity and redshift. In addition, we present follow-up UVES data of J0034+1639, a sightline containing three DLAs, including a metal-poor DLA with [Fe/H]$=-2.82$ (the third lowest [Fe/H] in DLAs identified to date) at \zabs{}$=4.25$. Lastly we study the dust-corrected [Zn/Fe], emphasizing that near-IR coverage of X-Shooter provides unprecedented access to Mg\sion{}, Ca\sion{} and Ti\sion{} lines (at redshifts 3--4) to provide additional evidence for subsolar [Zn/Fe] ratio in DLAs.
\end{abstract}


\section{Introduction}
\label{sec:intro}

Quasars (QSOs) exist at many different epochs, providing lines of sight through pockets of gas from the epoch of reionization to  the present day. One of the classes of intervening absorbers towards QSOs are damped Lyman-$\alpha$ systems \citep[DLAs;][]{Wolfe05}, defined by their large \HI{} column densities \citep[N(\HI{})$\geq 2\times{}10^{20}$ atoms cm$^{-2}$;][]{Wolfe86}. DLAs are common probes to study the evolution of neutral gas and metals in the interstellar medium (ISM) of galaxies from \zabs{}$\sim5$ to the present day \citep{Wolfe95,Pettini97,Prochaska04DR1,Rao06,Rafelski14,SanchezRamirez16}.

A large portion of DLA analyses has been concentrated on detailed abundance analyses of the host galaxies \citep[e.g.][]{Pettini94,Lu98,Centurion00,Wolfe03,Cooke11,Zafar14N, Berg15II}. As elements have unique physical properties and nucleosynthetic origins \citep{Woosley95,Mcwilliam97,Nomoto13}, different abundance ratios have been used to understand the star formation history and dust content of DLAs \citep{Ledoux02,Prochaska02II,Vladilo11}. The most common ratio to probe enrichment histories is \alphafe{}, which traces the star formation history due to the time-delayed contributions of Type II and Ia supernovae \citep{Tinsley79,Mcwilliam97,Venn04,Tolstoy09}. However elements such as Fe, Ni, and Cr are heavily depleted onto dust \citep{Savage96}, leading to overestimates of the measured gas-phase \alphafe{} in DLAs. These overestimates in \alphafe{} have led to the use of other undepleted elements that trace Fe (such as Zn) to better estimate the intrinsic \alphafe{} ratio \citep{Pettini97,Vladilo02b}. In the case of Zn, care must be taken as Zn does not necessarily trace Fe in all environments and metallicities \citep{Prochaska00disk,Chen04,Nissen07,Rafelski12,Berg15II}.

The physical nature of DLAs also influences their gas phase abundances, including the role of ionizing sources \citep{Dodorico07,Ellison10,Zafar14Ar} or the amount of dust \citep[e.g.][]{Pettini94,Kulkarni97,DLAcat50,Krogager16}. There are many sub-classes of DLAs that provide opportunities to probe these differing physical environments. Proximate DLAs (PDLAs) are DLAs defined to be within $\Delta v\leq 5000$ \kms{} of the host QSO, and more frequently seen than intervening systems \citep{Ellison02,Russell06,Prochaska08}. PDLAs have shown increasing metal abundances with increasing N(\HI{}), in particular both [S/H] and [Si/H] are $\sim3\times$ larger in four PDLAs with logN(\HI{})$>21.0$ \citep{Ellison10,Ellison11}. Multiple DLAs (MDLAs) along the same line of sight within $500\leq \Delta v \leq 10000$ \kms{} of eachother have also shown different metallicity effects, with a low \alphafe{} relative to the typical DLA \citep{Ellison01L,Lopez03}; an effect attributed to truncated star formation from environmental effects. However, the analyses of \cite{Lopez03} and \cite{Ellison10}  suffer from low numbers of MDLAs (seven absorbers) and PDLAs (16 absorbers). 

Recently there has been a significant effort to identify the  first stars and galaxies \citep{Cayrel04,Beers05,Suda08,Spite11,Frebel12,Norris13,Frebel15} to constrain Population III nucleosynthesis \citep{Umeda02,Greif07,Heger10,Cooke13}. In tandem with the search for metal-poor stars in the Galaxy and its nearby companions \citep[e.g.][]{Jacobson15,Skuladottir15MP}, work at higher redshifts focused the identification and measurement of abundances in the most metal-poor DLAs \citep[MPDLAs; {[Fe/H]$\leq-2.5$};][]{Penprase10,Cooke11,Becker12,Cooke14}. As the explosion mechanism of the supernovae models is very uncertain, chemical abundances in these metal-poor regimes are required to constrain the models of Population III nucleosynthesis. In particular the supernovae explosion energy influences the mass cut of the supernovae, and thus which elements escape into the ISM \citep{Umeda02,Nomoto13}. To date, abundances in the most MPDLAs reflect first generation stars that have undergone moderate to low energy core-collapse supernovae \citep{Cooke11,Cooke13}, but remains to be tested for a large sample of DLAs with [Fe/H]$\leq-3$.

The XQ-100 Large Programme survey \citep[PI: S. Lopez, ESO ID 189.A-0424;][]{Lopez16} has observed 100 QSOs at $z=$3.5--4.5 with the X-Shooter \citep{Vernet11} spectrograph on the Very Large Telescope (VLT). As the survey was primarily designed to study active galactic nuclei, the inter-galactic medium, and the Ly$\alpha$ forest; XQ-100 provides a near-random sample of intervening DLAs as the QSOs were selected without consideration of intervening absorbers. In this paper, we present the metal column densities for 14 species (O\textsc{i}, C\sion{}, Mg\textsc{i}, Mg\sion{}, Ca\sion{}, Si\sion{}, P\sion{}, S\sion{}, Ti\sion, Cr\sion{}, Mn\sion{}, Fe\sion{}, Ni\sion{}, Zn\sion{}) in the DLAs recently identified by \cite{SanchezRamirez16} in the XQ-100 survey. By combining the XQ-100 DLAs with a sample of DLA abundances in the literature \citep{Berg15II}, we investigate the elemental abundances of the XQ-100 sample and demonstrate the prospects of using X-Shooter to study absorption lines in the near infrared (NIR).

\section{Data}
In this section we present the DLA abundances derived in this work. The spectra used come from the XQ-100 survey \citep[][]{Lopez16}, as well as follow-up observations with VLT/UVES for one sightline (J0034+1639). 
 
\subsection{XQ-100 abundances}
\label{sec:XQdata}

The XQ-100 dataset consists of 100 QSOs observed with X-Shooter towards QSO sightlines at redshift \zem{}$\sim$ 3.5--4.5. The per-arm exposures were either $\sim0.5$ or $\sim1$ hour in length (depending whether the QSO was classified as `bright' or `faint'; respectively), providing signal-noise ratios ($snr$) of $\sim20$ pixel$^{-1}$ (median $\sim30$ pixel$^{-1}$) at resolution R$\sim$5000--9000 from the near UV ($3150$\AA{}) to the NIR ($25000$ \AA{}). The spectra were reduced using an internal IDL package. For more details, see \cite{Lopez16}.

The DLAs were identified and \HI{} column densities determined in \cite{SanchezRamirez16}. Table \ref{tab:DLAs} summarizes the DLAs identified in the XQ-100 survey, along with their redshifts\footnote{For J0034+1639, the redshifts of the DLAs have been tweaked based on the metal lines in the UVES spectra presented in Section \ref{sec:Udata}.}, logN(\HI{}) and line-of-sight velocity separations from the QSO ($\Delta$v). In \cite{SanchezRamirez16}, the \HI{} column densities were determined by simultaneously fitting the Lyman series (up to Ly$\epsilon$). Redshifts were primarily determined from the \HI{} fits, but in some cases were guided by the metal lines.

The XQ-100 spectra are released with two options for continuum fits. One option uses a power-law with select emission lines incorporated to fit each spectrum, providing complete wavelength coverage. However this power-law continuum does not provide a good fit around missing QSO emission lines, and therefore cannot be used for accurate DLA abundances in some cases. In this paper we use the alternative option, a by-hand fit to the continuum using a cubic spline\footnote{Code available at \url{https://github.com/trystynb/ContFit}.}, similar to the approach of \cite{SanchezRamirez16}. The error spectra were normalized by the same continuum fit.

\begin{table}
\begin{center}
\caption{XQ-100 DLA sample}
\label{tab:DLAs}
\begin{tabular}{lcccc}
\hline
QSO& \zem{}& \zabs{}& logN(\HI{})& $\Delta$v (\kms{})\\
\hline
J0003$-$2603& 4.125& 3.390& $21.40\pm0.10$& 46040\\
J0006$-$6208& 4.440& 3.203& $20.90\pm0.15$& 75670\\
 &  & 3.775& $21.00\pm0.20$& 38870\\
J0034+1639& 4.292& 3.752& $20.40\pm0.15$& 32110\\
 &  & 4.251& $20.60\pm0.10$& 2349$^{\rm P}$$^{\rm M}$\\
 &  & 4.283& $21.00\pm0.10$& 522$^{\rm P}$$^{\rm M}$\\
J0113$-$2803& 4.314& 3.106& $21.20\pm0.10$& 75650\\
\hline
\end{tabular}
$^{\rm P}$ -- Proximate DLA\\
$^{\rm M}$ -- Multiple DLA\\
\end{center}
\end{table}

\subsubsection*{Metal Column Densities}

Column densities were derived using the Apparent Optical Depth Method (AODM) presented in \cite{Savage91}. In brief, this method sums the optical depth of an unsaturated absorption feature \citep[at wavelength $\lambda$ with oscillator strength $f$; taken from][]{Morton03}, converting the total optical depth into a column density using

\begin{equation}
\label{eq:AODM}
 N=\frac{m_{e}c}{\pi e^{2} f \lambda}\int_{v_{min}}^{v_{max}} \tau dv.
\end{equation}

The integration limits $v_{min}$ and $v_{max}$ are selected such that the associated metal absorption profile are bounded by these velocity limits. We attempted to use the same velocity limits for all metal lines for a given DLA, however the velocity limits were sometimes adjusted on a line-by-line basis to exclude regions of contamination outside the metal absorption feature. If the weaker components of a shallow line are undetected, the velocity limits are purposefully made narrow to only contain associated absorption, and exclude flux contribution from  the continuum.  When the column density is derived using multiple clean lines for the same ion, we adopt the average value (logN$_{\rm adopt}$) and the standard error (with a minimum error of 0.05 dex). For features that exhibited obvious blending or saturation, we measure the AODM column density as a limit to the true value, adopting the most constraining limit as the final value. For cases where the absorption feature is severely blended, we do not consider adopting the measured column density.

The limiting spectral resolution of X-Shooter (R=5100, 8800, and 5300 for the UV, visible, and NIR arms; respectively) implies that there are cases where we cannot determine the presence of unseen contamination or saturation visually. Cases of hidden contamination or saturation were flagged when multiple lines are measured for a given species and resulted in significantly inconsistent column densities. Using multiple Si\sion{} and Fe\sion{} lines, we have noted that these cases of hidden saturation tend to occur when the strongest absorption reaches a relative flux of $\sim$0.3--0.5 \citep[similar to the $\sim0.25$ seen in the UVB arm by][]{Krogager16}. Lines with absorption stronger than $\sim0.5$ (in units of relative flux) were inspected for inconsistent column densities with other lines. Unless the derived column density is consistent with weaker lines, any discrepant column densities were flagged as saturated. We note that all of the lines with a peak absorption at $\sim0.5$ had additional lines to constrain potential saturation. However, there may still be cases with unresolved saturation or blending unaccounted for.

When weak lines are not detected and the spectrum is uncontaminated, we measure $3\sigma$ upper limits based on the ability to detect the strongest absorption feature within the signal-noise ratio of the spectrum using

\begin{equation}
\label{eq:limit}
N=\frac{3 m_{e} c ~ FWHM}{\pi e^{2} f \lambda^{2} (1+z_{abs}) snr}
\end{equation}

where FWHM is the full-width at half-maximum of the strongest absorption feature. As the components of the absorption profile are likely limited by the resolution of the instrument ($\Delta v \sim 55$\kms{}) rather than the DLA kinematics \citep[$\Delta v \sim 10 $\kms{}; e.g.][]{DZavadsky04,Jorgenson10,Tumlinson10,Cooke14}, we assume that the FWHM of this strongest absorption feature to be equal to the resolution limit of X-Shooter in the setup with a slit width of $\sim1$\arcsec{}.  The signal-noise ratio for the $3\sigma$ upper-limits was computed within the bounds of the AODM limits. 

\begin{landscape}
\begin{table*}
\tiny
\begin{center}
\caption{XQ-100 DLA Column Densities}
\label{tab:XQ100N}
\begin{tabular}{lccccccccccccccccc}
\hline
QSO& \zabs{}& logN(\HI{})& [M/H] (ion)& logN(C\textsc{ii})& logN(O\textsc{i})& logN(Mg\textsc{i})& logN(Mg\textsc{ii})& logN(Si\textsc{ii})& logN(P\textsc{ii})& logN(S\textsc{ii})& logN(Ca\textsc{ii})& logN(Ti\textsc{ii})& logN(Cr\textsc{ii})& logN(Mn\textsc{ii})& logN(Fe\textsc{ii})& logN(Ni\textsc{ii})& logN(Zn\textsc{ii})\\
\hline
J0003$-$2603& 3.390& $21.40\pm0.10$& $-1.93\pm0.11$ (ZnII)& \nodata{}& \nodata{}& $11.85\pm0.05$& $>13.91$& \nodata{}& $<13.73$& \nodata{}& $<11.81$& $<12.24$& $13.07\pm0.05$& $<12.39$& $14.75\pm0.05$& $13.42\pm0.05$& $12.10\pm0.05$\\
J0006$-$6208& 3.203& $20.90\pm0.15$& $-2.20\pm0.15$ (FeII)& \nodata{}& \nodata{}& $<12.39$& $13.33\pm0.05$& \nodata{}& \nodata{}& \nodata{}& $<12.38$& $<13.02$& $<13.08$& \nodata{}& $13.85\pm0.05$& $<13.86$& $<12.40$\\
J0006$-$6208& 3.775& $21.00\pm0.20$& $-0.94\pm0.20$ (ZnII)& \nodata{}& \nodata{}& \nodata{}& $>14.01$& \nodata{}& $<14.49$& \nodata{}& \nodata{}& $<12.30$& $<15.31$& $12.51\pm0.05$& \nodata{}& $<13.54$& $<12.69$\\
J0034+1639& 3.752& $20.40\pm0.15$& $-1.89\pm0.15$ (FeII)& $>14.72$& $>15.02$& \nodata{}& $>13.63$& \nodata{}& \nodata{}& $<14.70$& \nodata{}& $<12.31$& $<13.70$& $<12.49$& $13.66\pm0.05$& $<12.55$& $<12.39$\\
J0034+1639& 4.251& $20.60\pm0.10$& $-2.10\pm0.13$ (SII)& $>14.24$& $14.78\pm0.05$& \nodata{}& $>13.20$& $13.71\pm0.05$& $<13.66$& $13.65\pm0.08$& \nodata{}& $<12.36$& $<13.38$& \nodata{}& $13.23\pm0.06$& $<13.24$& $<12.47$\\
J0034+1639& 4.283& $21.00\pm0.10$& $-1.59\pm0.11$ (SII)& $>14.83$& $>15.29$& $12.13\pm0.05$& \nodata{}& $>14.83$& \nodata{}& $14.56\pm0.05$& \nodata{}& $<12.27$& $<13.21$& \nodata{}& $14.54\pm0.05$& $<13.05$& $<12.81$\\
J0113$-$2803& 3.106& $21.20\pm0.10$& $-1.11\pm0.10$ (SiII)& \nodata{}& \nodata{}& $12.60\pm0.05$& $>14.42$& $15.60\pm0.05$& \nodata{}& \nodata{}& $<11.94$& $<12.61$& $13.69\pm0.05$& $<12.63$& $<14.69$& $13.99\pm0.05$& $12.69\pm0.05$\\
\hline
\end{tabular}
\end{center}
\end{table*}
\end{landscape}

\begin{figure*}
\begin{center}
\includegraphics[width=1.1\textwidth]{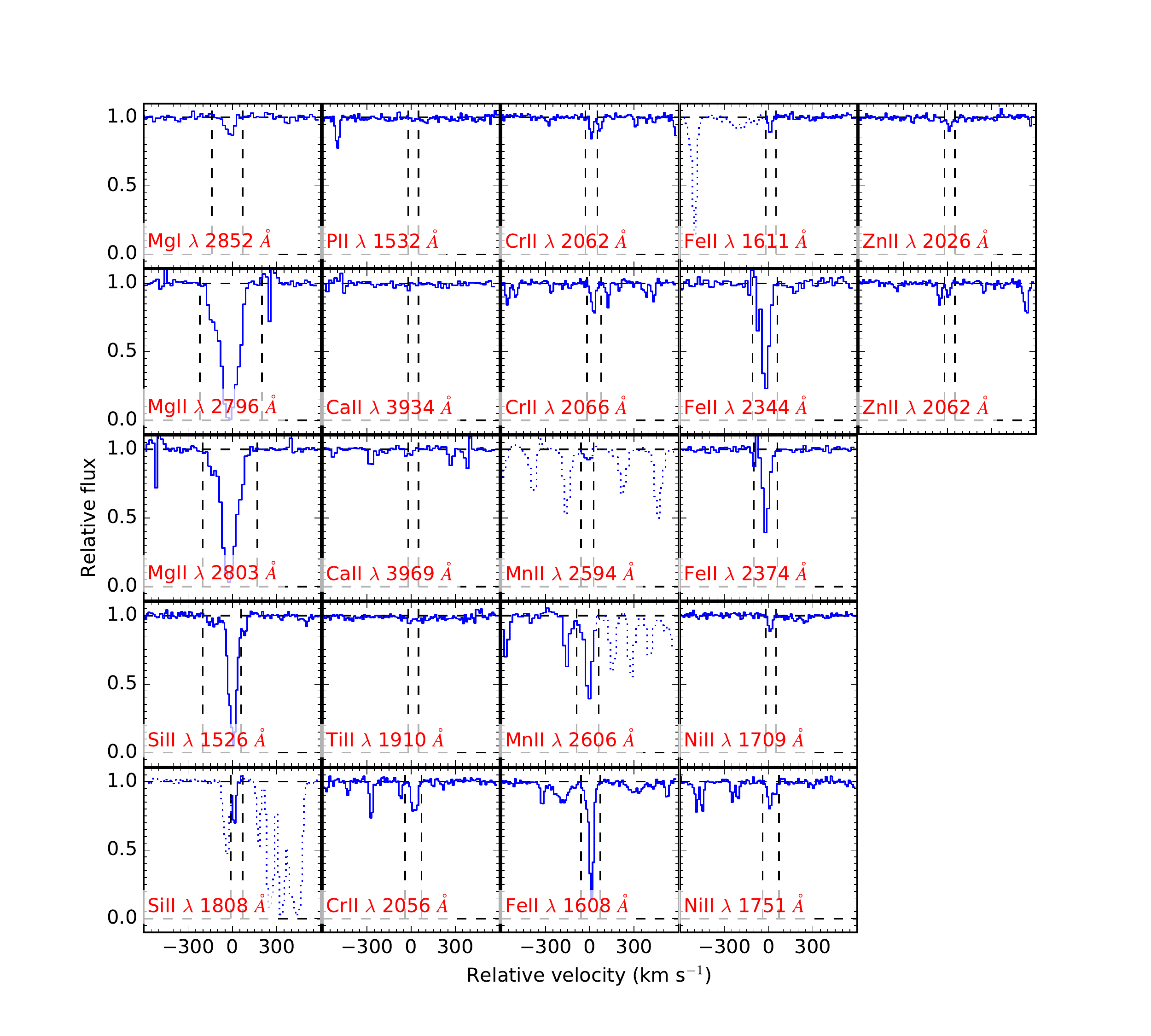}
\caption{Velocity profile of the XQ-100 spectrum towards J0003-2603 (\zabs{}=3.390). The vertical dashed lines indicated the AODM velocity bounds. Regions of the spectrum that are contaminated by strong, unassociated absorption are flagged as dotted lines.}
\label{fig:egvelprof}
\end{center}
\end{figure*}

As an example of data quality and coverage, Figure \ref{fig:egvelprof} shows the velocity profile for all the observed lines of the DLA in the spectrum of J0003-2603 (\zabs{}$=3.39$). The vertical dashed lines show the velocity limits used to integrate the spectrum for the AODM column density determinations. The column densities for all the measured lines for the DLA towards J0003-2603 are given in Table \ref{tab:J0003-2603,33900}. Table \ref{tab:J0003-2603,33900} displays the wavelength and oscillator strength of the absorption line \citep{Morton03}, the velocity integration bounds of the absorption feature used ($v_{min}$ and $v_{max}$), the measured column density logN(X) of the metal line, and whether the derived column density was included for the final computation of the adopted column density (logN$_{\rm adopt}$). The adopted logN$_{\rm adopt}$ is given for each species in the last row for the species.  Appendix \ref{app:xdata} contains the velocity profiles (Figures  \ref{fig:J0006-6208,32030}--\ref{fig:J0800+1920,39465}) and abundances (Tables \ref{tab:J0006-6208,32030}--\ref{tab:J0800+1920,39465}) for the remaining DLAs. If the adopted column density for a given species is best constrained by an upper and lower limit, the range of possible values is provided in Tables \ref{tab:J0006-6208,32030}--\ref{tab:J0800+1920,39465}. A summary of the final adopted column densities for all DLAs is listed in Table \ref{tab:XQ100N}\footnote{The adopted column densities in Table \ref{tab:XQ100N} for the DLAs towards J0034+1639 are preferentially adopted from the UVES data presented in Section \ref{sec:Udata}.}.

\begin{table*}
\begin{center}
\caption{X-Shooter metal column densities for J0003$-$2603 (\zabs{}=$3.390$)}
\label{tab:J0003-2603,33900}
\begin{tabular}{lcccccccc}
\hline
Ion& Line& $\lambda$& $f$& $v_{\rm min}$& $v_{\rm max}$& logN(X)& Included& logN$_{\rm adopt}$\\
 &  & \AA{}&  & \kms{}& \kms{}&  &  &  \\
\hline
Mg\textsc{i}& 2852& 2852.964& 1.81E+00& -140& 70& $11.85\pm0.03$& Y& --\\
Mg\textsc{i}& & & & & & & & $11.85\pm0.05$\\
\hline
Mg\textsc{ii}& 2796& 2796.352& 6.12E-01& -220& 200& $>13.91$& Y& --\\
Mg\textsc{ii}& 2803& 2803.531& 3.05E-01& -200& 170& $>14.04$& Y& --\\
Mg\textsc{ii}& & & & & & & & $>14.04$\\
\hline
Si\textsc{ii}& 1526& 1526.707& 1.27E-01& -200& 60& $>14.37$& Y& --\\
\hline
\end{tabular}
\end{center}
\end{table*}

\subsubsection*{\vninety{} parameter}

Following the analysis techniques described in \cite{Prochaska97}, we have assessed the kinematic characteristics of the XQ-100 DLA sample. In particular, we have measured the velocity width \vninety{} corresponding to 90\%\ of the integrated optical depth using one low-ion transition per DLA (listed in Table \ref{tab:v90}). We selected transitions according to the spectral S/N, avoiding saturated or weak absorption profiles. The optical depth per pixel was calculated from the normalized flux values and then smoothed by a boxcar with width of 22 \kms{} as in \cite{Prochaska97}.  From these smoothed optical depth arrays, we calculate the velocity width comprising 90\% of the integrated optical depth between the AODM integration bounds.

Adopting the results of \cite{Prochaska08vel} who also analysed a set of echellette observations ($R \approx 8,000$), we have reduced the raw \vninety{} values to correct for instrumental broadening.  Specifically, we lower the raw \vninety{} measurements by 30, 17, and 25 km/s for the UVB, VIS, and NIR arms respectively.  These corrections correspond to $\approx 40\%$ of the FWHM of the instrumental line-spread-function for our X-Shooter configuration. The final, reported \vninety{} values are presented in Table \ref{tab:v90}.    

\begin{table}
\begin{center}
\caption{\vninety{} measurements for the XQ-100 DLA sample}
\label{tab:v90}
\begin{tabular}{lccc}
\hline
QSO& \zabs{}& Transition&  \vninety{} \\
& & & \kms{}\\
\hline
J0003$-$2603& 3.390& Cr\sion{} 2066&  32  \\
J0006$-$6208& 3.203& Fe\sion{} 1608&  32\\
J0006$-$6208& 3.775& Fe\sion{} 1608&  54\\
J0034+1639& 3.755& Fe\sion{} 2586&  32\\
J0034+1639& 4.251& Si\sion{} 1526&  43\\
J0034+1639& 4.283& Fe\sion{} 1608&  109\\
J0113$-$2803& 3.106& Si\sion{} 1808&  219\\
J0124+0044& 2.261& Si\sion{} 1808&  142\\
J0132+1341& 3.936& Fe\sion{} 1608&  32\\
J0134+0400& 3.692& Fe\sion{} 1608&  10\\
J0134+0400& 3.772& Si\sion{} 1808&  109\\
J0234$-$1806& 3.693& Fe\sion{} 2344&  203\\
J0255+0048& 3.256& Fe\sion{} 1608&  208\\
J0255+0048& 3.914& S\sion{} 1253&  21\\
J0307$-$4945& 3.591& Fe\sion{} 2586&  51\\
J0307$-$4945& 4.466& Si\sion{} 1304&  219\\
J0415$-$4357& 3.808& Si\sion{} 1526&  131\\
J0424$-$2209& 2.982& Fe\sion{} 2600&  13\\
J0529$-$3552& 3.684& Si\sion{} 1526&  21\\
J0747+2739& 3.424& Fe\sion{} 1608&  120\\
J0747+2739& 3.901& Si\sion{} 1526&  153\\
J0800+1920& 3.946& Si\sion{} 1304&  43\\
J0818+0958& 3.306& Si\sion{} 1808&  76\\
J0835+0650& 3.955& Si\sion{} 1304&  21\\
J0920+0725& 2.238& Fe\sion{} 2374&  98\\
J0955$-$0130& 4.024& Fe\sion{} 2600&  336\\
J1020+0922& 2.592& Si\sion{} 1808&  76\\
J1024+1819& 2.298& Si\sion{} 1808&  54\\
J1057+1910& 3.373& Fe\sion{} 1608&  307\\
J1058+1245& 3.432& Mg\sion{} 2803&  127\\
J1108+1209& 3.397& C\sion{} 1334&  32 \\
J1108+1209& 3.546& Fe\sion{} 2374&  70 \\
J1312+0841& 2.660& Fe\sion{} 2344&  153 \\
J1421$-$0643& 3.449& Fe\sion{} 1608&  43\\
J1517+0511& 2.688& Fe\sion{} 1608&  43\\
J1552+1005& 3.601& S\sion{} 1253&  21\\
J1552+1005& 3.667& Si\sion{} 1304&  230\\
J1633+1411& 2.882& Mg\sion{} 2803&  13 \\
J1723+2243& 3.698& Fe\sion{} 2586&  374\\
J2239$-$0552& 4.080& Si\sion{} 1526&  131\\
J2344+0342& 3.220& Cr\sion{} 2056&  54\\
\hline
\end{tabular}
\end{center}
\end{table}

\subsubsection*{Comparison to literature measurements}

Of the DLAs studied in XQ-100, 15 DLAs have previously been observed in the literature sample of \citet[further referred to as the literature sample]{Berg15II}. The literature sample includes abundance determinations for a variety of elements\footnote{The elements are: N, O, Mg, Al, Si, S, Ca, Ti, Cr, Mn, Fe, Co, Ni, and Zn.} using high resolution spectrographs (predominantly VLT/UVES, Keck/HIRES, and Keck/ESI) for 341 DLAs between \zabs{}$\sim0$ and 5.

Similar to the comparison of N(\HI{}) to the literature in \cite{SanchezRamirez16}, Table \ref{tab:Ncompare} compares the metal column densities for the 15 DLAs common between the literature and XQ-100. $\Delta$N(X) represents the difference in column density between the value determined in this work and the value in the literature sample (where a positive difference implies the XQ-100 column density is larger than what is in the literature). When both the literature and XQ-100 measurements yield a clean detection, we compare difference in column densities ($\Delta$N(X)$_{\rm det.}$) to the combined column density errors. Consistent values of $\Delta$N(X)$_{\rm det.}$ should be of the order of the combined error, although the quoted errors in metal column densities do not generally account for sources of error other than photon noise (e.g. continuum fitting) and can be larger. For most of the detections, we see consistent values between XQ-100 and the high-resolution literature (to within 0.05 dex). The discrepant cases (the bolded ionic species in Table \ref{tab:Ncompare}) are discussed in Appendix \ref{sec:AppXQ100}. In all cases, the column densities derived in this work are adopted as they are consistent with the literature values (see Appendix \ref{sec:AppXQ100} for justification of discrepant cases).

For cases where both the XQ-100 and literature measurements are upper limits ($\Delta$N(X)$_{\rm lim.}$; positive values indicate that the XQ-100 limit is higher than the literature), 4 of 10 XQ-100 upper limits are more constraining than the literature column density derived from Keck/ESI data with comparable resolution. Naturally, the higher-resolution instruments provide more constraining upper limits due to their ability to resolve narrower features within the fluctuations of the continuum. For all cases where the literature provides better limits (or detections, as in the case for Ni\sion{} and Zn\sion{} in J2344+0342 at \zabs{}$=3.22$), we adopt the literature values.

\begin{table*}
\begin{center}
\caption{N(X) comparison between XQ-100 and literature}
\label{tab:Ncompare}
\begin{tabular}{lcccccccc}
\hline
QSO sightline& \zabs{}& Ion& logN(X)$_{\rm XQ100}$& logN(X)$_{\rm Lit.}$& $\Delta$N(X)$_{\rm det.}$& $\Delta$N(X)$_{\rm lim.}$& Lit. Instrument(s)& Ref.\\
\hline

J0003-2603& 3.390& Cr\sion{}& $13.07\pm0.05$& $13.09\pm0.03$& $-0.02\pm0.06$ & --& UVES,HIRES& 1,2\\
--& --& \textbf{FeII}& $14.75\pm0.05$& $14.87\pm0.03$& $-0.12\pm0.06$& --& UVES,HIRES& 1,2\\
--& --& Ni\sion{}& $13.42\pm0.05$& $13.39\pm0.03$& $+0.03\pm0.06$ & --& UVES,HIRES& 1,2\\
--& --& Zn\sion{}& $12.10\pm0.05$& $12.01\pm0.05$& $+0.09\pm0.07$ & --& UVES,HIRES& 1,2\\
\hline

J0134+0400& 3.692& Si\sion{}& $<14.36$& $>14.26$& --& --& ESI,UVES& 3,4\\
--& --& Fe\sion{}& $13.44\pm0.05$& $13.51\pm0.07$& $-0.07\pm0.09$& --& ESI,UVES& 3,4\\
\hline

J0134+0400& 3.772& \textbf{SiII}& $15.30\pm0.05$& $15.46\pm0.02$& $-0.16\pm0.05$ & --& ESI,UVES& 3,5\\
--& --& Cr\sion{}& $<12.83$& $<13.24$& --& $-0.41$& ESI,UVES& 3,5\\
--& --& Fe\sion{}& $14.97\pm0.05$& $>14.87$& --& --& ESI,UVES& 3,5\\
--& --& \textbf{NiII}& $13.84\pm0.05$& $13.98\pm0.03$& $-0.14\pm0.06$& --& ESI,UVES& 3,5\\
--& --& Zn\sion{}& $12.88\pm0.05$& $<13.10$& --& --& ESI,UVES& 3,5\\
\hline

J0255+0048& 3.256& Si\sion{}& $15.33\pm0.05$& $15.32\pm0.04$& $+0.01\pm0.06$ & --& HIRES& 2\\
--& --& Fe\sion{}& $>14.75$& $14.76\pm0.01$& --& --& HIRES& 2\\
--& --& \textbf{NiII}& $<13.44$& $13.61\pm0.07$& --& --& HIRES& 2\\
\hline

J0255+0048& 3.914& Si\sion{}& $15.02\pm0.05$& $>14.19$& --& --& HIRES& 2\\
--& --& S\sion{}& $14.73\pm0.05$& $14.72\pm0.01$& $0.01\pm0.05$& --& HIRES& 2\\
--& --& \textbf{NiII}& $13.50\pm0.05$& $13.27\pm0.04$& $+0.23\pm0.06$ & --& HIRES& 2\\
\hline

J0307-4945& 4.466& Si\sion{}& $14.59\pm0.05$& $14.68\pm0.07$& $-0.09\pm0.09$ & --& UVES& 6\\
--& --& Fe\sion{}& $14.11\pm0.05$& $14.21\pm0.17$& $-0.10\pm0.18$& --& UVES& 6\\
--& --& Ni\sion{}& $<13.43$& $<12.60$& --& $0.83$& UVES& 6\\
\hline

J0424-2209& 2.982& Cr\sion{}& $<13.01$& $<12.90$& --& $+0.11$& ESI& 3\\
--& --& Ni\sion{}& $<13.47$& $<13.37$& --& $0.10$& ESI& 3\\
--& --& Zn\sion{}& $<12.28$& $<12.17$& --& $0.11$& ESI& 3\\
\hline

J0747+2739& 3.424& Fe\sion{}& $14.47\pm0.05$& $>14.43$& --& --& ESI& 3\\
--& --& Ni\sion{}& $<13.46$& $<13.27$& --& $0.19$& ESI& 3\\
\hline

J0747+2739& 3.901& Si\sion{}& $14.08\pm0.05$& $14.03\pm0.01$& $+0.05\pm0.05$ & --& ESI& 3\\
--& --& Fe\sion{}& $14.03\pm0.05$& $<13.80$& --& --& ESI& 3\\
--& --& Ni\sion{}& $<13.45$& $<13.11$& --& $0.34$& ESI& 3\\
\hline

J0955-0130& 4.024& Fe\sion{}& $14.31\pm0.05$& $14.19\pm0.08$& $+0.12\pm0.09$ & --& HIRES,ESI& 2,7,8\\
--& --& Ni\sion{}& $<13.41$& $<13.44$& --& $-0.03$& HIRES,ESI& 2,7,8\\
\hline

J1108+1209& 3.397& SiII& $<13.77$& $>13.63$& --& --& ESI& 9\\
--& --& Fe\sion{}& $13.36\pm0.05$& $<13.72$& --& --& ESI& 9\\
\hline

J1421-0643& 3.449& Cr\sion{}& $<12.84$& $<12.71$& --& $0.13$& UVES& 5,10\\
--& --& \textbf{FeII}& $14.05\pm0.05$& $14.14\pm0.02$& $-0.09\pm0.05$ & --& UVES& 5,10\\
--& --& Zn\sion{}& $<13.25$& $<11.98$& --& $1.19$& UVES& 5,10\\
\hline

J1723+2243& 3.698& Fe\sion{}& $14.62\pm0.05$& $>14.57$& --& --& ESI& 3\\
--& --& Ni\sion{}& $<13.63$& $<13.95$& --& $-0.32$& ESI& 3\\
\hline

J2239-0552& 4.080& Fe\sion{}& $14.00\pm0.05$& $13.88\pm0.12$& $+0.12\pm0.13$ & --& HIRES,ESI,UVES& 8,11,12\\
--& --& Ni\sion{}& $<13.02$& $<13.17$& --& $-0.15$& HIRES,ESI,UVES& 8,11,12\\
\hline

J2344+0342& 3.220& Cr\sion{}& $13.38\pm0.05$& $13.34\pm0.10$& $+0.04\pm0.11$ & --& UVES,ESI& 3,13\\
--& --& Ni\sion{}& $<13.58$& $13.59\pm0.11$& --& --& UVES,ESI& 3,13\\
--& --& Zn\sion{}& $<12.85$& $12.23\pm0.30$& --& --& UVES,ESI& 3,13\\
\hline

\end{tabular}
\textbf{Bolded} species indicate cases of column density discrepancies between XQ-100 and the literature, and are discussed in Section \ref{sec:AppXQ100}.\\
$\Delta$N(X) is positive when the XQ-100 column density is larger than the 
literature value.\\
\textsc{References} -- 
	(1) \cite{DLAcat21}.
	(2) \cite{DLAcat23}.
	(3) \cite{Prochaska03ApJS147}.
	(4) \cite{DLAcat69}.
	(5) \cite{DLAcat70}.
	(6) \cite{DLAcat28}.
	(7) \cite{DLAcat22}.
	(8) \cite{DLAcat37}.
	(9) \cite{Penprase10}.
	(10) \cite{DLAcat50}.
	(11) \cite{DLAcat3}.
	(12) \cite{DLAcat65}.
	(13) Dessauges-Zavadsky et al. (unpublished).
\end{center}
\end{table*}

\subsubsection*{Continuum-fitting errors}
As a baseline for comparison, the spline continuum-fitting was done two times to assess the random errors of the continuum fit. The median difference in column density for all lines in the two iterations is 0.0 dex, with an interquartile range of 0.02 dex and 67\% of column densities being within $\pm0.01$ dex. Although the 3$\sigma$ upper limits are not included in this calculation, it is worth noting that derived column densities for these limits are sensitive to the continuum placement.

\subsection{UVES data for J0034+1639}
\label{sec:Udata}

The X-Shooter spectrum of the \zabs{}$=4.25$ DLA (Figure \ref{fig:J0034+1639,42523}) towards J0034+1639 suggests a very low metallicity  system \citep[{[Fe/H]}$=-2.82\pm0.11$, third lowest {[Fe/H]} to date;][see Table \ref{tab:J0034+1639,42523}]{Cooke15}. The sightline also contains two other DLAs, as well as a number of other \HI{} absorbers (which will be discussed in a future paper).

To confirm the column densities of this metal-poor DLA are free of undetected contamination and saturation, as well as attempting to measure Fe-peak elements such as Ni\sion{} to pinpoint the underlying supernovae population \citep{Cooke13}, we obtained VLT/UVES \citep{Dekker00} data towards the sightline J0034+1639. The observing programme (PI: T. Berg; Programme number 094.A-0223A) was organized in five observing blocks (OBs) of 48 min each for a total of 4 hours of on-target integration to obtain a signal-noise of 20--25 pixel$^{-1}$. We used the dichroic DIC-2 at the standard setting of 437+760 to obtain wavelength coverage between 6000\AA{} and 9000\AA{} (slit width 0.8$''$; $R\sim80000$). The data were reduced using the standard UVES data-reduction software in \textsc{Reflex}\footnote{\url{http://www.eso.org/sci/software/reflex/}}. After correcting for the heliocentric velocity, the five OBs were median combined with IRAF's \textsc{scombine}, whilst the error spectra were combined in quadrature. The combined spectrum was then continuum normalized using the cubic spline software presented in Section \ref{sec:XQdata}.

\begin{figure*}
\begin{center}
\includegraphics[width=1.1\textwidth, height=0.95\textheight, keepaspectratio]{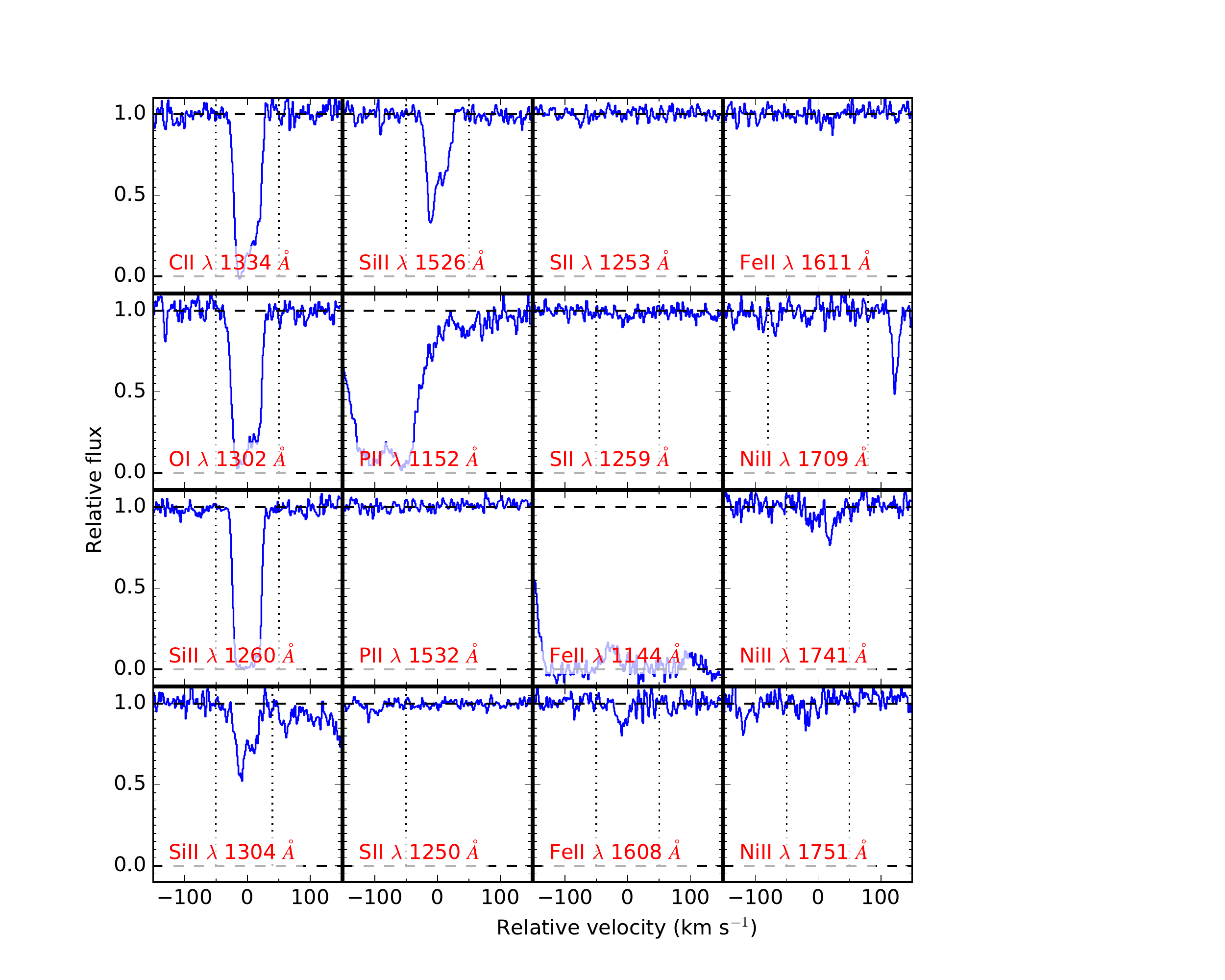}
\caption{Velocity profiles of the UVES spectrum towards J0034+1639 (\zabs{}$=4.25$). Vertical dashed lines represent the velocity limits of the AODM integration.}
\label{fig:Uvesvelprof} 
\end{center}
\end{figure*}

Metal column densities for the three DLAs along the sightline (see Table \ref{tab:DLAs}) were derived using the AODM, as in Section \ref{sec:XQdata}. As UVES can resolve down to the kinematic structure of the DLA clouds, we use the measured FWHM of the strongest component of an unsaturated metal line profile as the minimum equivalent width observable. This FWHM is used in Equation \ref{eq:limit} to determine the $3\sigma$ limits in the absorber.  The signal-noise ratio for the $3\sigma$ upper-limits was computed within the bounds of the AODM limits. Figure \ref{fig:Uvesvelprof} and Table \ref{tab:UVESJ0034+1639,42507} present the velocity profiles of the absorption lines towards the MPDLA and the corresponding column densities.  The data for the other absorbers are presented in the same format in Appendix \ref{app:udata}. 

\begin{table*}
\begin{center}
\caption{UVES metal column densities for J0034+1639 (\zabs{}=$4.251$)}
\label{tab:UVESJ0034+1639,42507}
\begin{tabular}{lcccccccc}
\hline
Ion& Line& $\lambda$& $f$& $v_{\rm min}$& $v_{\rm max}$& logN(X)& Included& logN$_{\rm adopt}$\\
 &  & \AA{}&  & \kms{}& \kms{}&  &  &  \\
\hline
C\textsc{ii}& 1334& 1334.532& 1.28E-01& -50& 50& $>14.24$& Y& --\\
C\textsc{ii}& & & & & & & & $>14.24$\\
\hline
O\textsc{i}& 1302& 1302.168& 4.89E-02& -50& 50& $14.78\pm0.02$& Y& --\\
O\textsc{i}& & & & & & & & $14.78\pm0.05$\\
\hline
Si\textsc{ii}& 1260& 1260.422& 1.01E+00& -50& 50& $>13.68$& N& --\\
Si\textsc{ii}& 1304& 1304.370& 9.40E-02& -50& 40& $13.71\pm0.02$& Y& --\\
Si\textsc{ii}& 1526& 1526.707& 1.27E-01& -50& 50& $13.71\pm0.01$& Y& --\\
Si\textsc{ii}& & & & & & & & $13.71\pm0.05$\\
\hline
P\textsc{ii}& 1532& 1532.533& 7.61E-03& -80& 80& $<13.66$& Y& --\\
P\textsc{ii}& & & & & & & & $<13.66$\\
\hline
S\textsc{ii}& 1250& 1250.584& 5.43E-03& -80& 80& $<13.79$& N& --\\
S\textsc{ii}& 1253& 1253.811& 1.09E-02& -80& 80& $<13.64$& N& --\\
S\textsc{ii}& 1259& 1259.519& 1.66E-02& -80& 80& $13.65\pm0.08$& Y& --\\
S\textsc{ii}& & & & & & & & $13.65\pm0.08$\\
\hline
Fe\textsc{ii}& 1608& 1608.451& 5.80E-02& -50& 50& $13.23\pm0.06$& Y& --\\
Fe\textsc{ii}& 1611& 1611.200& 1.36E-03& -80& 80& $<14.56$& N& --\\
Fe\textsc{ii}& & & & & & & & $13.23\pm0.06$\\
\hline
Ni\textsc{ii}& 1709& 1709.604& 3.24E-02& -80& 80& $<13.33$& N& --\\
Ni\textsc{ii}& 1741& 1741.553& 4.27E-02& -50& 50& $<13.48$& N& --\\
Ni\textsc{ii}& 1751& 1751.916& 2.77E-02& -50& 50& $<13.24$& Y& --\\
Ni\textsc{ii}& & & & & & & & $<13.24$\\
\hline
\end{tabular}
\end{center}
\end{table*}

In general, we preferentially adopt the UVES column density as the higher resolution provides an accurate determination of the absorption profile and the data has a similar or higher signal-noise ratio. The one exception where we adopt the XQ-100 value is due to a lack of strong, uncontaminated Fe\textsc{ii} lines covered in the UVES spectrum required to derive an accurate column density. We note that the absorption in the  Ni\sion{} 1741 \AA{} panel of Figure \ref{fig:Uvesvelprof} is likely contaminated as it does not produce a consistent measurement with the derived column density limits for the Ni\sion{} 1709\AA{} and 1751\AA{} lines.

\subsection{XQ-100 sample properties}
\begin{figure}
\begin{center}
\includegraphics[width=0.5\textwidth]{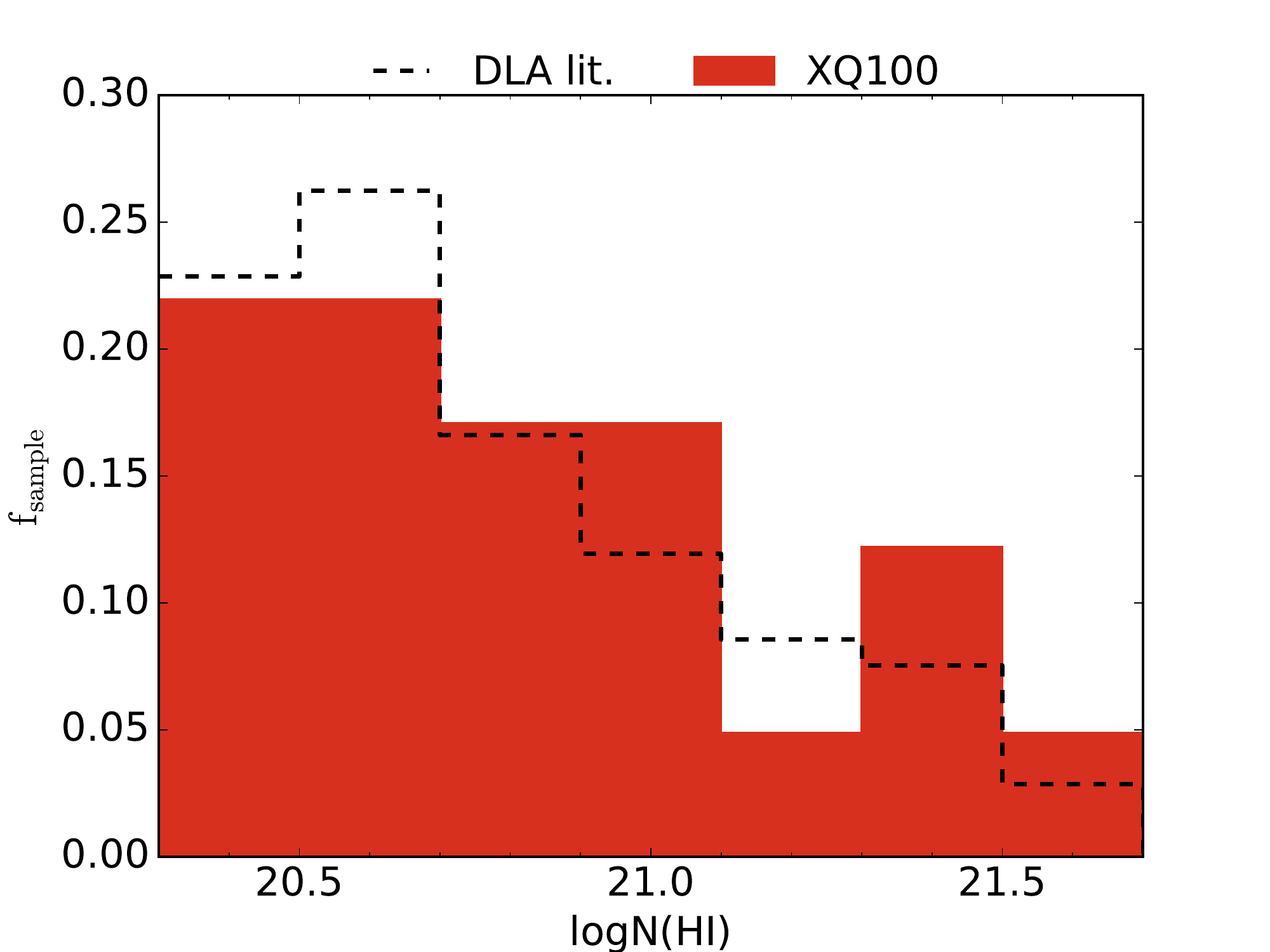}
\caption{The normalized logN(\HI{}) distribution of the XQ-100 (red bars) and literature DLA \citep[][black dashed line]{Berg15II} samples.}
\label{fig:HIdist}
\end{center}
\end{figure}

\begin{figure}
\begin{center}
\includegraphics[width=0.5\textwidth]{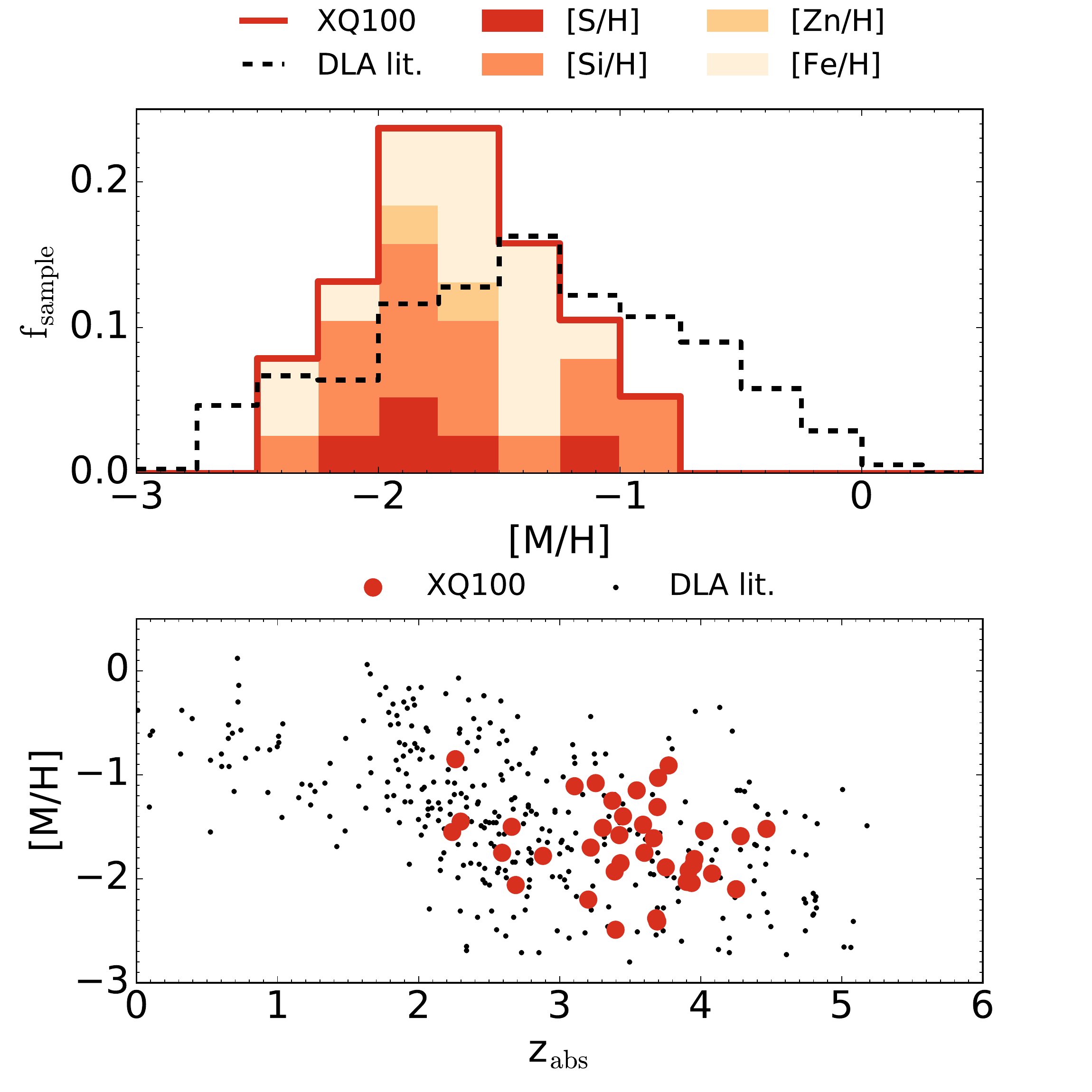}
\caption{\emph{Top Panel:}  The metallicity distribution of the XQ-100 DLAs. The solid red line outlines the total metallicity distribution of the XQ-100 sample, whereas the shaded regions within each bin show the distribution of elements used as the metallicity indicator for the DLAs. The metallicity distribution of the literature DLA sample is shown by the black dashed line. \emph{Bottom Panel:} The distribution of the XQ-100 (red circles)  and literature DLAs (black points) in metallicity-redshift space.}
\label{fig:zZ}
\end{center}
\end{figure}

To briefly highlight the properties of the 41 DLAs observed in the XQ-100 survey, Figures \ref{fig:HIdist} and \ref{fig:zZ} show the logN(\HI{}) and metallicity-redshift distribution (respectively) of the XQ-100 sample relative to the DLA literature sample compiled in \cite{Berg15II}. The metallicities ([M/H]; see Table \ref{tab:XQ100N}) were derived following a similar scheme outlined in \cite{Rafelski12}, where they choose (in order of decreasing preference) S, Si, Zn, and Fe as the metallicity tracer. They add a $+0.3$ dex correction to Fe to account for the difference in observed \alphafe{}. We adopt the \cite{Asplund09} solar scale for all metallicity calculations. The logN(\HI{}) distribution shows no significant deviation from what is probed by the literature DLAs; a  Kolmogorov-Smirnov (KS) test of the two distributions rejects the null hypothesis at 99.2\% confidence.

Figure \ref{fig:zZ} shows the metallicity distribution of the XQ-100 sample (solid red line; top panel) and the metallicity distribution as a function of redshift (bottom panel). Figure \ref{fig:zZ} emphasizes that the XQ-100 DLAs are predominantly between redshifts \zabs{}$=$3--4.5, and as a result appear more metal-poor than the literature \citep[which includes the lower redshift and thus higher metallicity DLAs;][]{Rafelski12}. A KS test rejects the null hypothesis that the metallicity distributions of the literature and XQ-100 DLA samples are drawn from the same distribution at 0.3\% confidence. However, when both the literature and XQ-100 samples are restricted to \zabs{}$=$3--4.5, the KS test rejects the null hypothesis at 90.6\% confidence.

Using the measured \zabs{} and \vninety{} for the XQ-100 DLAs, we compared the derived metallicities from the [M/H]-\zabs{}-\vninety{} relation in \cite{Neeleman13} to our directly-measured [M/H].  We find good agreement between the two metallicities (mean difference of 0.02 dex), with a root-mean-squared scatter between the two metallicities (0.32 dex) that is consistent with the scatter observed in \citet[0.37 dex]{Neeleman13}. 

\section{Discussion}
\subsection{The metal-poor DLA towards J0034+1639}
\label{sec:J0034}

Within the XQ-100 sample, we identified a MPDLA candidate towards J0034+1639 (\zabs{}$=4.2507$), for which we followed up with higher resolution observations with VLT/UVES  to confirm the derived metal column densities are free from contamination (see Section \ref{sec:Udata} for details on observations). The column densities for this MPDLA are presented in Table \ref{tab:UVESJ0034+1639,42507}. It is interesting to note that, although the [Fe/H] abundance ($-2.82\pm0.11$) is the third lowest observed to date in a DLA \citep[][and references therein]{Cooke15}, its metallicity using the \cite{Rafelski12} scheme is much higher ([M/H]$=-2.40\pm0.11$; 23\ts{rd} lowest). 

Table \ref{tab:MPDLA} shows the abundances for the metal-poor DLA towards J0034+1639, compared to the mean abundance derived from the metal-poor DLAs ($\langle$[X/O]$_{\rm MPDLA}\rangle$) from the MPDLA sample in \cite{Cooke11}. The abundance ratios (relative to O) of C, Si, and Fe are consistent\footnote{We suspect the [C/O] ratio is likely consistent with the typical MPDLA values, as the C\sion{} absorption is mildly saturated (see Figure \ref{fig:Uvesvelprof}).} with the typical MPDLA abundance pattern (although we note that [Si/O] is almost 0.2 dex larger than the typical MPDLA measurement), implying consistency with yields from low-energy supernovae \citep{Heger10,Cooke11}. Unfortunately, we are unable to place strong constraints on Ni and other Fe-peak abundances ([Ni/Fe]$<1.26$), which \citet[and references therein]{Cooke13} have demonstrated to be a key discriminator of the supernovae energy. 

\begin{table}
\begin{center}
\caption{J0034+1639 MPDLA (\zabs{}$=4.25$) abundances}
\label{tab:MPDLA}
\begin{tabular}{lccc}
\hline
Element & [X/H] & [X/O] & $\langle$[X/O]$_{\rm MPDLA}\rangle^{1}$ \\
\hline
O & $-2.51\pm0.12$ & -- & -- \\
C & $>-2.79$& $>-0.28$& $-0.28\pm0.12$ \\
Si & $-2.40\pm0.11$& $0.11\pm0.16$ & $-0.08\pm0.10$\\
Fe & $-2.82\pm0.11$& $-0.31\pm0.16$ & $-0.39\pm 0.12$\\
\hline
\end{tabular}
$^1$Mean MPDLA abundance from \cite{Cooke11}.\\
\end{center}
\end{table}

\subsection{Multiple DLAs}
\label{sec:MDLA}
Previous work by \cite{Lopez03} on three MDLA systems found a slight deficit of \alphafe{} relative to the typically observed DLA. They suggested that the low \alphafe{} is due to environmental effects truncating star formation. With the large increase in DLA abundance measurements over the past decade, the robustness of these results can be tested. Following \cite{Lopez03}, we identified MDLAs in the XQ-100 and in the \cite{Berg15II} literature DLA samples as systems of two or more absorbers within 500\kms{}$\leq \Delta v \leq 10,000$\kms{} of each other. \nxmdla{}  and \nlmdla{} MDLAs from the XQ-100  and the literature samples (respectively) were identified (Table \ref{tab:DLAs}). Using the measured \alphafe{} from literature and XQ-100 DLAs, we tested the potential enhancement by comparing each MDLA galaxy with a control-matched sample of intervening DLAs from the literature sample. The control matching technique accounts for the intrinsic evolution of similar DLAs, allowing a comparative test of different environments.

The control matching was completed for each MDLA by selecting all DLAs within the pool of literature DLAs with a redshifts and metallicity identical to the MDLA, within a prescribed tolerance. We impose that the DLA control pool 
are not classified as MDLAs or PDLAs ($\Delta v<5000$\kms{}). The matching tolerance for metallicity was set by the error in the MDLA's metallicity. The matching tolerance for redshift was adopted based on the known redshift evolution of mean metallicity \citep[{[M/H]}$\propto-0.2$ \zabs{};][]{Pettini97,Rafelski12,Rafelski14}. The error in metallicity provides a relative spread in redshifts for a DLA to have undergone a similar metal-enrichment history. Using the slope of the redshift-metallicity evolution, we calculate the redshift tolerance ($\delta$\zabs{}) using $\delta$\zabs{}$=\frac{\delta{\rm [M/H]}}{0.2}$, where $\delta{\rm [M/H]}$ is the error in the metallicity of the MDLA.

To ensure a representative control-matched sample, we required that each MDLA had at least five matched control DLAs. Each of the matched control DLAs must have at least five detections of both Si and Fe to compute \alphafe{}. If these criteria were not met within  the initial tolerances, the size of the two matching criteria are repeatedly increased in small increments until a sufficient number of control matched DLAs were obtained. The metallicity bin grew by $\pm0.025$ dex increments (15--25\% of the typical error in metallicity), while the redshift criteria grew by $\pm0.125$ ($\sim$20\% of the associated error due to metallicity evolution). However, expansions were stopped after 10 iterations to ensure that the control-matched sample still resembles the MDLA, effectively matching within $\sim3\sigma$ of \zabs{} and [M/H]. These criteria result in controls samples being identified for all MDLAs, with only three MDLAs requiring more than two expansions. The median number of matched control DLAs is 14 per MDLA, with 25\% of MDLAs having 19 or more (up to 37) control DLAs matched.

\begin{figure}
\begin{center}
\includegraphics[width=0.5\textwidth]{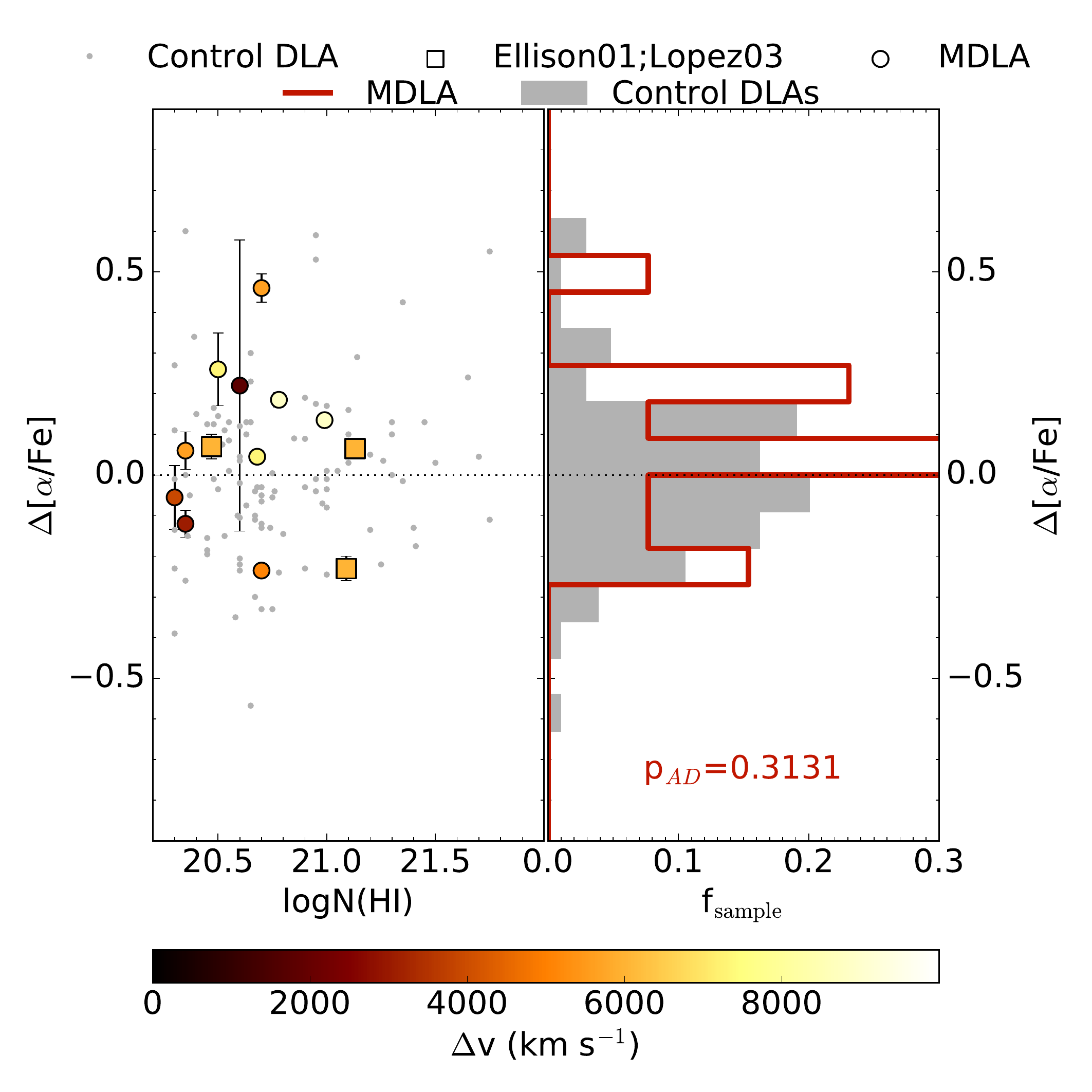}
\caption{The relative change in \alphafe{} for MDLAs ($\Delta$\alphafe{}) as a function of logN(\HI{}). The left panel show $\Delta$\alphafe{} in MDLA sightlines compared to the control-matched sample (small grey points). For reference, the MDLAs observed in \citet{Lopez03} are shown as large squares. The colour of the MDLA points represent the velocity separation of the DLAs along the sightline. The right panel shows the distribution of $\Delta$\alphafe{} for the MDLA (red) and control (grey) samples. The p-value from the Anderson-Darling test for the two MDLA samples relative to the control sample is displayed; suggesting that the MDLA and control DLAs are likely drawn from the same parent sample.}
\label{fig:MDLAalpha}
\end{center}
\end{figure}

The left panel of Figure \ref{fig:MDLAalpha} shows the relative change of \alphafe{} ($\Delta$\alphafe{}) in MDLAs compared to their control-matched counterparts, as a function of N(\HI{}). $\Delta$\alphafe{} is computed as the difference in [Si/Fe] between the MDLA absorber and median [Si/Fe] of the control-matched DLAs, such that a positive $\Delta$\alphafe{} implies that the MDLA has a higher \alphafe{} than its control sample. Note that, although all MDLAs are matched, only 14 of the \nmdla{} MDLAs have measured $\Delta$\alphafe{} due to lack of Si and Fe column densities. Although Fe is depleted onto dust, the relative comparison of \alphafe{} of MDLAs and control DLAs at a fixed metallicity should remove the effects of dust depletion, as the depletion of refractory elements appears to scale with metallicity in DLAs \citep[e.g.][]{Pettini97,Ledoux02,Prochaska02II,Berg15II}. For reference, the MDLAs from \cite{Lopez03} are shown as large squares, and are included in the analysis. The grey points are the measured $\Delta$\alphafe{} from repeating the matching procedure on all control DLAs. The errors on the MDLA points were derived from the spread in the \alphafe{} of the control sample using the jackknife technique; 

\begin{equation}
\sigma_{jack}=\sum\limits_{i=1}^{N}\sqrt{\frac{N-1}{N}(med(N)-med(N-1))^{2}}
\end{equation}

where N is the number of DLAs, $med(N)$ is the median of the entire sample, and $med(N-1)$ is the median of the sample with the i\ts{th} DLA removed. The colour of the MDLA points indicates the velocity separation of the MDLAs from their counterpart (Table \ref{tab:DLAs}).  The median  $\Delta$\alphafe{} (and $\sigma_{jack}$) for the MDLAs and control sample are $0.06\pm0.01$ dex and $-0.01\pm0.00$\footnote{$\sigma_{jack}$ is much smaller than 0.01 dex, and is thus quoted to be 0 dex.} dex (respectively).

The right panel of Figure \ref{fig:MDLAalpha} shows the distribution of $\Delta$\alphafe{} for both the MDLA and control samples. The p-value from the Anderson-Darling (AD) test (p$_{\rm AD}$)\footnote{We preferentially use the AD test over a KS test as the AD test is independent of the shape of the distribution, and is more sensitive to discrepancies in the tails of the distribution.} rejects the null hypothesis that the two samples are drawn from the same parent sample at $\sim31$\% confidence, suggesting MDLAs likely do not show any deficit (or enhancement) in \alphafe{} relative to the typical DLA.

We note that the \cite{Berg15II} literature sample includes surveys that have purposefully observed DLAs of specific properties and may have a biased representation of these properties, such as high or low metallicity. As a test of robustness of these results, we limited the control-matched samples to DLAs from the relatively unbiased \cite{Rafelski12} literature sample, and re-computed $\Delta$\alphafe{}. Only 6 of the \nmdla{} MDLAs were matched, and still showed no deficit in \alphafe{} (p$_{\rm AD}\sim0.86$).

We note that the upper velocity limit of MDLAs  defined in \citet[10000\kms{}]{Lopez03} is much larger than the typical galaxy cluster velocity dispersion \citep[values are typically smaller than 2000 \kms{};][]{Ruel14}. Reducing the definition of an MDLA to a separation of 2000 \kms{} limits the sample to one MDLA system, which shows a $\Delta$\alphafe{} of $\sim+0.2$dex. Unfortunately more MDLAs are required to further test if a smaller velocity separation does have an impact on the MDLA abundances.

\subsection{Proximate DLAs}
\label{sec:PDLA}

The XQ-100 DLA catalogue from \cite{SanchezRamirez16} showed that five DLAs are PDLAs ($\Delta v \leq 5000$ \kms{} from the host QSO). The properties of the complete set of associated absorbers in XQ-100, not just limited to DLAs, can be found in \cite{Perrotta16}. With the addition of \nlpdla{} PDLAs in the literature sample \citep{Berg15II}, we test the effect of proximity of a DLA to its host QSO for \npdla{} absorbers. Improving upon the analysis of 16 PDLAs in \cite{Ellison10}, we compare the relative abundance of a given PDLA to a control-matched sample of intervening DLAs. 

The control matching was undertaken in an identical manner as for the MDLAs, with the exception that DLAs were matched by logN(\HI{}) and redshift. In addition, we require the control sample to have at least five measured abundances for each of the following elements: S, Si, Fe, and Zn\footnote{The requirement of having five abundances for each element is the most restrictive criterion; but is required to ensure that DLAs are compared to the same control pool for each element. Without this criterion, all matching is completed within at most four expansions.}. The tolerance for the  logN(\HI{}) match was set by the error in the PDLA's logN(\HI{}). If required, the logN(\HI{}) selection tolerance grew by $\pm0.025$ dex increments (15--25\% of the typical error). The same criteria for redshift used for the MDLAs was imposed for the PDLAs as well. This matching procedure resulted in successful matches for all \npdla{} PDLAs, with 25 PDLAs being matched within three expansions, and six PDLAs requiring 9 expansions. Each PDLA had at least 12 control matches. The median number of matched controls is 35.5 per PDLA, with 25\% of PDLAs having 52 or more controls matches.

\begin{figure*}
\begin{center}
\includegraphics[width=\textwidth, height=0.8\textheight, keepaspectratio]{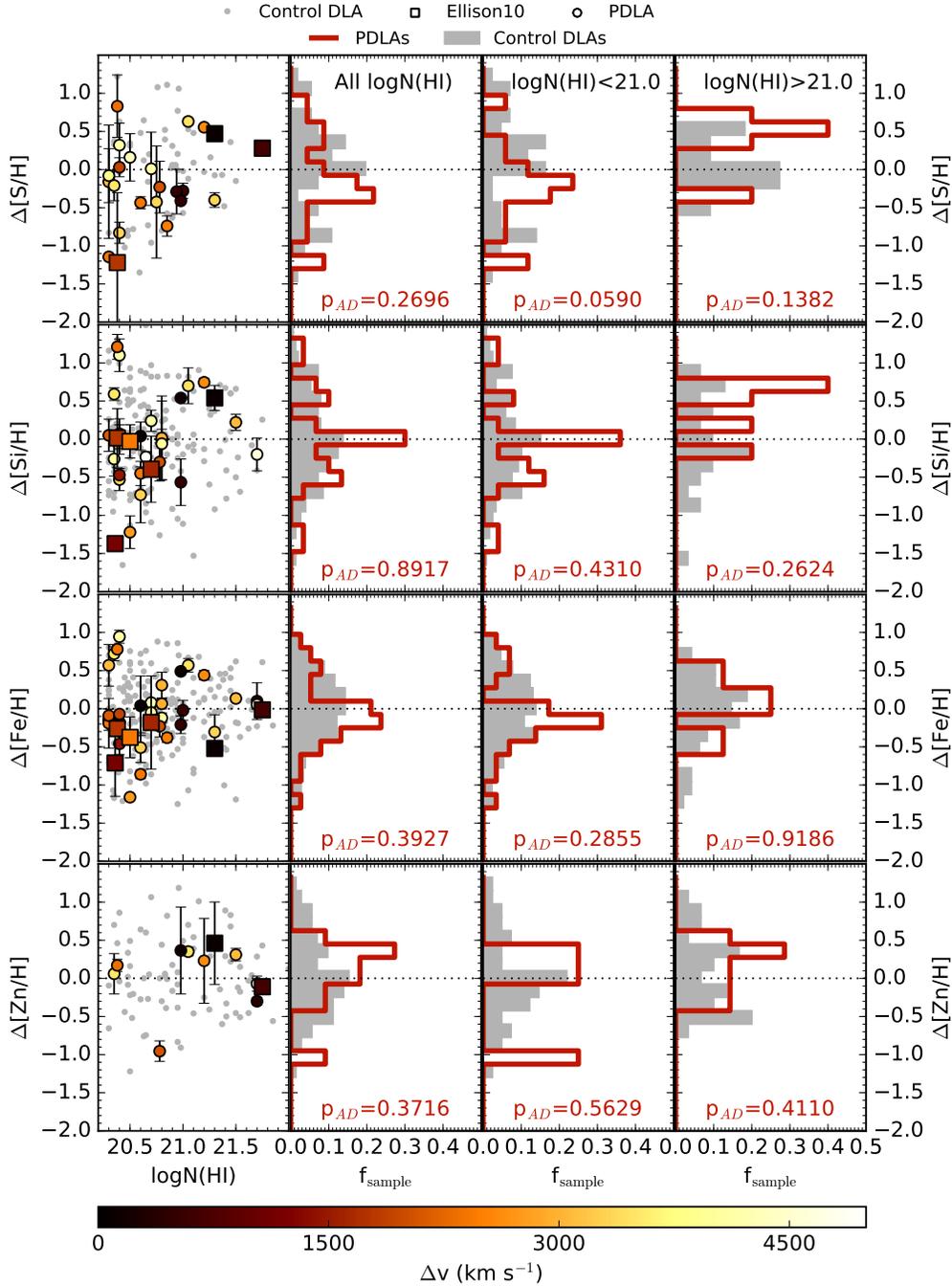}
\caption{\emph{First column:} The relative change in PDLA abundance compared to a sample of DLAs matched in redshift and logN(\HI{}) ($\Delta$[X/H]) as a function of logN(\HI{}). Each row is $\Delta$[X/H] for S, Si, Fe, and Zn (top to bottom). The notation is the same as Figure \ref{fig:MDLAalpha}. In all panels, we highlight the data from \citet{Ellison10} as large squares for comparison. \emph{Second column:} The fractional distribution of $\Delta$[X/H] for each element in PDLAs (red line) and the control sample (grey shaded region).  \emph{Third column:} The fractional distribution of $\Delta$[X/H] for each element, restricting both samples to DLAs with only logN(\HI{})$<21.0$. \emph{Fourth column:} The fractional distribution of $\Delta$[X/H] for each element, restricting both samples to DLAs with logN(\HI{})$>21.0$. All PDLA distributions are compared to the control-sample distribution using the Anderson-Darling test, showing the p-value (p$_{\rm AD}$) that the two distributions are drawn from the same parent sample.}
\label{fig:PDLAX}
\end{center}
\end{figure*}

The main science result from the control-matching analysis for PDLAs is presented in Figure \ref{fig:PDLAX}. We define $\Delta$[X/H] as the difference in the PDLA abundance [X/H] relative to the median abundance of control-matched sample; such that a negative $\Delta$[X/H] would imply that the PDLA has a lower [X/H] than the control-matched sample.  The left-most panels of Figure \ref{fig:PDLAX} show $\Delta$[X/H] as a function of logN(\HI{}) for elements (in order from top to bottom rows) S, Si, Fe, and Zn. The error bars represent $\sigma_{jack}$ of the control-matched sample. The colour of the points indicate the velocity separation ($\Delta$v; \kms{}) from the host QSO. PDLAs from \cite{Ellison10} are shown as large squares, and are included in the analysis. For comparison to the DLAs in the control pool, the smaller grey points are the $\Delta$[X/H] obtained from repeating the control match on all control DLAs.  The left panels of Figure \ref{fig:PDLAX} visually hint at a possible deficit of [X/H] for PDLAs with logN(\HI{})$\lesssim 21.0$, and an enhanced [X/H] at logN(\HI{})$\gtrsim21.0$ \citep[the enhancement was first identified for {[S/H]} and {[Si/H]} in][]{Ellison10}.

To quantitatively test the possibility of a correlation between $\Delta$[X/H] and logN(\HI{}) in PDLAs, we performed the Pearson-r test. However, the test did not show any significant correlation of $\Delta$[X/H] with logN(\HI{}) ($r=$0.33, 0.18, 0.08, 0.00 for S, Si, Fe, and Zn; respectively). Despite the lack of correlation, the AD statistic will additionally constrain whether the PDLAs are drawn from a different distribution of $\Delta$[X/H] compared to the intervening control DLAs. The AD test was performed in three regimes: all logN(\HI{}), logN(\HI{})$<21.0$ , and logN(\HI{})$>21.0$. The distributions of $\Delta$[X/H] (along with the associated p$_{\rm AD}$) for each regime is displayed in the second, third, and fourth column of Figure \ref{fig:PDLAX} (respectively). The p$_{\rm AD}$ measurements suggest that PDLAs are likely drawn from a different distribution than the control DLAs for [S/H] (null hypothesis rejected at $\sim6$\% confidence) and [Fe/H] ($\sim32$\% confidence) for logN(\HI{})$<21.0$, as well as [S/H]  ($\sim14$\% confidence) and [Si/H] ($\sim26$\% confidence) for logN(\HI{})$>21.0$.  

As with the MDLAs, we attempted to assess the potential biases of using the \cite{Berg15II} literature DLAs by only adopting control DLAs from the \cite{Rafelski12} sample. Unfortunately only two PDLAs were matched at logN(\HI{})$>21.0$, and are unable to test how robust the results are. Similarly for logN(\HI{})$<21.0$, only seven PDLAs had measured $\Delta$[S/H], six of which had $\Delta$[S/H]$\leq0$. Again, we note that by matching logN(\HI{}) should remove any effects caused by \HI{} selection of control sample DLAs; whilst taking the median value of the control sample should remove any outlaying DLAs. A larger unbiased sample of DLAs is required to further evaluate the robustness of our results.

\begin{figure*}
\begin{center}
\includegraphics[width=\textwidth, height=0.8\textheight, keepaspectratio]{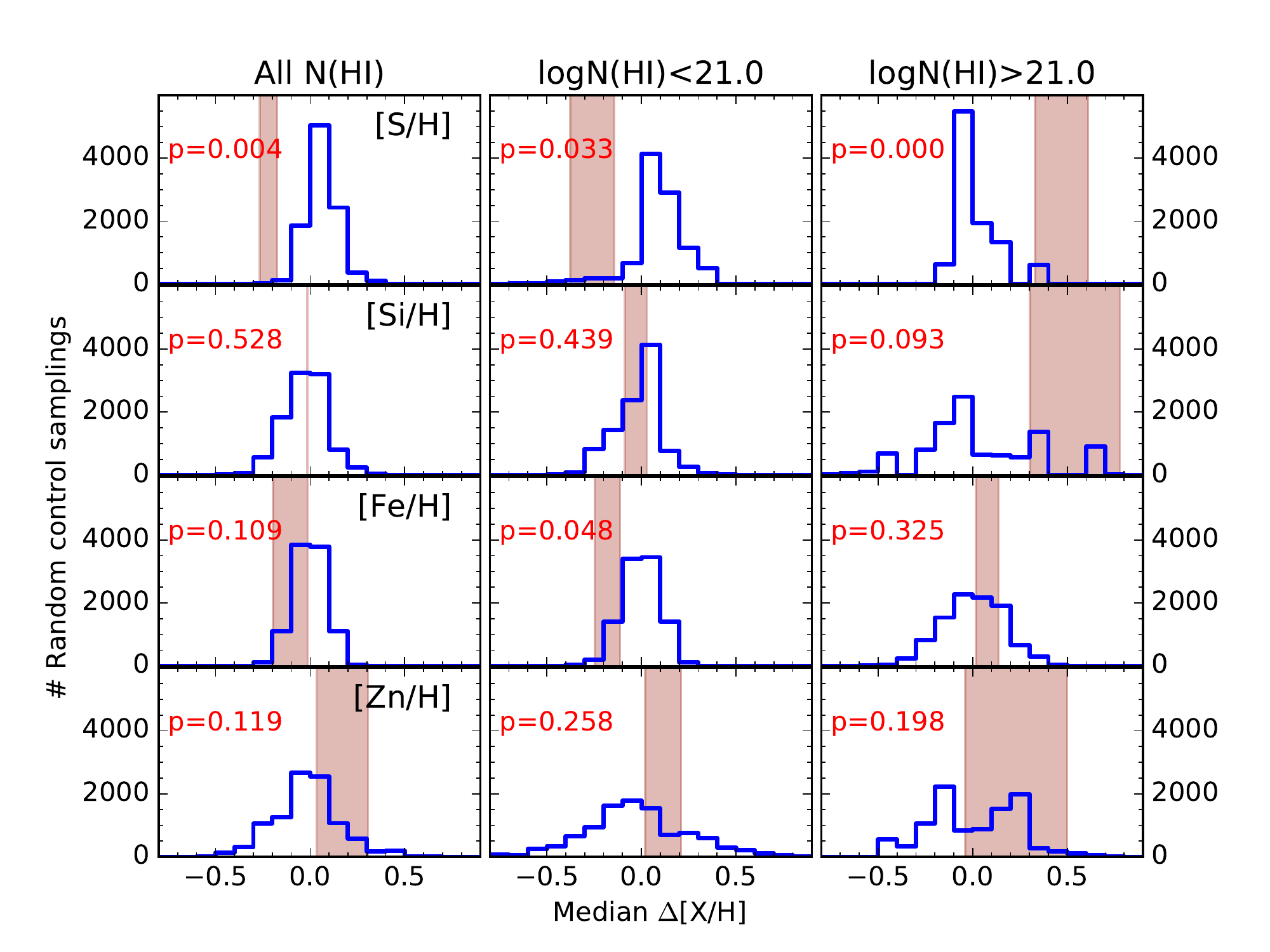}
\caption{The frequency (of 10 000 iterations) of observing a median $\Delta$[X/H] in a randomly-selected control sample of the same size as the PDLA sample. Each row is for the elements S, Si, Fe, and Zn (top to bottom; respectively), while each column is the logN(\HI{}) cut on the sample (all N(\HI{}), logN(\HI{})$<21.0$, and logN(\HI{})$>21.0$; left to right, respectively). The red band indicates the median $\Delta$[X/H] of the PDLA sample, whose half-width is given by $\sigma_{jack}$ of the PDLA $\Delta$[X/H]. The probability (p) of observing a median $\Delta$[X/H] of the same magnitude (or further offset from 0) as the PDLA median $\Delta$[X/H] is displayed.}
\label{fig:MCsamp}
\end{center}
\end{figure*}

To test whether the $\Delta$[X/H] offsets seen in PDLAs could occur by chance in a sample of this modest size, we ran a Monte Carlo simulation with 10 000 iterations, drawing a random sample of control DLAs\footnote{The randomly generated sample in each iteration does not contain the same DLA more than once; however the same DLA can appear in the samples of other iterations.} of the same size as the PDLAs sample for each element and logN(\HI{}) cut. At each iteration, we computed the obtained median $\Delta$[X/H] for each logN(\HI{}) criteria. The distribution of obtained medians is shown in Figure \ref{fig:MCsamp}. To understand the likelihood of observing such an offset by chance, we calculated the frequency of observing a median $\Delta$[X/H] that is identical or further offset from zero than the median $\Delta$[X/H] observed for the PDLAs. Table \ref{tab:MCsamp} provides the median $\Delta$[X/H] of the PDLA sample, and summarizes the frequencies of this simulation for all four elements\footnote{To estimate the uncertainty on the simulation, we recomputed the frequencies of observing a $\Delta$[X/H], taking into account the $\sigma_{jack}$ of the PDLA $\Delta$[X/H] to make the offset weaker. These recomputed frequencies are provided in parentheses in Table \ref{tab:MCsamp}.}. The deficit of [S/H] and [Fe/H] for logN(\HI{})$<21.0$ was seen 3.2\% and 5.4\% of the time (respectively), suggesting it is unlikely the deficit in the observations is observed by random chance. The excess of [S/H] at logN(\HI{})$>21.0$ is also unlikely to be caused by random chance, since \emph{none} of the iterations demonstrated a similar median of $\Delta$[S/H]. The excess of [Si/H] is seen slightly frequently 9.7\% of the time. Therefore there is observational evidence that PDLAs exhibit different abundances than intervening DLAs.

\begin{table*}
\begin{center}
\caption{Frequencies of obtaining median PDLA $\Delta$[X/H] from control DLA resampling.}
\label{tab:MCsamp}
\begin{tabular}{lcccccc}
\hline
& \multicolumn{2}{c}{All N(\HI{})}& \multicolumn{2}{c}{logN(\HI{})$<21.0$}& \multicolumn{2}{c}{logN(\HI{})$>21.0$}\\
X& Median {$\Delta$[X/H]}& Freq. (\%)& Median {$\Delta$[X/H]}& Freq. (\%)& Median {$\Delta$[X/H]}& Freq. (\%)\\
\hline
{[S/H]}& $-0.22\pm0.05$& 0.4 (0.5)$^{1}$& $-0.26\pm0.12$& 3.3 (5.8)$^{1}$& $0.47\pm0.14$& 0.0 (6.2)$^{1}$\\
{[Si/H]}& $-0.02\pm0.00$& 52.8 (53.0)$^{1}$& $-0.03\pm0.06$& 43.9 (62.4)$^{1}$& $0.54\pm0.24$& 9.3 (23.1)$^{1}$\\
{[Fe/H]}& $-0.10\pm0.09$& 10.9 (43.6)$^{1}$& $-0.18\pm0.07$& 4.8 (14.0)$^{1}$& $0.08\pm0.06$& 32.5 (47.6)$^{1}$\\
{[Zn/H]}& $0.17\pm0.14$& 11.9 (38.4)$^{1}$& $0.12\pm0.10$& 25.8 (39.6)$^{1}$& $0.23\pm0.27$& 19.8 (58.5)$^{1}$\\
\hline
\end{tabular}
\\ $^{1}$Frequency including PDLA $\sigma_{jack}$ in median $\Delta$[X/H].
\end{center}
\end{table*}

We now consider the possible source of distinct abundances between PDLAs and intervening systems. The deficit of S and Fe in PDLA abundances at low logN(\HI{}) might arise from ionization effects as the hard ionizing spectrum of the quasar is more likely to penetrate the self-shielding effect of DLAs at low logN(\HI{}). A natural assumption might be that PDLAs closer to the QSO \citep[$\Delta v \leq 1500$\kms{}][]{Ellison10} may be more influenced by the higher radiation field. The mean and scatter in $\Delta$[S/H] and $\Delta$[Si/H] for logN(\HI{})$<21.0$ is the same across all $\Delta v$, suggesting that the velocity separation of a PDLA has little effect.

It is worth noting that the uncertainty in QSO's systemic \zem{} can be large \citep[depending on which emission lines are used, e.g.][]{Gaskell82,Tytler92}, leading to a scatter in $\Delta$v determinations from tens to a couple of hundred \kms{} \citep[for QSOs identified with the Sloan Digital Sky Survey; see][]{Ellison10,Hewett10,Paris11}. In addition, the error in \zabs{} for the DLAs will also contribute to the precision of the $\Delta$v measurements. For the XQ-100 sample, the scatter in \zabs{} derived from \HI{} absorption (Table \ref{tab:DLAs}) and the metal lines is $\pm47$ \kms{}. However, the lack of a strong correlation between $\Delta$[X/H] and $\Delta v$ suggests that higher precision measurements  $\Delta v$ are not required.

To further test ionization effects at logN(\HI{})$<21.0$, the observed $\Delta$[X/H] were compared to abundance discrepancies observed in QSO-like radiation models from the literature. Both \cite{Vladilo01} and \cite{Rix07} modelled the effect of radiation fields (for a variety of ionization strengths $U$) on abundances in DLAs; with their ionization abundance corrections  shown in Table \ref{tab:ioncorr} (where positive corrections imply the observed abundances are underestimated relative to the true value). As a reference, the soft stellar-like ionizing spectrum (denoted by $^{\rm S}$ in Table \ref{tab:ioncorr}) from \cite{Vladilo01} is also included. For a hard, QSO-like ionizing field (flagged by $^{\rm H}$ in Table \ref{tab:ioncorr}); \cite{Vladilo01} found that [Fe/H] is \emph{over-predicted} by a factor of  0.01--0.04 dex; whereas we are finding that [Fe/H] is \emph{underestimated} in PDLAs. The corrections for the single PDLA modelled by \cite{Rix07} are qualitatively similar. Both sets of models suggest that the deficit of [Fe/H] in PDLAs \emph{cannot} be explained by ionization; and we rule out ionization as a possible explanation. However, the deficit for [S/H] could be explained  by ionization corrections for logN(\HI{})$<21.0$; but not for the excess seen at logN(\HI{})$>21.0$ \citep{Ellison10}.

\begin{table*}
\begin{center}
\caption{Literature PDLA ionization corrections}
\label{tab:ioncorr}
\begin{tabular}{l|ccc|ccc|cc}
\hline
 & \multicolumn{3}{c|}{logN(\HI{})$=20.35^{1}$} & \multicolumn{3}{c|}{logN(\HI{})$=20.75^{1}$} & \multicolumn{2}{c}{logN(\HI{})$=20.8^{2}$}\\
 {[X/H]}& $U=-4.2^{\rm H}$ & $U=-2.2^{\rm S}$ & $U=-1.7^{\rm S}$ & $U=-4.8^{\rm H}$ & $U=-2.2^{\rm S}$ & $U=-1.7^{\rm S}$ & $U=-4.0^{\rm H}$ & $U=-2.0^{\rm H}$\\
  & (dex) & (dex) & (dex) & (dex) & (dex) & (dex) & (dex) & (dex)\\
\hline
{[Si/H]} & $-0.04$ & $-0.09$ & $-0.16$ & $-0.01$ & $-0.03$ & $-0.07$ &$-0.02$ &$-0.09$\\
{[S/H]} & $+0.15$ & $-0.20$ & $-0.15$ & $+0.09$ & $-0.07$&$-0.09$ &$+0.03$ & $+0.26$ \\
{[Fe/H]} & $-0.04$ & $-0.02$ &$-0.02$ &$-0.01$ &$-0.01$ &$-0.01$ &$-0.01$ &$-0.05$ \\
{[Zn/H]} & $+0.32$ & $-0.35$ & $-0.73$ & $+0.16$ & $-0.17$ & $-0.44$ & $+0.15$ & $+0.58$ \\
\hline
\end{tabular}
\\$^{1}$\cite{Vladilo01}.\\
$^{2}$\cite{Rix07}. \\
$^{\rm S}$Soft, stellar-type ionization spectrum.\\
$^{\rm H}$Hard, QSO-like ionization spectrum.\\
\end{center}
\end{table*}

A suggestion in \cite{Ellison10} is that PDLAs are typically probing more massive galaxies (relative to intervening systems). To test this explanation for the metallicity effect we are seeing, we checked whether the \vninety{} parameter \citep[a proxy for mass in DLAs;][]{Prochaska97,Ledoux06,Moller13,Neeleman13} is substantially larger in PDLAs than in the control sample. In addition to the \vninety measurements  of the XQ-100 sample (Table \ref{tab:v90}), we also adopt the 139 \vninety{} measurements from HIRES data for DLAs in the literature \citep[][]{Neeleman13,Berg15} for the control-matched sample.

\begin{figure}
\begin{center}
\includegraphics[width=0.5\textwidth]{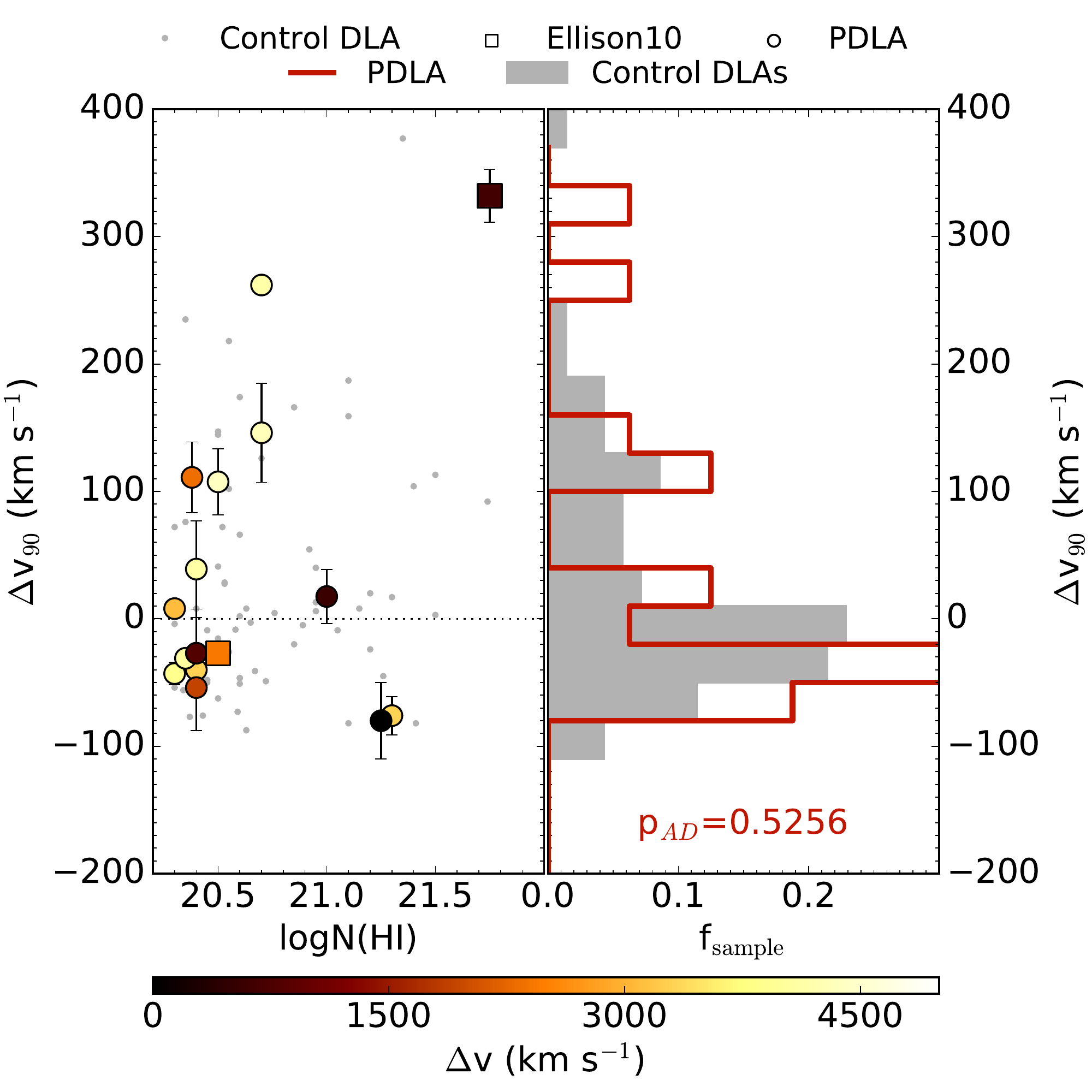}
\caption{\emph{Left Panel:} The difference in \vninety{} between a given PDLA and  control-matched sample ($\Delta$\vninety{}), as a function of log(N\HI{}). The notation is the same as Figure \ref{fig:MDLAalpha}. \emph{Right Panel:} distributions of $\Delta$\vninety{} for the PDLA sample (red line) compared to the control sample (grey bars). The Anderson-Darling p-value is shown for the comparison of the two distributions.}
\label{fig:PDLAv}
\end{center}
\end{figure}

Figure \ref{fig:PDLAv} shows the difference in \vninety{} for a PDLA relative to its control matched sample ($\Delta$\vninety{}; a positive value implying the PDLA has a large \vninety{} than the matched controls). Again, the smaller grey points show the scatter within the control sample. There is no significant discrepancy between the \vninety{} in PDLAs relative to the control, suggesting no dependence on mass; assuming \vninety{} is a good proxy for mass.

\subsection{{[Zn/Fe]} and dust depletion}
\label{sec:dust}

Dust depletion plagues DLA abundance analysis work, making it difficult to disentangle nucleosynthetic patterns from dust depletion effects. In particular, a dust-free Fe peak tracer is necessary to measure the intrinsic [$\alpha$/Fe] ratio, an important tracer of star formation history \citep{Tinsley79,Venn04,Tolstoy09}. In general, the community has either used elements that are relatively unaffected by depletion \citep[such as S and Zn; e.g.][]{Pettini97,Centurion00,Vladilo11} or modelled dust depletions based on chemical evolution assumptions \citep[e.g.][]{Vladilo11}.  One of the most prominent assumptions is that Zn is a dust-free proxy of the Fe peak \citep[][]{Pettini94,Pettini97,Vladilo02a, Vladilo02b}.   The assumption that Zn traces Fe was motivated by early observations of Galactic stellar abundances showing [Zn/Fe]$\sim0$ over a range of metallicities \citep[][]{Sneden88,Sneden91,Chen04,Nissen04}. 

However, with the search for metal-poor stars extending to lower metallicities \citep[e.g.][]{Prochaska00disk,Nissen07}, stellar [Zn/Fe] showed deviations from solar suggesting that Zn does not necessarily trace Fe; leading some authors to question whether Zn traces Fe in DLAs \citep[such as][]{DLAcat22,Nissen11,Rafelski12}. Recently, \cite{Berg15II} compared [Zn/Fe] in stars in the Milky Way and local dwarf spheroidal (dSph) galaxies, finding a large scatter of [Zn/Fe] at $-2.0<$[Zn/H]$<-0.5$. In particular, many (but not all) Local Group dwarf galaxy stars exhibit subsolar [Zn/Fe] indicating that Zn is not necessarily a nucleosynthetic tracer of the Fe-peak elements at [Zn/H]$\geq-1.5$. In order to investigate the possibility of intrinsically non-solar [Zn/Fe] ratios, \cite{Berg15II} infer the dust-corrected values of [Zn/Fe] using a novel method. Rather than using modelled corrections which require assumptions about the nature of DLAs \citep[such as][]{Vladilo11}, \cite{Berg15II} assume that all $\alpha$-elements trace each other in DLAs, and that the measured ratio of two $\alpha$-elements may be a tracer of dust depletion. Using the relative depletions seen towards Galactic ISM sightlines \citep{Savage96}, they suggest that [Si/Ti] should have the same depletion correction as [Zn/Fe], and can be used to correct the [Zn/Fe] abundance for depletion to within 0.1 dex.

Although the \cite{Berg15II} analysis of dust corrected [Zn/Fe] is strongly suggestive of sub-solar ratios in DLAs, their analysis was limited to the 43 systems (of which only 10 were detections) for which Ti measurements are available.  This limitation is driven by typically only covering the weak Ti\sion{} 1910 \AA{} line in optical spectra for the majority of the DLA sample. Previous surveys of Mg\sion{} abundances with high resolution spectroscopy have been limited to \zabs{}$\sim$2--3 DLAs \citep[e.g.][]{Prochaska03ApJS147,DZavadsky06,Prochaska07,DLAcat70}. With the additional IR coverage that X-Shooter provides, we have unprecedented access to absorption lines of $\alpha$-elements that are not typically observed in the optical; particularly in absorbers at redshifts 3--4. These lines include: Mg, Ca, as well as stronger Ti lines (relative to Ti\sion{} 1910\AA{}; Ti\sion{} $\lambda$ 3073\AA{}, 3242\AA{}, and 3384\AA{}).

\begin{figure*}
\begin{center}
\includegraphics[width=\textwidth]{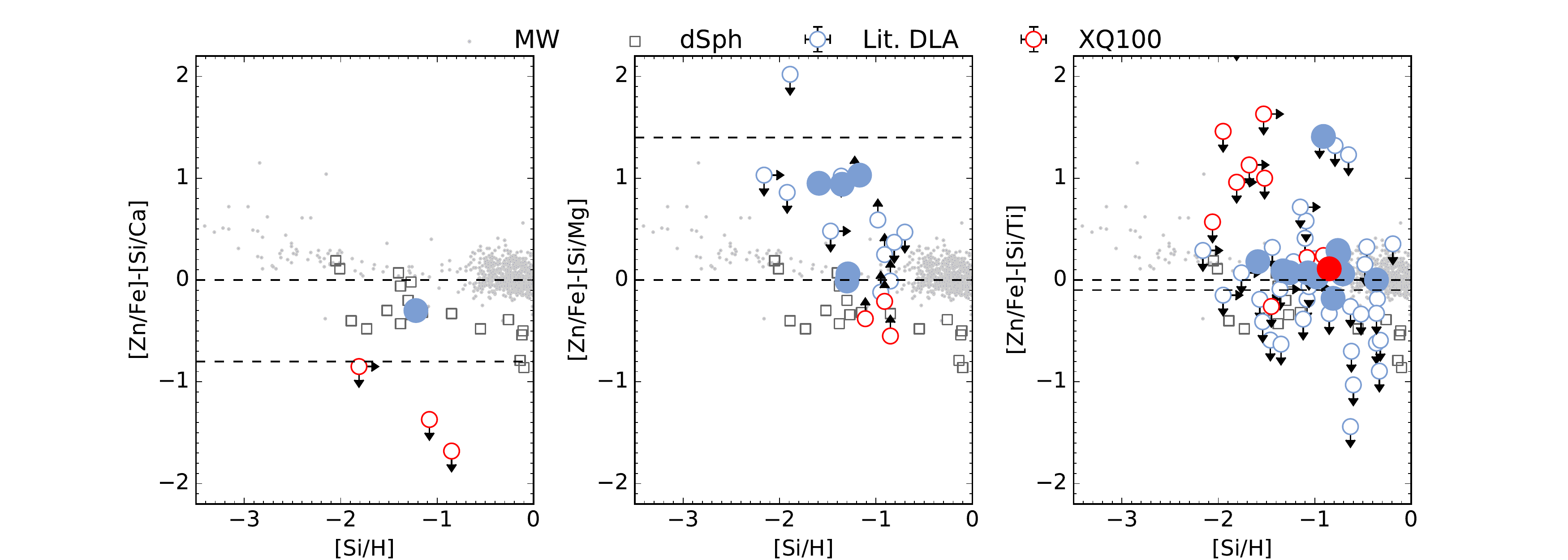}
\caption{Dust-corrected [Zn/Fe] DLA abundances (XQ-100 are in red, literature DLAs in blue) relative to stellar [Zn/Fe] (grey points are the Milky Way, dark grey squares are dSphs) from \citet{Berg15II}. Larger, filled DLA circles represent detections, whereas the unfilled circles are limits. The dust correction is determined by subtracting the ratio of two $\alpha$-elements of different relative dust depletions ([Si/Ca] left panel; [Si/Mg] middle panel; [Si/Ti] right panel). The two dashed lines bound the region where [Zn/Fe] may be consistent with solar, such that measurements below the line are consistent with subsolar [Zn/Fe]. In addition to [Si/Ti], [Si/Ca] can also constrain the intrinsic [Zn/Fe] in DLAs. Both of these ratios indicate that DLAs are consistent with the subsolar values seen in dSphs \citep[][]{Berg15II}. This subsolar [Zn/Fe] is likely nucleosynthetic in origin. }
\label{fig:dustcorr}
\end{center}
\end{figure*}

With the XQ-100 DLA sample we are thus able to further test the intrinsic [Zn/Fe] ratio of DLAs, by using the new Ti\sion{} abundances, as well as other $\alpha$-elements  ratios ([Mg/Si] and [Ca/Si]) as additional dust-depletion proxies. Figure \ref{fig:dustcorr} shows the DLA dust-corrected [Zn/Fe] abundances (coloured circles) as a function of metallicity using [Si/Ca] (left panel), [Si/Mg] (middle panel), and [Si/Ti] (right panel). For reference, the Galactic (grey points) and dSph (grey squares) stellar trends of [Zn/Fe] are shown in each panel. 

The difficulty of using Mg\sion{} and Ca\sion{} is that the differential depletion of [Si/Ca] and [Si/Mg] is much different than [Zn/Fe] \citep[differences of $+0.8$ and $-1.4$ dex in Galactic ISM sightlines, respectively][]{Savage96}. The two horizontal dashed lines in each panel constrain the region where the dust-corrected [Zn/Fe] could be consistent with solar [Zn/Fe] based on the differential depletion of [$\alpha$/Si]$-$[Zn/Fe] in the Milky Way; where DLA points below the lowest line are consistent with intrinsically subsolar [Zn/Fe], and DLA above the higher line likely have supersolar [Zn/Fe].

The addition of Ti abundances from the XQ-100 DLA sample (red circles) does not provide any additional information on the intrinsic nature of [Zn/Fe] in DLAs previously identified by \cite{Berg15II} (blue circles). One of the XQ-100 DLAs has a dust-corrected [Zn/Fe]$\sim0.3$ dex. Although supersolar, this one system is consistent with both Milky Way and dSphs stellar [Zn/Fe] abundances \citep{Berg15II}. Although the Ti\sion{} lines accessible in the NIR are nominally stronger than those in the optical, the typical detection limits of Ti\sion{} from the lower resolution X-Shooter spectra are not as sensitive as the measurements available from UVES or HIRES. It is worth noting that of the 38 cases Ti\sion{} abundances were measured in the XQ-100 DLAs, 28 abundances were preferentially adopted based on the Ti\sion{} lines in the NIR arm rather than the Ti\sion{} $\lambda$ 1910\AA{} line. Therefore the use of X-Shooter to observe lines in the NIR can be a useful tool to obtain (more constraining) Ti\sion{} abundances. 

The additional five Ca\sion{} robust upper limits (to the single detection in the literature) in the left panel of Figure \ref{fig:dustcorr} further support that DLAs span the same range of [Zn/Fe] as dSphs seen by \cite{Berg15II}, suggesting that DLAs share a similar nucleosynthetic history as Galactic dSphs. However, we note that Ca\sion{}  (excitation potential of 11.87eV) is not the dominant ionization state of Ca in DLAs \citep[e.g.][]{Wild06}, suggesting that the measured N(Ca\sion{}) may require significant ionization corrections.

\section{Summary and conclusions}
The sample of XQ-100 DLAs provides coverage of the relatively moderately-sampled redshift range \zabs{}= 3--4. We have computed the column densities for a variety of metals in the 41 DLAs in the XQ-100 sample. The additional coverage from the NIR arm of X-Shooter provides coverage of rarely detected lines at redshifts 3--4 in abundance studies such as Mg\sion{}, Ca\sion{}, and strong Ti\sion{} lines. With the addition of dust-depleted $\alpha$-elements, we are able to test the dust-corrected [Zn/Fe] to see if [Zn/Fe] is solar in DLAs. We have shown in Section \ref{sec:dust} that [Zn/Fe] is not necessarily solar in DLAs, and that [Zn/Fe] shows the same range of values as seen in the dSphs of the Local Group \citep[in agreement with][]{Berg15II}.

In combination with a sample of DLAs drawn from the literature, we have provided a statistical analysis of PDLAs (within $\Delta v\leq 5000$ \kms{} of the host QSO) and MDLAs (two or more DLAs separated by $500\leq \Delta v \leq 10000$ \kms{}) by comparing to a control-matched sample of individual, intervening absorbers. We do not find any suppression in \alphafe{} in MDLAs, suggesting that there is no evidence for truncated star formation between nearby DLAs on their abundance. Relative to a control sample of DLAs, we note a mildly elevated [S/H] and [Si/H] for high logN(\HI{})$>21.0$ PDLAs at (AD test rejects the null hypothesis at 14\% and 26\% confidence; respectively), as previously seen by \cite{Ellison10}; however, we also detect a deficit in [S/H] and [Fe/H] (null hypothesis rejected at 6\% and 32\% confidence, respectively) for PDLAs with logN(\HI{})$<21.0$. These abundance discrepancies appear to be independent of velocity separation of the host QSO and the mass proxy \vninety{}. It is possible to explain the deficit of [S/H] at low logN(\HI{}) through ionization corrections, but not the deficit of [Fe/H].

We have also presented UVES observations of three DLAs towards J0034+1639 in order to investigate an MPDLA candidate at \zabs{}$\sim4.25$ with a [Fe/H]$=-2.82\pm0.11$. This MPDLA is consistent with abundances in the typical MPDLA \citep{Cooke11}. MPDLAs such as the one towards J0034+1639 prime targets for easily detecting Ni and other discriminating elements in future follow-up observations with 30-m class telescopes.

\section*{Acknowledgements}
We thank the anonymous referee for their useful comments that clarified the text. TAMB is grateful for Ryan Cooke's assistance with developing a science case for the UVES observations of the MPDLA, and discussing the results. SLE acknowledges the receipt of an NSERC Discovery Grant which supported this research. JXP is supported by NSF grant AST-1109447. SL has been supported by FONDECYT grant number 1140838 and partially by PFB-06 CATA. KDD is supported by an NSF AAPF fellowship awarded under
NSF grant AST-1302093.
\bibliography{bibref}
\appendix
\section{Velocity Profiles and AODM tables}
\subsection{XQ-100 sample}
\label{app:xdata}
\clearpage
\begin{table*}
\begin{center}
\caption{X-Shooter metal column densities for J0006$-$6208 (\zabs{}=$3.203$)}
\label{tab:J0006-6208,32030}
\begin{tabular}{lcccccccc}
\hline
Ion& Line& $\lambda$& $f$& $v_{\rm min}$& $v_{\rm max}$& logN(X)& Included& logN$_{\rm adopt}$\\
 &  & \AA{}&  & \kms{}& \kms{}&  &  &  \\
\hline
Mg\textsc{i}& 2852& 2852.964& 1.81E+00& -60& 60& $<12.39$& Y& --\\
Mg\textsc{i}& & & & & & & & $<12.39$\\
\hline
Mg\textsc{ii}& 2796& 2796.352& 6.12E-01& -90& 70& $13.33\pm0.00$& Y& --\\
Mg\textsc{ii}& 2803& 2803.531& 3.05E-01& -90& 70& $<13.63$& N& --\\
Mg\textsc{ii}& & & & & & & & $13.33\pm0.05$\\
\hline
Ca\textsc{ii}& 3934& 3934.777& 6.50E-01& -40& 40& $<12.38$& Y& --\\
Ca\textsc{ii}& 3969& 3969.591& 3.22E-01& -60& 60& $<12.14$& N& --\\
Ca\textsc{ii}& & & & & & & & $<12.38$\\
\hline
Ti\textsc{ii}& 1910& 1910.750& 2.02E-01& -40& 40& $<13.02$& Y& --\\
Ti\textsc{ii}& & & & & & & & $<13.02$\\
\hline
Cr\textsc{ii}& 2056& 2056.254& 1.05E-01& -50& 50& $<13.08$& Y& --\\
Cr\textsc{ii}& 2062& 2062.234& 7.80E-02& -50& 50& $<13.45$& N& --\\
Cr\textsc{ii}& 2066& 2066.161& 5.15E-02& -50& 50& $<13.61$& N& --\\
Cr\textsc{ii}& & & & & & & & $<13.08$\\
\hline
Fe\textsc{ii}& 1608& 1608.451& 5.80E-02& -60& 60& $13.85\pm0.03$& Y& --\\
Fe\textsc{ii}& 1611& 1611.200& 1.36E-03& -60& 60& $<14.99$& N& --\\
Fe\textsc{ii}& 2586& 2586.650& 6.91E-02& -70& 80& $13.74\pm0.00$& Y& --\\
Fe\textsc{ii}& 2600& 2600.173& 2.39E-01& -70& 80& $>13.67$& Y& --\\
Fe\textsc{ii}& & & & & & & & $13.74\pm0.05$\\
\hline
Ni\textsc{ii}& 1709& 1709.604& 3.24E-02& -20& 80& $<13.86$& Y& --\\
Ni\textsc{ii}& 1741& 1741.553& 4.27E-02& -60& 60& $<13.61$& Y& --\\
Ni\textsc{ii}& 1751& 1751.916& 2.77E-02& -60& 60& $<13.77$& Y& --\\
Ni\textsc{ii}& & & & & & & & $<13.61$\\
\hline
Zn\textsc{ii}& 2026& 2026.136& 4.89E-01& -50& 50& $<12.40$& Y& --\\
Zn\textsc{ii}& & & & & & & & $<12.40$\\
\hline
\end{tabular}
\end{center}
\end{table*} 

\begin{figure*}
\begin{center}
\includegraphics[width=1.1\textwidth]{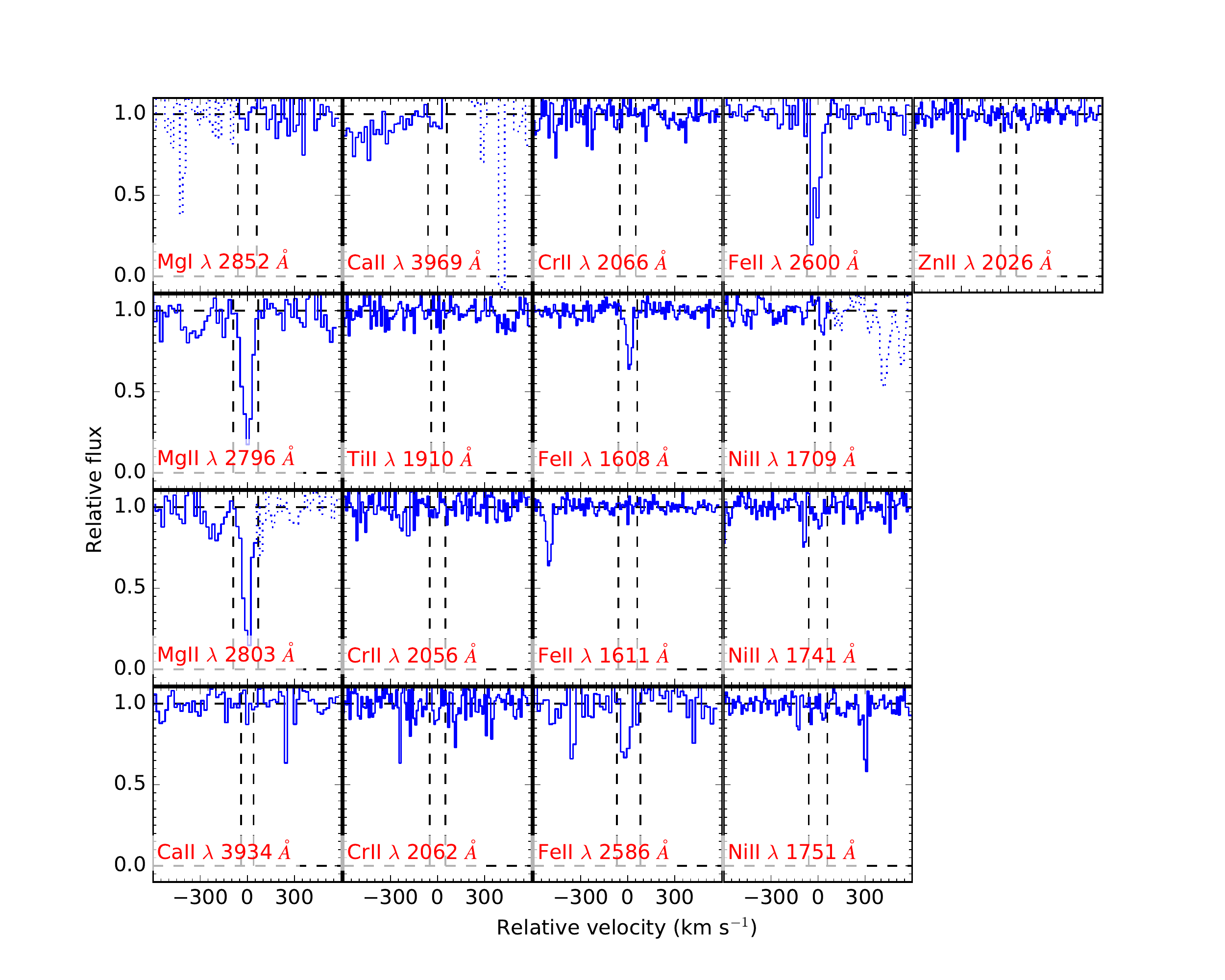}
\caption{Velocity profile of the XQ100 spectrum towards J0006-6208 (\zabs{}=3.203).}
\label{fig:J0006-6208,32030}
\end{center}
\end{figure*}

\begin{table*}
\begin{center}
\caption{X-Shooter metal column densities for J0006$-$6208 (\zabs{}=$3.775$)}
\label{tab:J0006-6208,37750}
\begin{tabular}{lcccccccc}
\hline
Ion& Line& $\lambda$& $f$& $v_{\rm min}$& $v_{\rm max}$& logN(X)& Included& logN$_{\rm adopt}$\\
 &  & \AA{}&  & \kms{}& \kms{}&  &  &  \\
\hline
Mg\textsc{ii}& 2796& 2796.352& 6.12E-01& -40& 180& $>14.01$& N& --\\
Mg\textsc{ii}& 2803& 2803.531& 3.05E-01& -60& 200& $>14.01$& Y& --\\
Mg\textsc{ii}& & & & & & & & $>14.01$\\
\hline
Si\textsc{ii}& 1260& 1260.422& 1.01E+00& -20& 180& $>13.95$& N& --\\
Si\textsc{ii}& 1526& 1526.707& 1.27E-01& 0& 200& $>14.47$& Y& --\\
Si\textsc{ii}& 1808& 1808.013& 2.19E-03& 0& 160& $<15.16$& Y& --\\
Si\textsc{ii}& & & & & & & & $14.46$--$15.16$\\
\hline
P\textsc{ii}& 1532& 1532.533& 7.61E-03& 0& 180& $<14.49$& Y& --\\
P\textsc{ii}& & & & & & & & $<14.49$\\
\hline
Ti\textsc{ii}& 1910& 1910.750& 2.02E-01& 0& 160& $<12.75$& N& --\\
Ti\textsc{ii}& 3242& 3242.929& 2.32E-01& 0& 160& $<12.75$& N& --\\
Ti\textsc{ii}& 3384& 3384.740& 3.58E-01& 0& 160& $<12.30$& Y& --\\
Ti\textsc{ii}& & & & & & & & $<12.30$\\
\hline
Cr\textsc{ii}& 2056& 2056.254& 1.05E-01& -20& 180& $<15.31$& Y& --\\
Cr\textsc{ii}& & & & & & & & $<15.31$\\
\hline
Mn\textsc{ii}& 2594& 2594.499& 2.71E-01& 0& 140& $12.51\pm0.00$& Y& --\\
Mn\textsc{ii}& 2606& 2606.462& 1.93E-01& 0& 110& $>12.25$& N& --\\
Mn\textsc{ii}& & & & & & & & $12.51\pm0.05$\\
\hline
Fe\textsc{ii}& 1608& 1608.451& 5.80E-02& 0& 160& $>14.39$& Y& --\\
Fe\textsc{ii}& 1611& 1611.200& 1.36E-03& 15& 160& $<15.12$& Y& --\\
Fe\textsc{ii}& 2586& 2586.650& 6.91E-02& 0& 180& $>14.26$& Y& --\\
Fe\textsc{ii}& 2600& 2600.173& 2.39E-01& -30& 150& $>13.96$& N& --\\
Fe\textsc{ii}& & & & & & & & $14.40$--$15.12$\\
\hline
Ni\textsc{ii}& 1741& 1741.553& 4.27E-02& 0& 160& $<13.54$& Y& --\\
Ni\textsc{ii}& & & & & & & & $<13.54$\\
\hline
Zn\textsc{ii}& 2026& 2026.136& 4.89E-01& 0& 160& $<12.69$& Y& --\\
Zn\textsc{ii}& & & & & & & & $<12.69$\\
\hline
\end{tabular}
\end{center}
\end{table*} 

\begin{figure*}
\begin{center}
\includegraphics[width=1.1\textwidth]{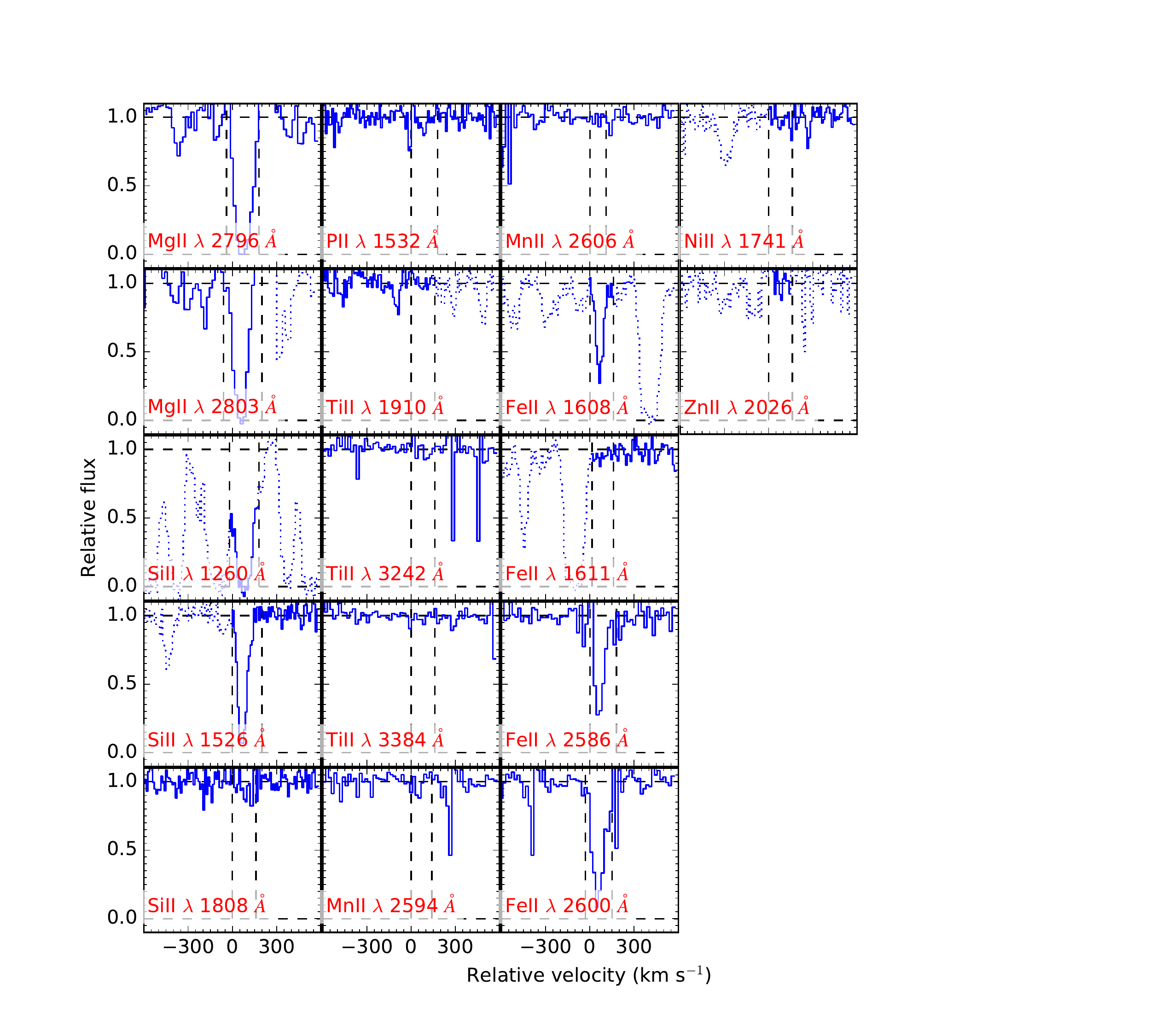}
\caption{Velocity profile of the XQ100 spectrum towards J0006-6208 (\zabs{}=3.775).}
\label{fig:J0006-6208,37750}
\end{center}
\end{figure*}

\begin{table*}
\begin{center}
\caption{X-Shooter metal column densities for J0034+1639 (\zabs{}=$3.755$)}
\label{tab:J0034+1639,37550}
\begin{tabular}{lcccccccc}
\hline
Ion& Line& $\lambda$& $f$& $v_{\rm min}$& $v_{\rm max}$& logN(X)& Included& logN$_{\rm adopt}$\\
 &  & \AA{}&  & \kms{}& \kms{}&  &  &  \\
\hline
Mg\textsc{ii}& 2796& 2796.352& 6.12E-01& -180& 0& $>13.36$& N& --\\
Mg\textsc{ii}& 2803& 2803.531& 3.05E-01& -140& 0& $>13.63$& Y& --\\
Mg\textsc{ii}& & & & & & & & $>13.63$\\
\hline
Si\textsc{ii}& 1526& 1526.707& 1.27E-01& -140& 0& $>14.05$& Y& --\\
Si\textsc{ii}& 1808& 1808.013& 2.19E-03& -140& 0& $<14.55$& Y& --\\
Si\textsc{ii}& & & & & & & & $14.05$--$14.55$\\
\hline
Ti\textsc{ii}& 1910& 1910.750& 2.02E-01& -140& 0& $<12.83$& N& --\\
Ti\textsc{ii}& 3242& 3242.929& 2.32E-01& -140& 0& $<12.31$& Y& --\\
Ti\textsc{ii}& 3384& 3384.740& 3.58E-01& -140& 0& $<12.50$& N& --\\
Ti\textsc{ii}& & & & & & & & $<12.31$\\
\hline
Cr\textsc{ii}& 2062& 2062.234& 7.80E-02& -140& 0& $<13.70$& Y& --\\
Cr\textsc{ii}& 2066& 2066.161& 5.15E-02& -140& 0& $<15.03$& N& --\\
Cr\textsc{ii}& & & & & & & & $<13.70$\\
\hline
Mn\textsc{ii}& 2576& 2576.877& 3.51E-01& -140& 0& $<12.49$& Y& --\\
Mn\textsc{ii}& 2594& 2594.499& 2.71E-01& -140& 0& $<12.48$& Y& --\\
Mn\textsc{ii}& & & & & & & & $<12.48$\\
\hline
Fe\textsc{ii}& 2249& 2249.877& 1.82E-03& -140& 0& $<14.96$& N& --\\
Fe\textsc{ii}& 2344& 2344.214& 1.14E-01& -140& 0& $13.66\pm0.03$& Y& --\\
Fe\textsc{ii}& 2586& 2586.650& 6.91E-02& -140& -20& $13.69\pm0.04$& Y& --\\
Fe\textsc{ii}& 2600& 2600.173& 2.39E-01& -140& 5& $>13.57$& Y& --\\
Fe\textsc{ii}& & & & & & & & $13.67\pm0.05$\\
\hline
Ni\textsc{ii}& 1709& 1709.604& 3.24E-02& -140& 0& $<13.38$& N& --\\
Ni\textsc{ii}& 1741& 1741.553& 4.27E-02& -140& 0& $<13.38$& Y& --\\
Ni\textsc{ii}& 1751& 1751.916& 2.77E-02& -140& 0& $<13.44$& N& --\\
Ni\textsc{ii}& & & & & & & & $<13.38$\\
\hline
Zn\textsc{ii}& 2026& 2026.136& 4.89E-01& -140& 0& $<12.39$& Y& --\\
Zn\textsc{ii}& 2062& 2062.664& 2.56E-01& -140& 0& $<12.69$& N& --\\
Zn\textsc{ii}& & & & & & & & $<12.39$\\
\hline
\end{tabular}
\end{center}
\end{table*} 

\begin{figure*}
\begin{center}
\includegraphics[width=1.1\textwidth]{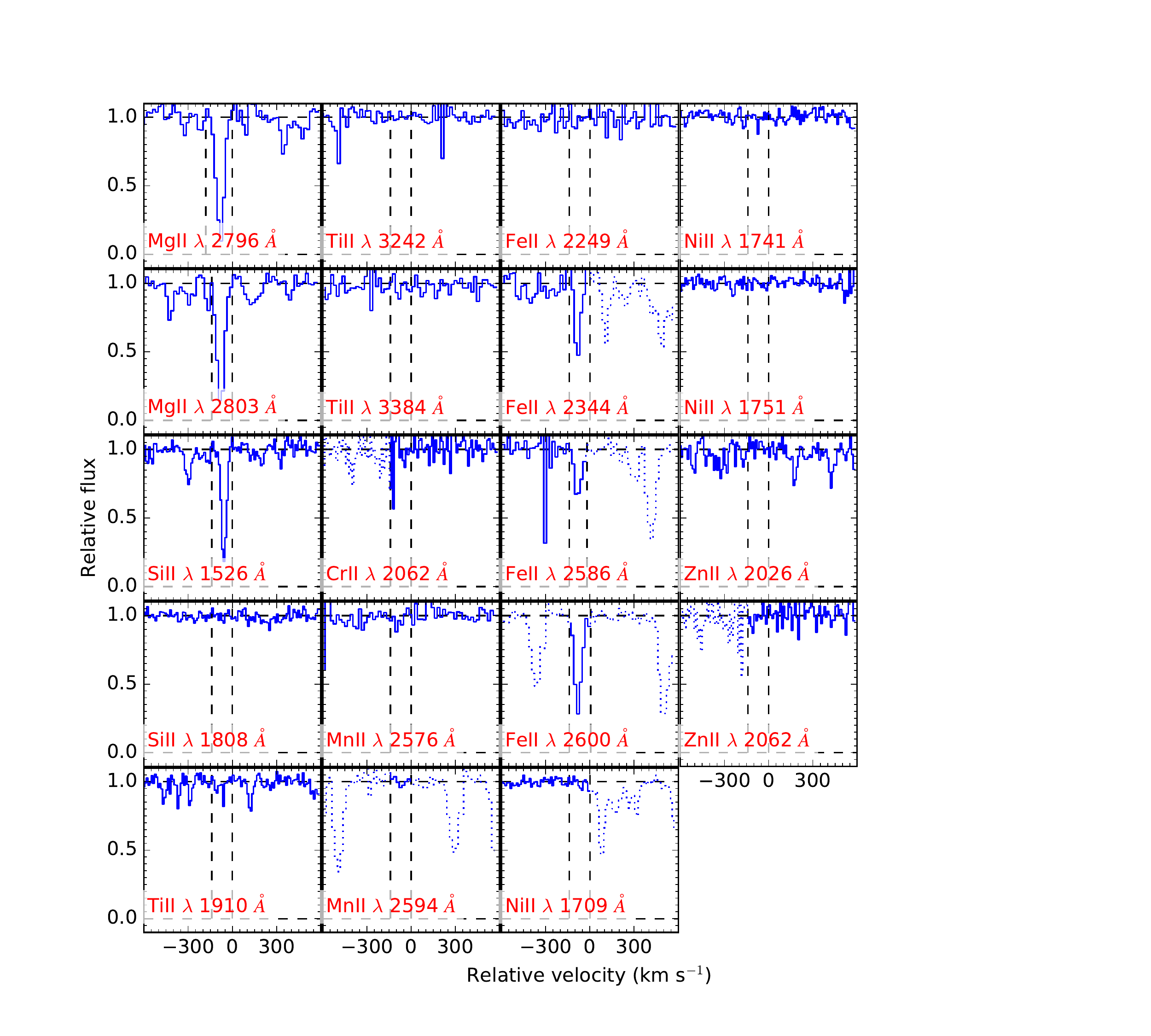}
\caption{Velocity profile of the XQ100 spectrum towards J0034+1639 (\zabs{}=3.755).}
\label{fig:J0034+1639,37550}
\end{center}
\end{figure*}
\begin{table*}
\begin{center}
\caption{X-Shooter metal column densities for J0034+1639 (\zabs{}=$4.284$)}
\label{tab:J0034+1639,42835}
\begin{tabular}{lcccccccc}
\hline
Ion& Line& $\lambda$& $f$& $v_{\rm min}$& $v_{\rm max}$& logN(X)& Included& logN$_{\rm adopt}$\\
 &  & \AA{}&  & \kms{}& \kms{}&  &  &  \\
\hline
C\textsc{ii}& 1334& 1334.532& 1.28E-01& -60& 110& $>14.87$& Y& --\\
C\textsc{ii}& & & & & & & & $>14.87$\\
\hline
O\textsc{i}& 1302& 1302.168& 4.89E-02& -60& 110& $>15.17$& N& --\\
O\textsc{i}& & & & & & & & $0.00\pm0.05$\\
\hline
Mg\textsc{i}& 2852& 2852.964& 1.81E+00& -70& 90& $12.13\pm0.05$& Y& --\\
Mg\textsc{i}& & & & & & & & $12.13\pm0.05$\\
\hline
\end{tabular}
\end{center}
\end{table*} 

\begin{table*}
\begin{center}
\caption{X-Shooter metal column densities for J0034+1639 (\zabs{}=$4.252$)}
\label{tab:J0034+1639,42523}
\begin{tabular}{lcccccccc}
\hline
Ion& Line& $\lambda$& $f$& $v_{\rm min}$& $v_{\rm max}$& logN(X)& Included& logN$_{\rm adopt}$\\
 &  & \AA{}&  & \kms{}& \kms{}&  &  &  \\
\hline
C\textsc{ii}& 1334& 1334.532& 1.28E-01& -60& 70& $>14.29$& Y& --\\
C\textsc{ii}& & & & & & & & $>14.29$\\
\hline
O\textsc{i}& 1302& 1302.168& 4.89E-02& -60& 70& $>14.71$& Y& --\\
O\textsc{i}& & & & & & & & $>14.71$\\
\hline
Mg\textsc{ii}& 2796& 2796.352& 6.12E-01& -100& 70& $>13.20$& Y& --\\
Mg\textsc{ii}& 2803& 2803.531& 3.05E-01& -130& 60& $>13.82$& N& --\\
\hline
\end{tabular}
\end{center}
\end{table*} 

\clearpage

\begin{figure*}
\begin{center}
\includegraphics[width=1.1\textwidth]{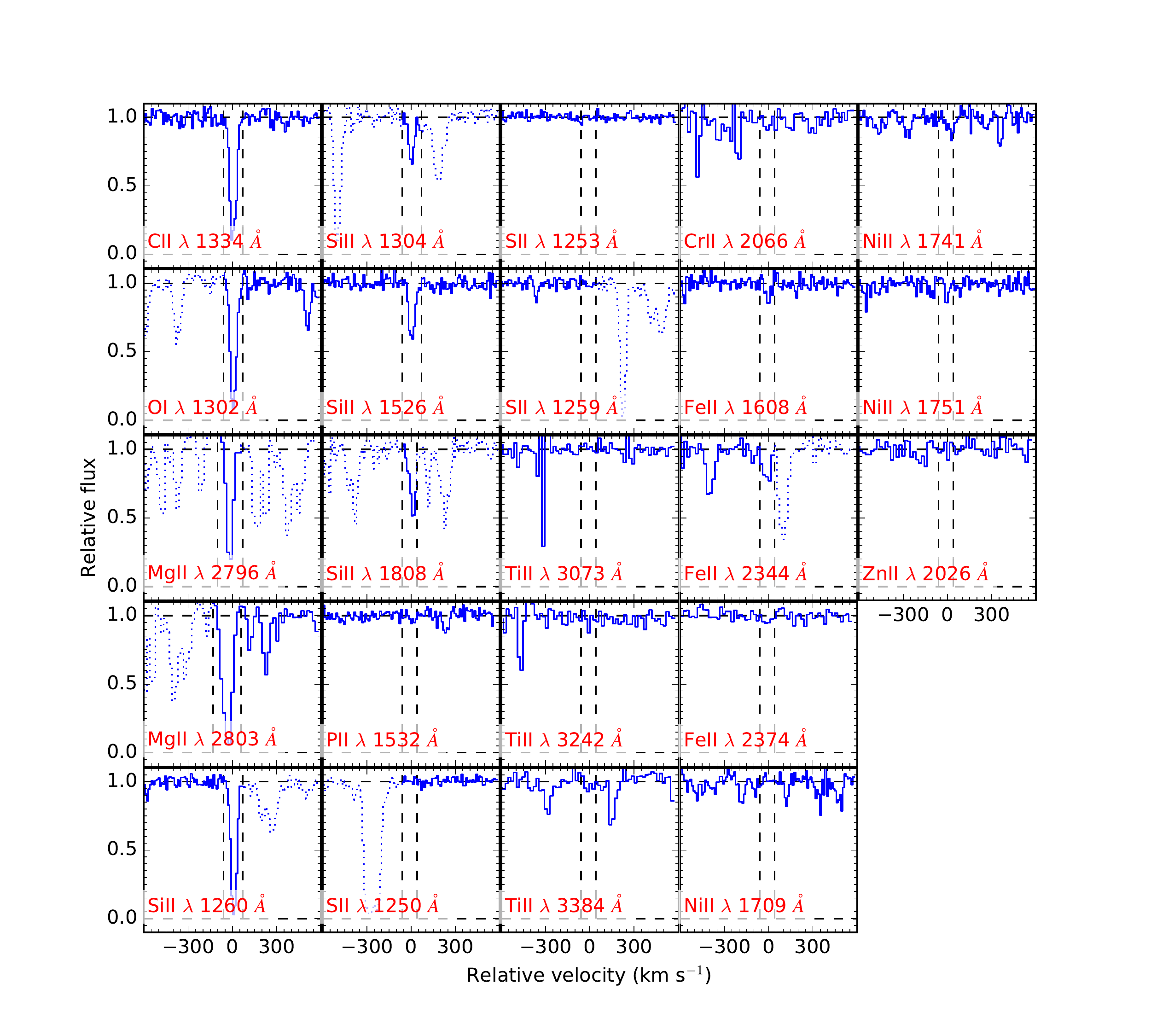}
\caption{Velocity profile of the XQ100 spectrum towards J0034+1639 (\zabs{}=4.251).}
\label{fig:J0034+1639,42523}
\end{center}
\end{figure*}

\begin{figure*}
\begin{center}
\includegraphics[width=1.1\textwidth]{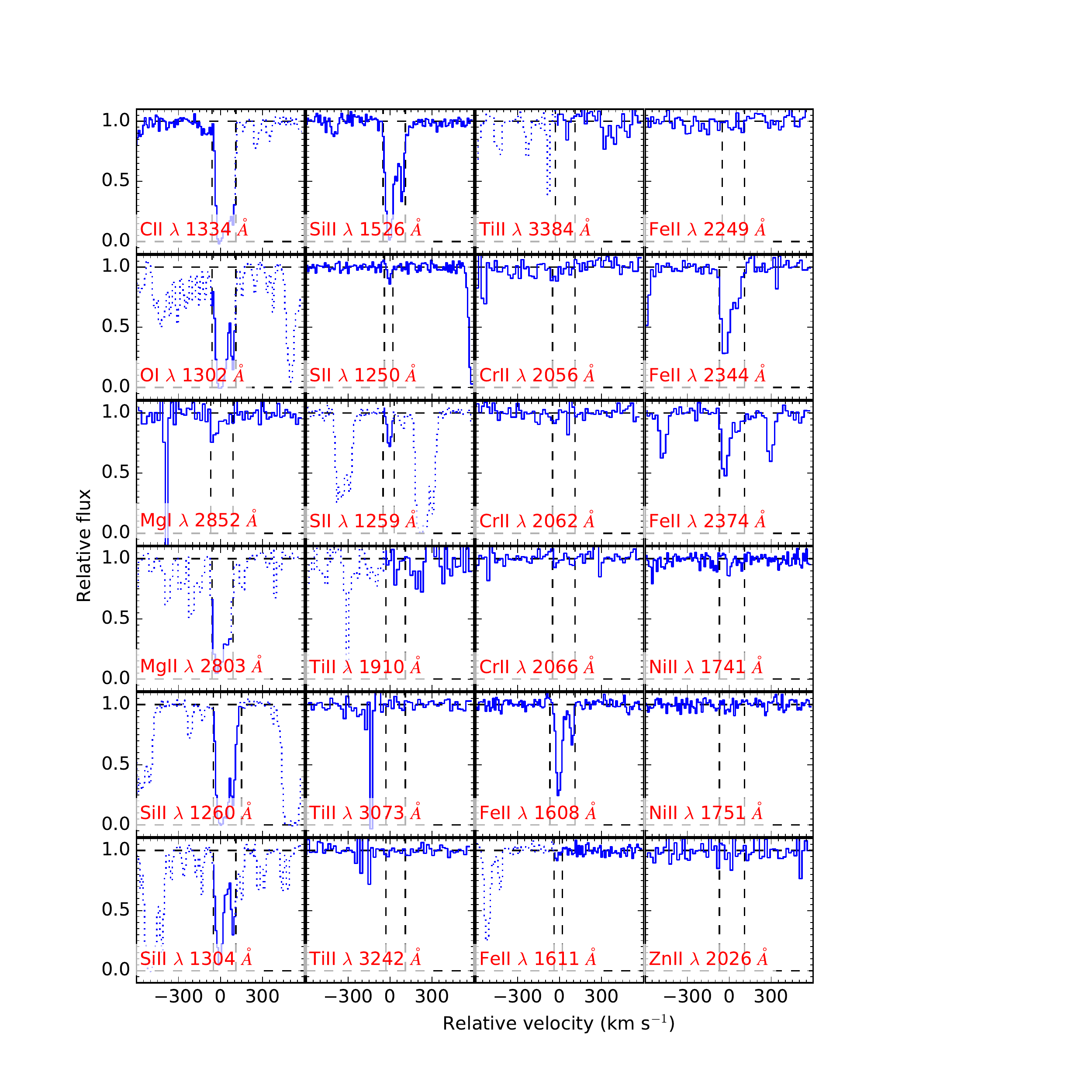}
\caption{Velocity profile of the XQ100 spectrum towards J0034+1639 (\zabs{}=4.283).}
\label{fig:J0034+1639,42835}
\end{center}
\end{figure*}

\clearpage
\begin{table*}
\begin{center}
\caption{X-Shooter metal column densities for J0113$-$2803 (\zabs{}=$3.106$)}
\label{tab:J0113-2803,31060}
\begin{tabular}{lcccccccc}
\hline
Ion& Line& $\lambda$& $f$& $v_{\rm min}$& $v_{\rm max}$& logN(X)& Included& logN$_{\rm adopt}$\\
 &  & \AA{}&  & \kms{}& \kms{}&  &  &  \\
\hline
Mg\textsc{i}& 2852& 2852.964& 1.81E+00& -400& 110& $12.60\pm0.06$& Y& --\\
Mg\textsc{i}& & & & & & & & $12.60\pm0.06$\\
\hline
Mg\textsc{ii}& 2796& 2796.352& 6.12E-01& -550& 110& $>14.42$& Y& --\\
Mg\textsc{ii}& 2803& 2803.531& 3.05E-01& -500& 130& $>14.55$& Y& --\\
Mg\textsc{ii}& & & & & & & & $>14.55$\\
\hline
Si\textsc{ii}& 1808& 1808.013& 2.19E-03& -400& 110& $15.60\pm0.02$& Y& --\\
\hline
\end{tabular}
\end{center}
\end{table*} 

\begin{figure*}
\begin{center}
\includegraphics[width=1.1\textwidth]{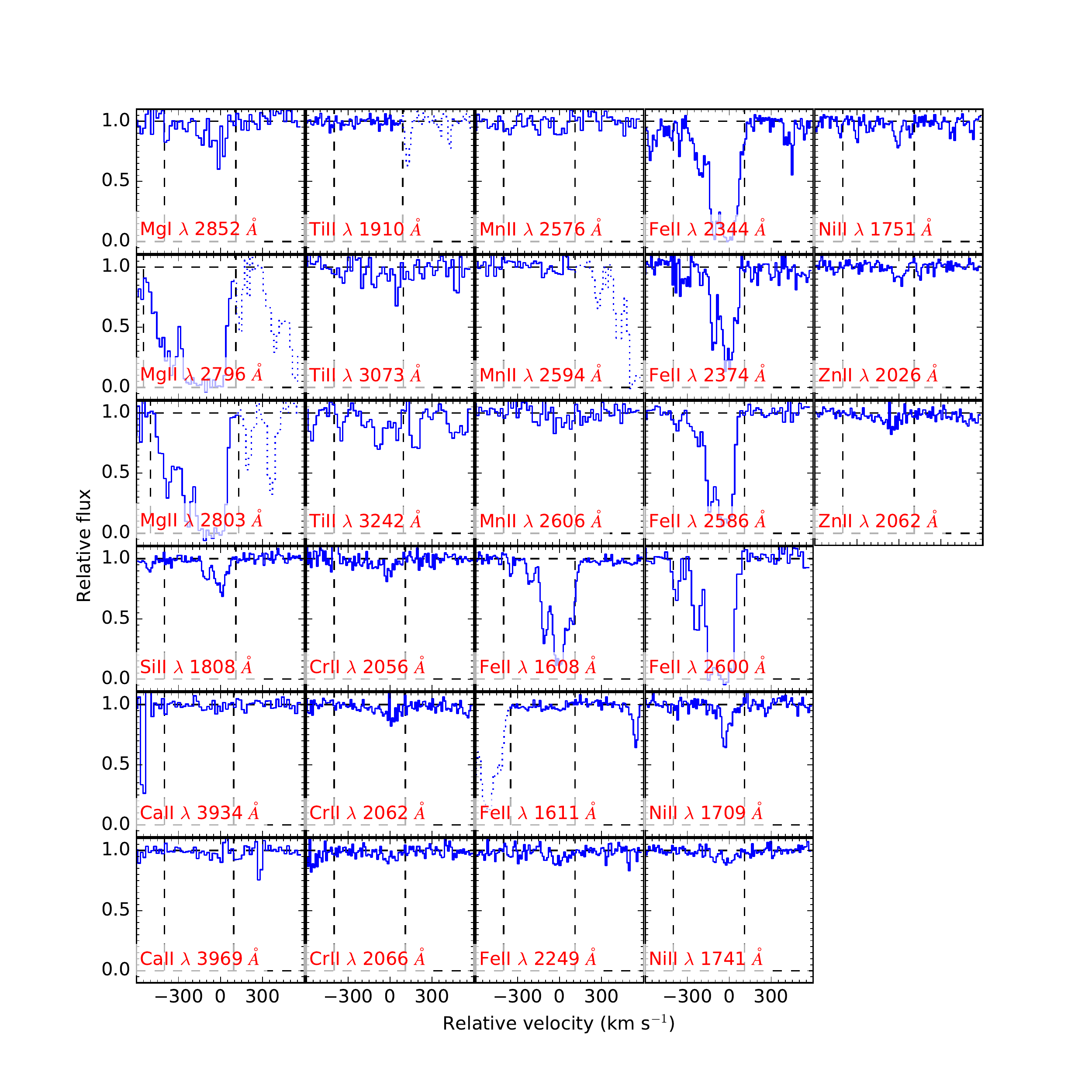}
\caption{Velocity profile of the XQ100 spectrum towards J0113-2803 (\zabs{}=3.106).}
\label{fig:J0113-2803,31060}
\end{center}
\end{figure*}

\begin{table*}
\begin{center}
\caption{X-Shooter metal column densities for J0124+0044 (\zabs{}=$2.261$)}
\label{tab:J0124+0044,22610}
\begin{tabular}{lcccccccc}
\hline
Ion& Line& $\lambda$& $f$& $v_{\rm min}$& $v_{\rm max}$& logN(X)& Included& logN$_{\rm adopt}$\\
 &  & \AA{}&  & \kms{}& \kms{}&  &  &  \\
\hline
Mg\textsc{i}& 2852& 2852.964& 1.81E+00& -80& 190& $12.50\pm0.02$& Y& --\\
Mg\textsc{i}& & & & & & & & $12.50\pm0.05$\\
\hline
Mg\textsc{ii}& 2796& 2796.352& 6.12E-01& -80& 190& $>14.00$& N& --\\
Mg\textsc{ii}& 2803& 2803.531& 3.05E-01& -80& 190& $>14.31$& Y& --\\
Mg\textsc{ii}& & & & & & & & $>14.31$\\
\hline
Si\textsc{ii}& 1808& 1808.013& 2.19E-03& -80& 190& $15.36\pm0.02$& Y& --\\
\hline
\end{tabular}
\end{center}
\end{table*} 

\begin{figure*}
\begin{center}
\includegraphics[width=1.1\textwidth]{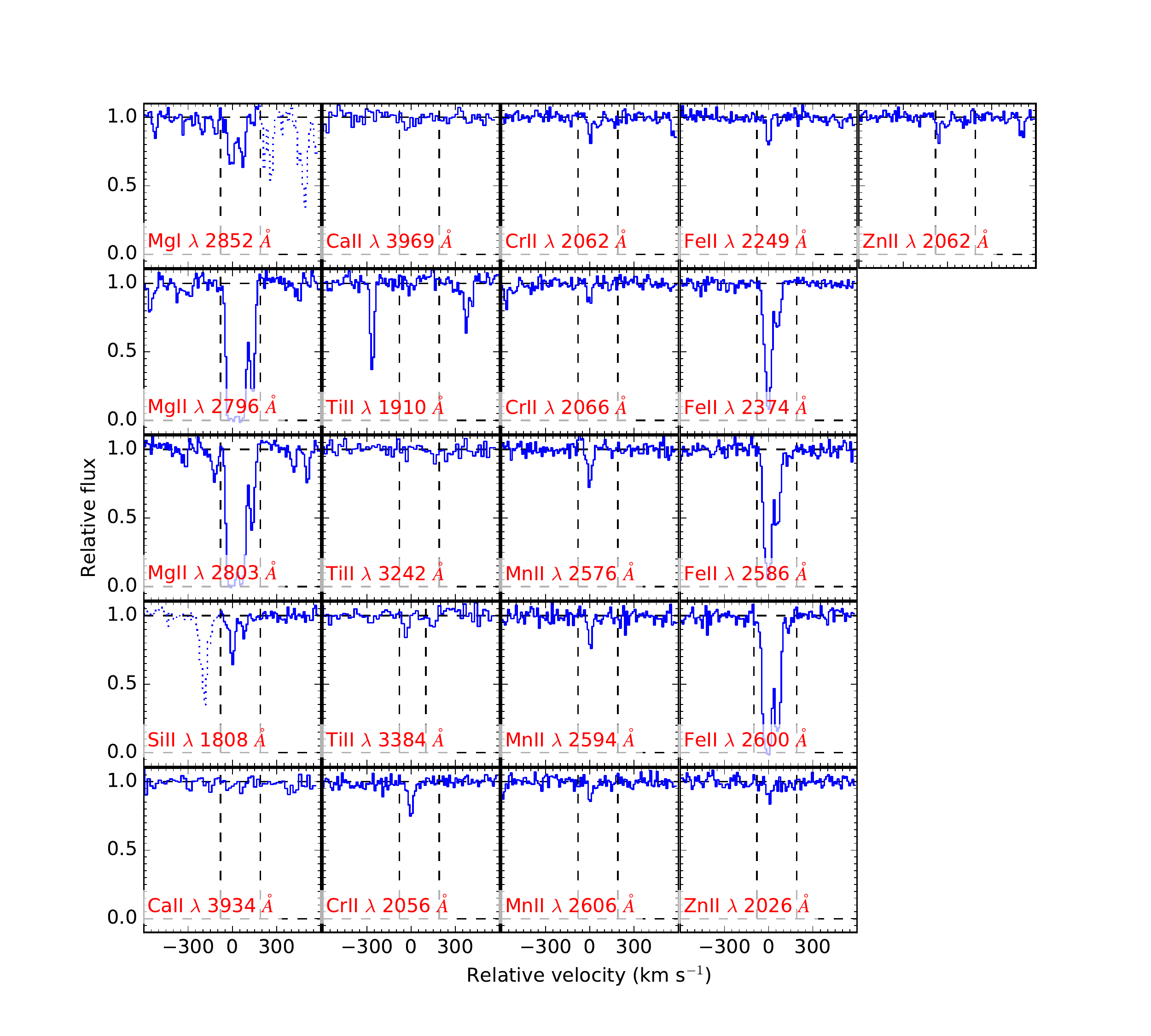}
\caption{Velocity profile of the XQ100 spectrum towards J0124+0044 (\zabs{}=2.261).}
\label{fig:J0124+0044,22610}
\end{center}
\end{figure*}

\begin{table*}
\begin{center}
\caption{X-Shooter metal column densities for J0132+1341 (\zabs{}=$3.936$)}
\label{tab:J0132+1341,39360}
\begin{tabular}{lcccccccc}
\hline
Ion& Line& $\lambda$& $f$& $v_{\rm min}$& $v_{\rm max}$& logN(X)& Included& logN$_{\rm adopt}$\\
 &  & \AA{}&  & \kms{}& \kms{}&  &  &  \\
\hline
C\textsc{ii}& 1334& 1334.532& 1.28E-01& -70& 110& $>14.39$& Y& --\\
C\textsc{ii}& & & & & & & & $>14.39$\\
\hline
O\textsc{i}& 1302& 1302.168& 4.89E-02& -50& 65& $>14.77$& Y& --\\
O\textsc{i}& & & & & & & & $>14.77$\\
\hline
Si\textsc{ii}& 1304& 1304.370& 9.40E-02& -50& 60& $13.87\pm0.02$& Y& --\\
Si\textsc{ii}& 1526& 1526.707& 1.27E-01& -40& 70& $13.87\pm0.02$& Y& --\\
\hline
\end{tabular}
\end{center}
\end{table*} 

\begin{figure*}
\begin{center}
\includegraphics[width=1.1\textwidth]{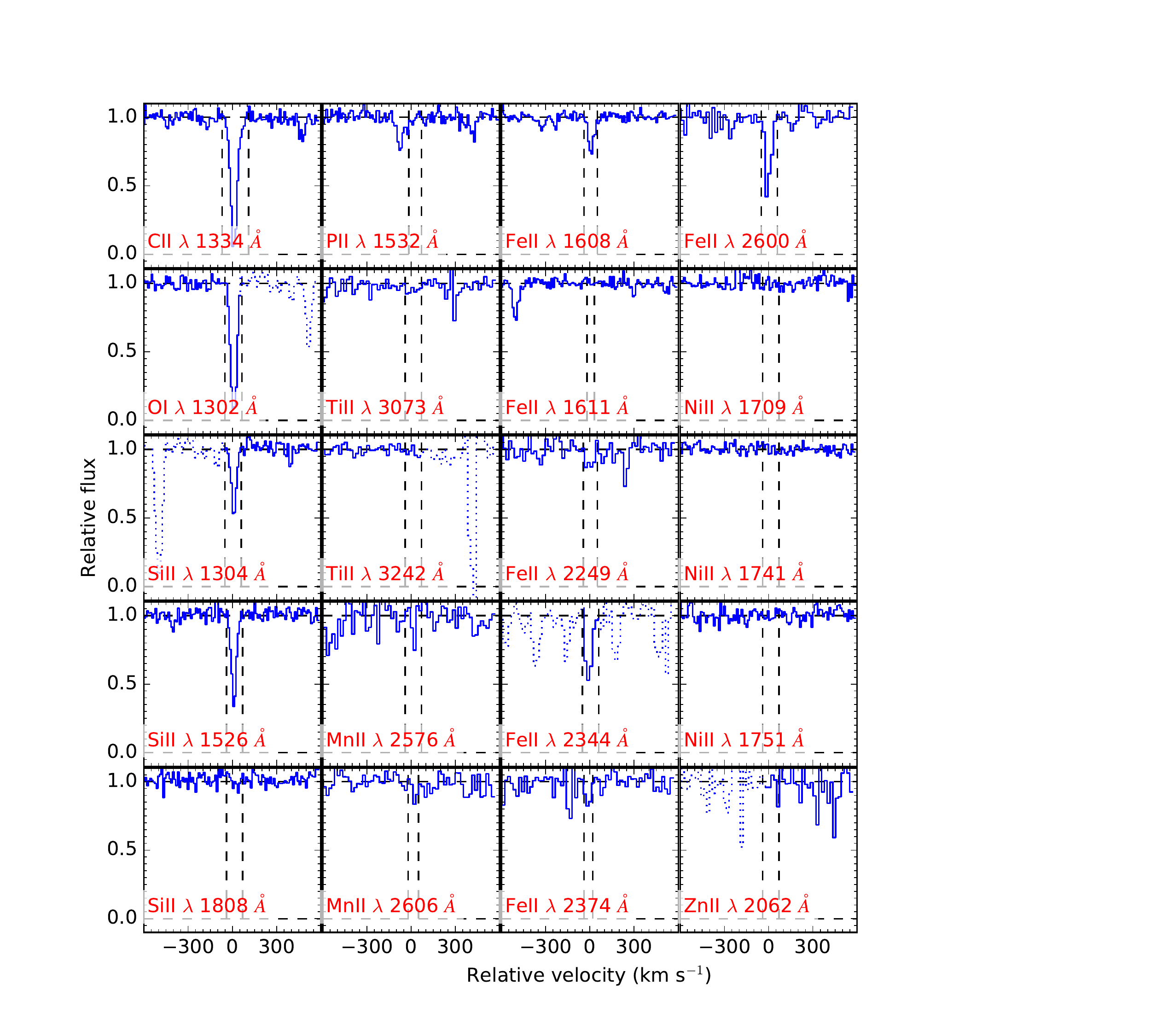}
\caption{Velocity profile of the XQ100 spectrum towards J0132+1341 (\zabs{}=3.936).}
\label{fig:J0132+1341,39360}
\end{center}
\end{figure*}

\clearpage
\begin{table*}
\begin{center}
\caption{X-Shooter metal column densities for J0134+0400 (\zabs{}=$3.692$)}
\label{tab:J0134+0400,36920}
\begin{tabular}{lcccccccc}
\hline
Ion& Line& $\lambda$& $f$& $v_{\rm min}$& $v_{\rm max}$& logN(X)& Included& logN$_{\rm adopt}$\\
 &  & \AA{}&  & \kms{}& \kms{}&  &  &  \\
\hline
Mg\textsc{ii}& 2796& 2796.352& 6.12E-01& -100& 40& $>13.26$& N& --\\
Mg\textsc{ii}& 2803& 2803.531& 3.05E-01& -100& 20& $>13.49$& Y& --\\
Mg\textsc{ii}& & & & & & & & $>13.49$\\
\hline
Si\textsc{ii}& 1808& 1808.013& 2.19E-03& -100& 35& $<14.36$& Y& --\\
Si\textsc{ii}& & & & & & & & $<14.36$\\
\hline
P\textsc{ii}& 1532& 1532.533& 7.61E-03& -30& 70& $<13.73$& Y& --\\
\hline
\end{tabular}
\end{center}
\end{table*} 

\begin{figure*}
\begin{center}
\includegraphics[width=1.1\textwidth]{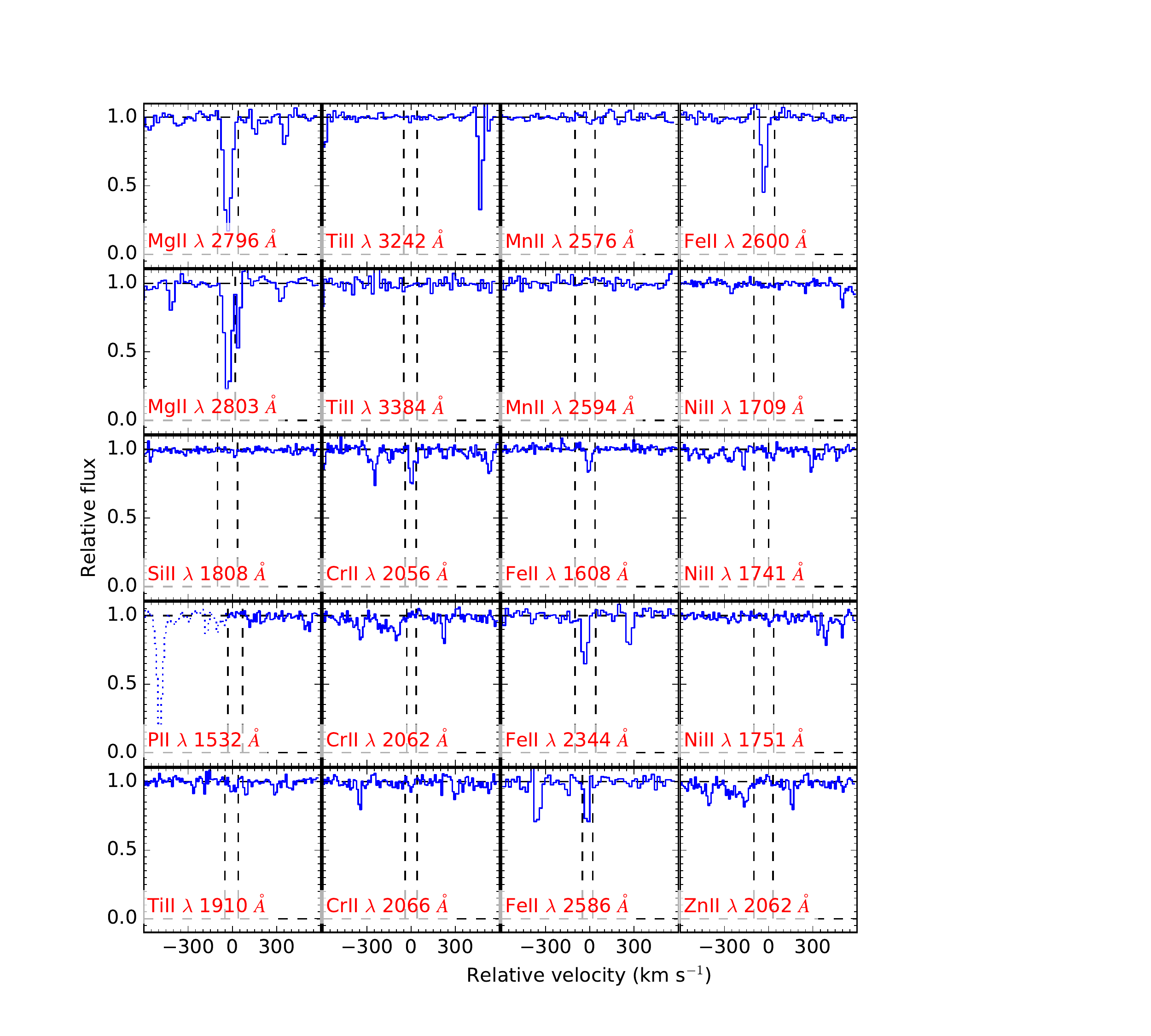}
\caption{Velocity profile of the XQ100 spectrum towards J0134+0400 (\zabs{}=3.692).}
\label{fig:J0134+0400,36920}
\end{center}
\end{figure*}

\begin{table*}
\begin{center}
\caption{X-Shooter metal column densities for J0134+0400 (\zabs{}=$3.772$)}
\label{tab:J0134+0400,37725}
\begin{tabular}{lcccccccc}
\hline
Ion& Line& $\lambda$& $f$& $v_{\rm min}$& $v_{\rm max}$& logN(X)& Included& logN$_{\rm adopt}$\\
 &  & \AA{}&  & \kms{}& \kms{}&  &  &  \\
\hline
C\textsc{ii}& 1334& 1334.532& 1.28E-01& -160& 150& $>15.21$& Y& --\\
C\textsc{ii}& & & & & & & & $>15.21$\\
\hline
Mg\textsc{ii}& 2796& 2796.352& 6.12E-01& -170& 230& $>14.15$& N& --\\
Mg\textsc{ii}& 2803& 2803.531& 3.05E-01& -200& 230& $>14.38$& Y& --\\
Mg\textsc{ii}& & & & & & & & $>14.38$\\
\hline
Si\textsc{ii}& 1526& 1526.707& 1.27E-01& -170& 230& $>14.95$& N& --\\
\hline
\end{tabular}
\end{center}
\end{table*} 

\begin{figure*}
\begin{center}
\includegraphics[width=1.1\textwidth]{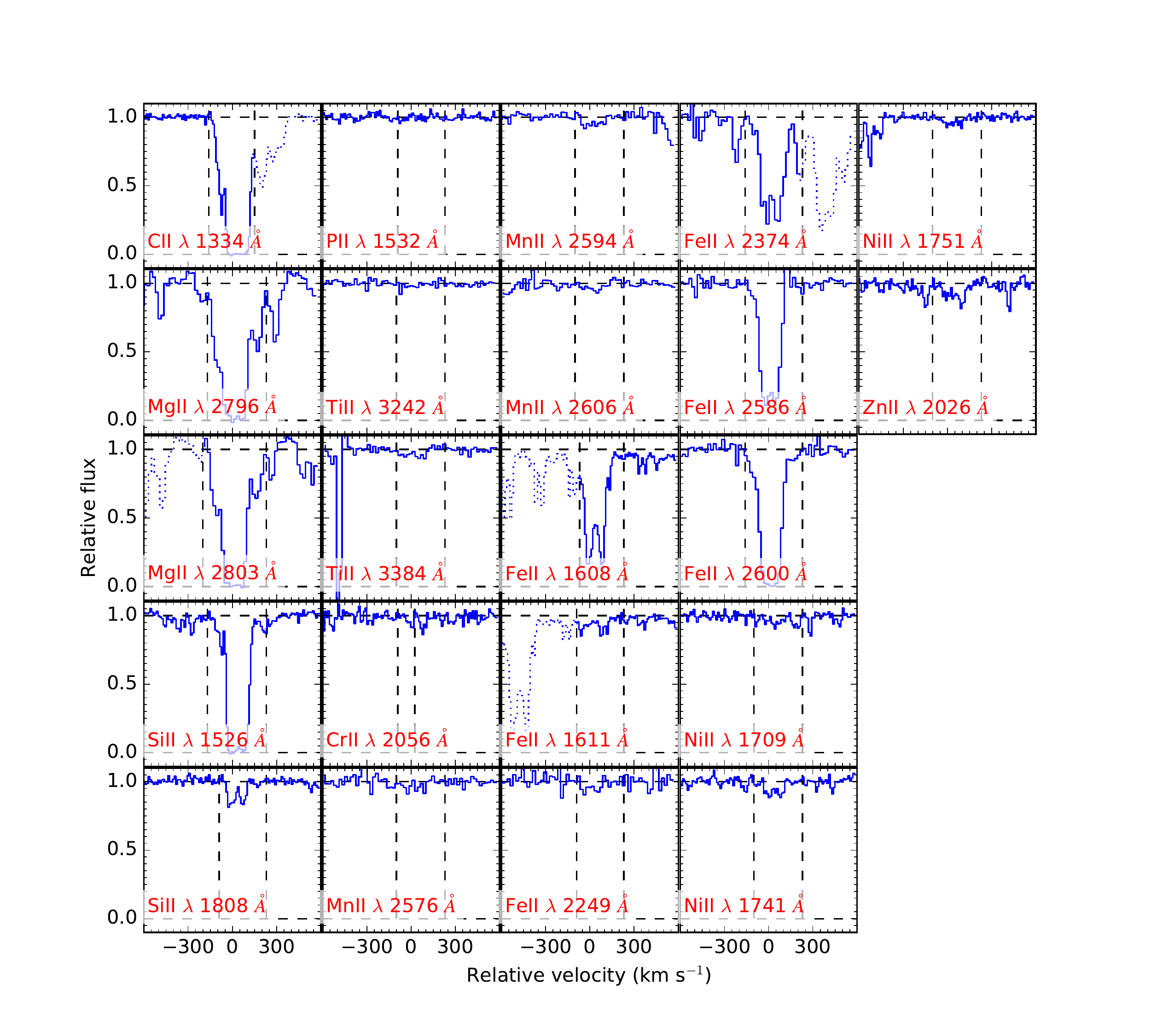}
\caption{Velocity profile of the XQ100 spectrum towards J0134+0400 (\zabs{}=3.772).}
\label{fig:J0134+0400,37725}
\end{center}
\end{figure*}

\begin{table*}
\begin{center}
\caption{X-Shooter metal column densities for J0234$-$1806 (\zabs{}=$3.693$)}
\label{tab:J0234-1806,36930}
\begin{tabular}{lcccccccc}
\hline
Ion& Line& $\lambda$& $f$& $v_{\rm min}$& $v_{\rm max}$& logN(X)& Included& logN$_{\rm adopt}$\\
 &  & \AA{}&  & \kms{}& \kms{}&  &  &  \\
\hline
Mg\textsc{i}& 2852& 2852.964& 1.81E+00& -200& 200& $12.67\pm0.02$& Y& --\\
Mg\textsc{i}& & & & & & & & $12.67\pm0.05$\\
\hline
Mg\textsc{ii}& 2796& 2796.352& 6.12E-01& -200& 330& $>14.04$& N& --\\
Mg\textsc{ii}& 2803& 2803.531& 3.05E-01& -175& 330& $>14.30$& Y& --\\
Mg\textsc{ii}& & & & & & & & $>14.30$\\
\hline
Si\textsc{ii}& 1526& 1526.707& 1.27E-01& -200& 330& $>14.55$& Y& --\\
Si\textsc{ii}& 1808& 1808.013& 2.19E-03& -200& 330& $<14.62$& Y& --\\
Si\textsc{ii}& & & & & & & & $14.54$--$14.64$\\
\hline
Ti\textsc{ii}& 3242& 3242.929& 2.32E-01& -200& 330& $<12.47$& N& --\\
Ti\textsc{ii}& 3384& 3384.740& 3.58E-01& -200& 330& $<12.44$& Y& --\\
Ti\textsc{ii}& & & & & & & & $<12.44$\\
\hline
Cr\textsc{ii}& 2056& 2056.254& 1.05E-01& 0& 330& $<13.11$& Y& --\\
Cr\textsc{ii}& 2062& 2062.234& 7.80E-02& -150& 90& $<13.20$& N& --\\
Cr\textsc{ii}& 2066& 2066.161& 5.15E-02& -200& 330& $<14.35$& N& --\\
Cr\textsc{ii}& & & & & & & & $<13.11$\\
\hline
Mn\textsc{ii}& 2576& 2576.877& 3.51E-01& -200& 330& $<12.62$& Y& --\\
Mn\textsc{ii}& 2594& 2594.499& 2.71E-01& -130& 330& $<12.81$& N& --\\
Mn\textsc{ii}& & & & & & & & $<12.62$\\
\hline
Fe\textsc{ii}& 1608& 1608.451& 5.80E-02& -200& 330& $14.27\pm0.02$& Y& --\\
Fe\textsc{ii}& 1611& 1611.200& 1.36E-03& -200& 330& $15.24\pm0.09$& N& --\\
Fe\textsc{ii}& 2249& 2249.877& 1.82E-03& -200& 330& $14.97\pm0.14$& N& --\\
Fe\textsc{ii}& 2344& 2344.214& 1.14E-01& -200& 330& $14.18\pm0.03$& Y& --\\
Fe\textsc{ii}& 2600& 2600.173& 2.39E-01& -200& 330& $>14.03$& N& --\\
Fe\textsc{ii}& & & & & & & & $14.24\pm0.05$\\
\hline
Ni\textsc{ii}& 1709& 1709.604& 3.24E-02& -200& 330& $<13.59$& Y& --\\
Ni\textsc{ii}& 1751& 1751.916& 2.77E-02& -200& 100& $<13.62$& N& --\\
Ni\textsc{ii}& & & & & & & & $<13.59$\\
\hline
\end{tabular}
\end{center}
\end{table*} 

\begin{figure*}
\begin{center}
\includegraphics[width=1.1\textwidth]{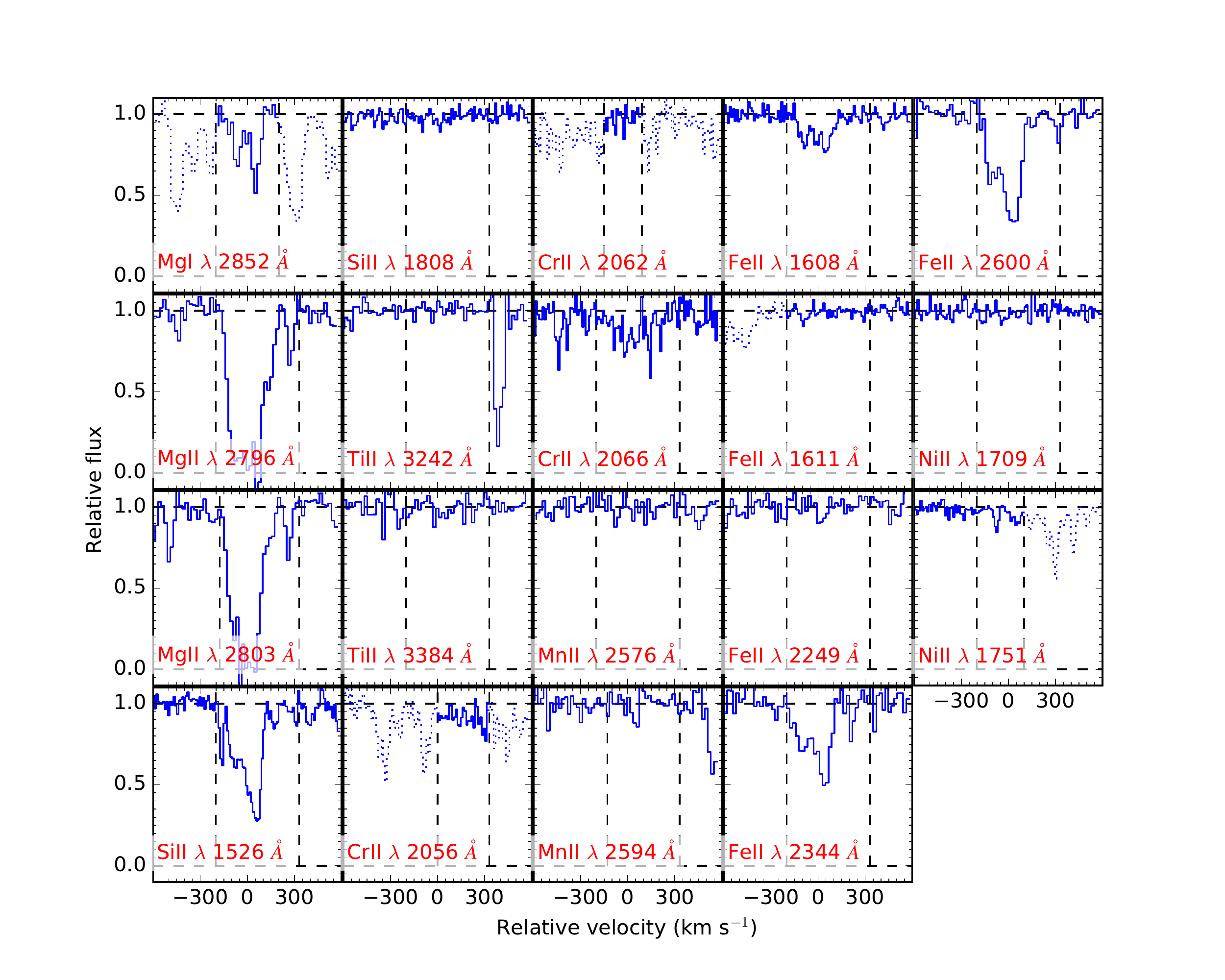}
\caption{Velocity profile of the XQ100 spectrum towards J0234-1806 (\zabs{}=3.693).}
\label{fig:J0234-1806,36930}
\end{center}
\end{figure*}

\clearpage

\begin{table*}
\begin{center}
\caption{X-Shooter metal column densities for J0255+0048 (\zabs{}=$3.256$)}
\label{tab:J0255+0048,32555}
\begin{tabular}{lcccccccc}
\hline
Ion& Line& $\lambda$& $f$& $v_{\rm min}$& $v_{\rm max}$& logN(X)& Included& logN$_{\rm adopt}$\\
 &  & \AA{}&  & \kms{}& \kms{}&  &  &  \\
\hline
Mg\textsc{ii}& 2796& 2796.352& 6.12E-01& -300& 110& $>14.09$& N& --\\
Mg\textsc{ii}& 2803& 2803.531& 3.05E-01& -300& 80& $>14.41$& Y& --\\
Mg\textsc{ii}& & & & & & & & $>14.41$\\
\hline
Si\textsc{ii}& 1526& 1526.707& 1.27E-01& -300& 100& $>15.01$& N& --\\
Si\textsc{ii}& 1808& 1808.013& 2.19E-03& -300& 100& $15.33\pm0.03$& Y& --\\
Si\textsc{ii}& & & & & & & & $15.33\pm0.05$\\
\hline
\end{tabular}
\end{center}
\end{table*} 

\begin{figure*}
\begin{center}
\includegraphics[width=1.1\textwidth]{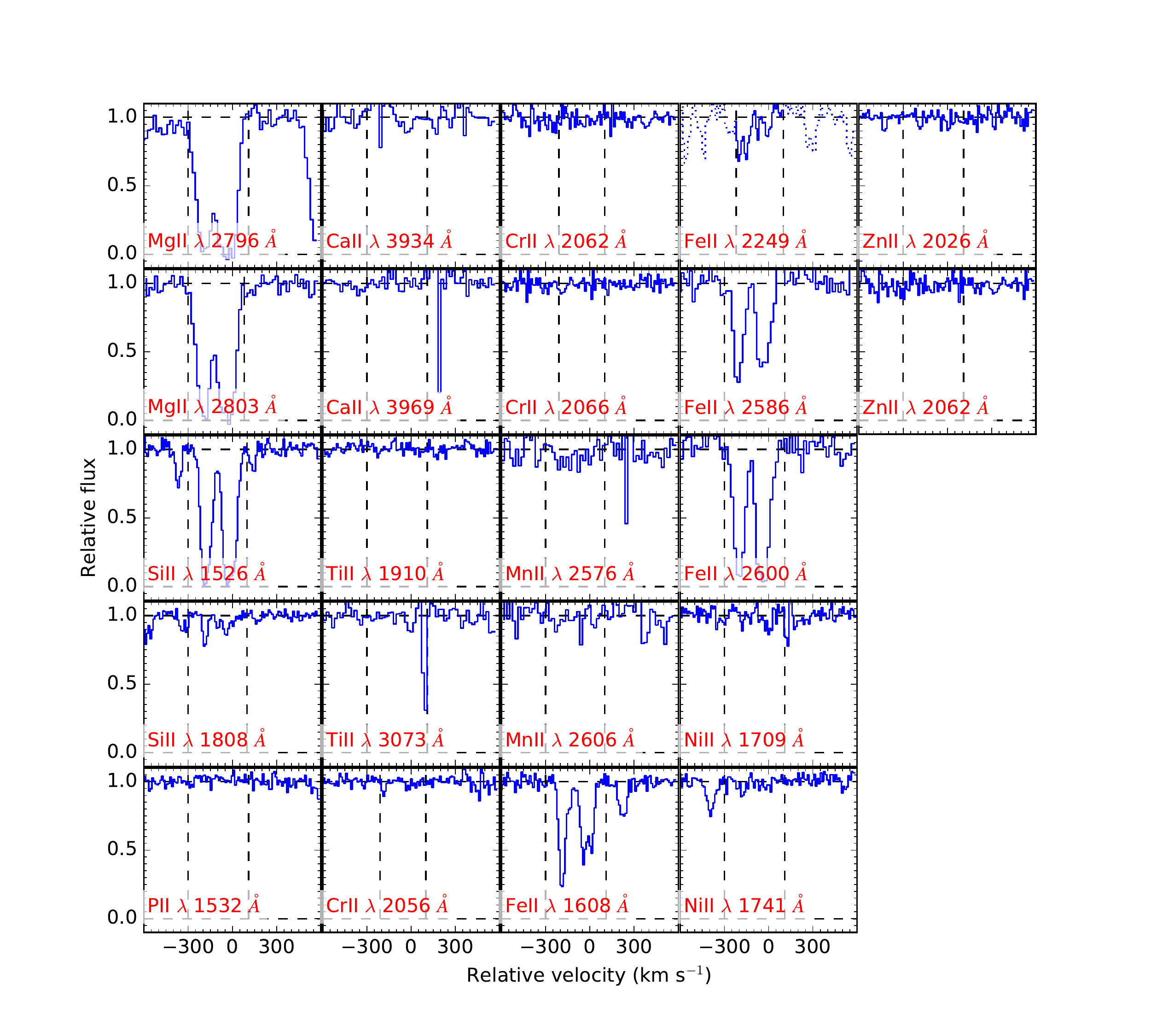}
\caption{Velocity profile of the XQ100 spectrum towards J0255+0048 (\zabs{}=3.256).}
\label{fig:J0255+0048,32555}
\end{center}
\end{figure*}

\begin{table*}
\begin{center}
\caption{X-Shooter metal column densities for J0255+0048 (\zabs{}=$3.914$)}
\label{tab:J0255+0048,39145}
\begin{tabular}{lcccccccc}
\hline
Ion& Line& $\lambda$& $f$& $v_{\rm min}$& $v_{\rm max}$& logN(X)& Included& logN$_{\rm adopt}$\\
 &  & \AA{}&  & \kms{}& \kms{}&  &  &  \\
\hline
C\textsc{ii}& 1334& 1334.532& 1.28E-01& -70& 70& $>14.55$& Y& --\\
C\textsc{ii}& & & & & & & & $>14.55$\\
\hline
O\textsc{i}& 1302& 1302.168& 4.89E-02& -70& 70& $>14.95$& Y& --\\
O\textsc{i}& & & & & & & & $>14.95$\\
\hline
Si\textsc{ii}& 1260& 1260.422& 1.01E+00& -70& 90& $>13.61$& N& --\\
Si\textsc{ii}& 1304& 1304.370& 9.40E-02& -70& 90& $>14.58$& N& --\\
\hline
\end{tabular}
\end{center}
\end{table*} 

\begin{figure*}
\begin{center}
\includegraphics[width=1.1\textwidth]{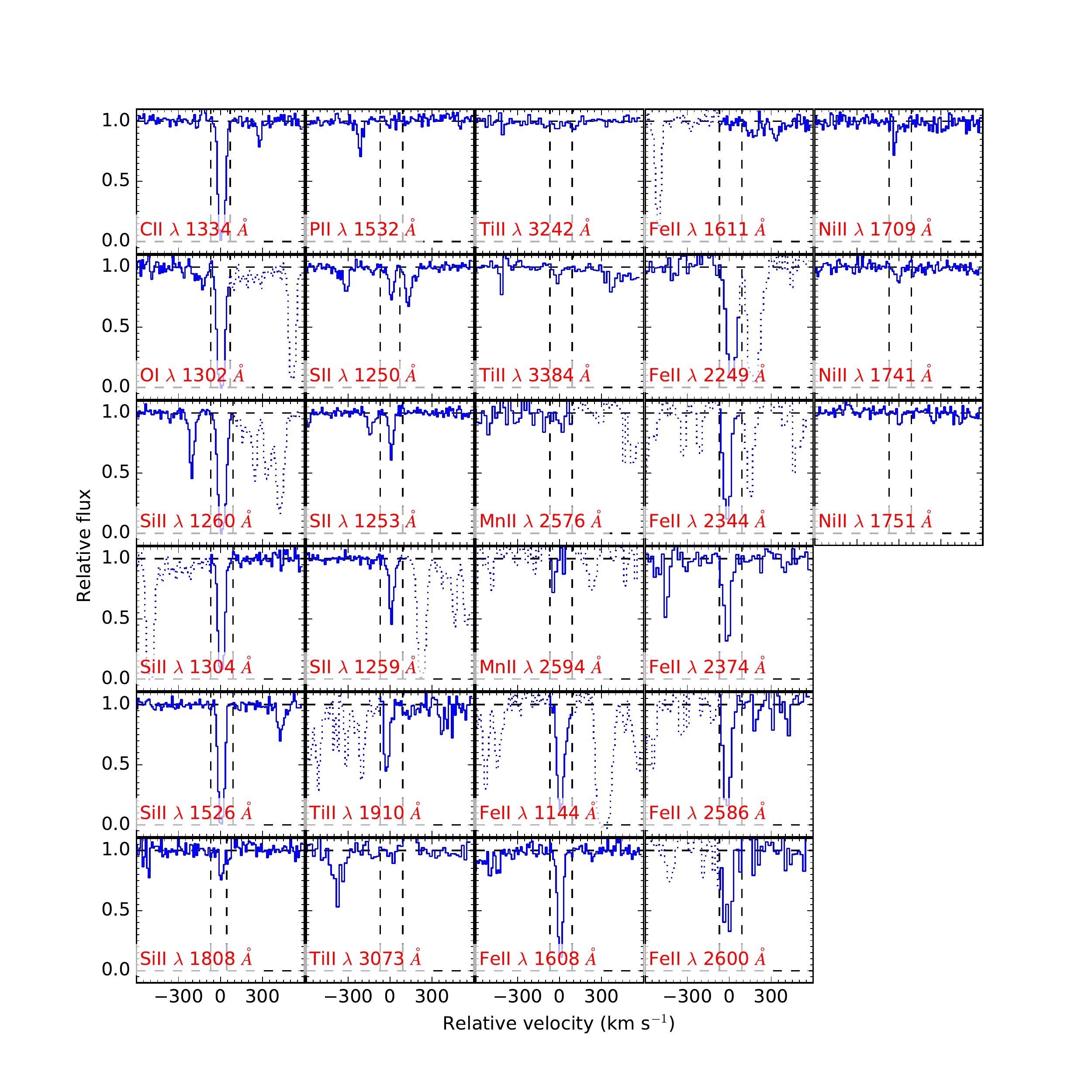}
\caption{Velocity profile of the XQ100 spectrum towards J0255+0048 (\zabs{}=3.914).}
\label{fig:J0255+0048,39145}
\end{center}
\end{figure*}

\begin{table*}
\begin{center}
\caption{X-Shooter metal column densities for J0307$-$4945 (\zabs{}=$3.591$)}
\label{tab:J0307-4945,35910}
\begin{tabular}{lcccccccc}
\hline
Ion& Line& $\lambda$& $f$& $v_{\rm min}$& $v_{\rm max}$& logN(X)& Included& logN$_{\rm adopt}$\\
 &  & \AA{}&  & \kms{}& \kms{}&  &  &  \\
\hline
O\textsc{i}& 1302& 1302.168& 4.89E-02& -170& 100& $>15.17$& Y& --\\
O\textsc{i}& & & & & & & & $>15.17$\\
\hline
Mg\textsc{i}& 2852& 2852.964& 1.81E+00& -170& 100& $12.21\pm0.02$& Y& --\\
Mg\textsc{i}& & & & & & & & $12.21\pm0.05$\\
\hline
Mg\textsc{ii}& 2796& 2796.352& 6.12E-01& -170& 210& $>13.98$& N& --\\
Mg\textsc{ii}& 2803& 2803.531& 3.05E-01& -170& 200& $>14.26$& Y& --\\
Mg\textsc{ii}& & & & & & & & $>14.26$\\
\hline
Si\textsc{ii}& 1808& 1808.013& 2.19E-03& -60& 25& $<14.26$& Y& --\\
Si\textsc{ii}& & & & & & & & $<14.26$\\
\hline
Ti\textsc{ii}& 1910& 1910.750& 2.02E-01& -90& 60& $<12.41$& N& --\\
Ti\textsc{ii}& 3384& 3384.740& 3.58E-01& -90& 60& $<12.18$& Y& --\\
Ti\textsc{ii}& & & & & & & & $<12.18$\\
\hline
Mn\textsc{ii}& 2576& 2576.877& 3.51E-01& -40& 50& $<12.05$& N& --\\
Mn\textsc{ii}& 2594& 2594.499& 2.71E-01& -40& 50& $<11.93$& Y& --\\
Mn\textsc{ii}& 2606& 2606.462& 1.93E-01& -40& 50& $<11.87$& Y& --\\
Mn\textsc{ii}& & & & & & & & $<11.87$\\
\hline
Fe\textsc{ii}& 1608& 1608.451& 5.80E-02& -30& 100& $14.16\pm0.01$& Y& --\\
Fe\textsc{ii}& 1611& 1611.200& 1.36E-03& -40& 50& $<14.74$& N& --\\
Fe\textsc{ii}& 2249& 2249.877& 1.82E-03& -70& 100& $<14.68$& N& --\\
Fe\textsc{ii}& 2344& 2344.214& 1.14E-01& -170& 100& $14.12\pm0.01$& Y& --\\
Fe\textsc{ii}& 2374& 2374.461& 3.13E-02& -90& 100& $14.19\pm0.02$& Y& --\\
Fe\textsc{ii}& 2586& 2586.650& 6.91E-02& -120& 100& $14.20\pm0.01$& Y& --\\
Fe\textsc{ii}& 2600& 2600.173& 2.39E-01& -180& 110& $>14.02$& N& --\\
Fe\textsc{ii}& & & & & & & & $14.17\pm0.05$\\
\hline
Ni\textsc{ii}& 1709& 1709.604& 3.24E-02& -90& 80& $<13.21$& Y& --\\
Ni\textsc{ii}& 1741& 1741.553& 4.27E-02& -90& 80& $<13.23$& N& --\\
Ni\textsc{ii}& 1751& 1751.916& 2.77E-02& -90& 80& $<13.88$& N& --\\
Ni\textsc{ii}& & & & & & & & $<13.21$\\
\hline
\end{tabular}
\end{center}
\end{table*} 

\begin{figure*}
\begin{center}
\includegraphics[width=1.1\textwidth]{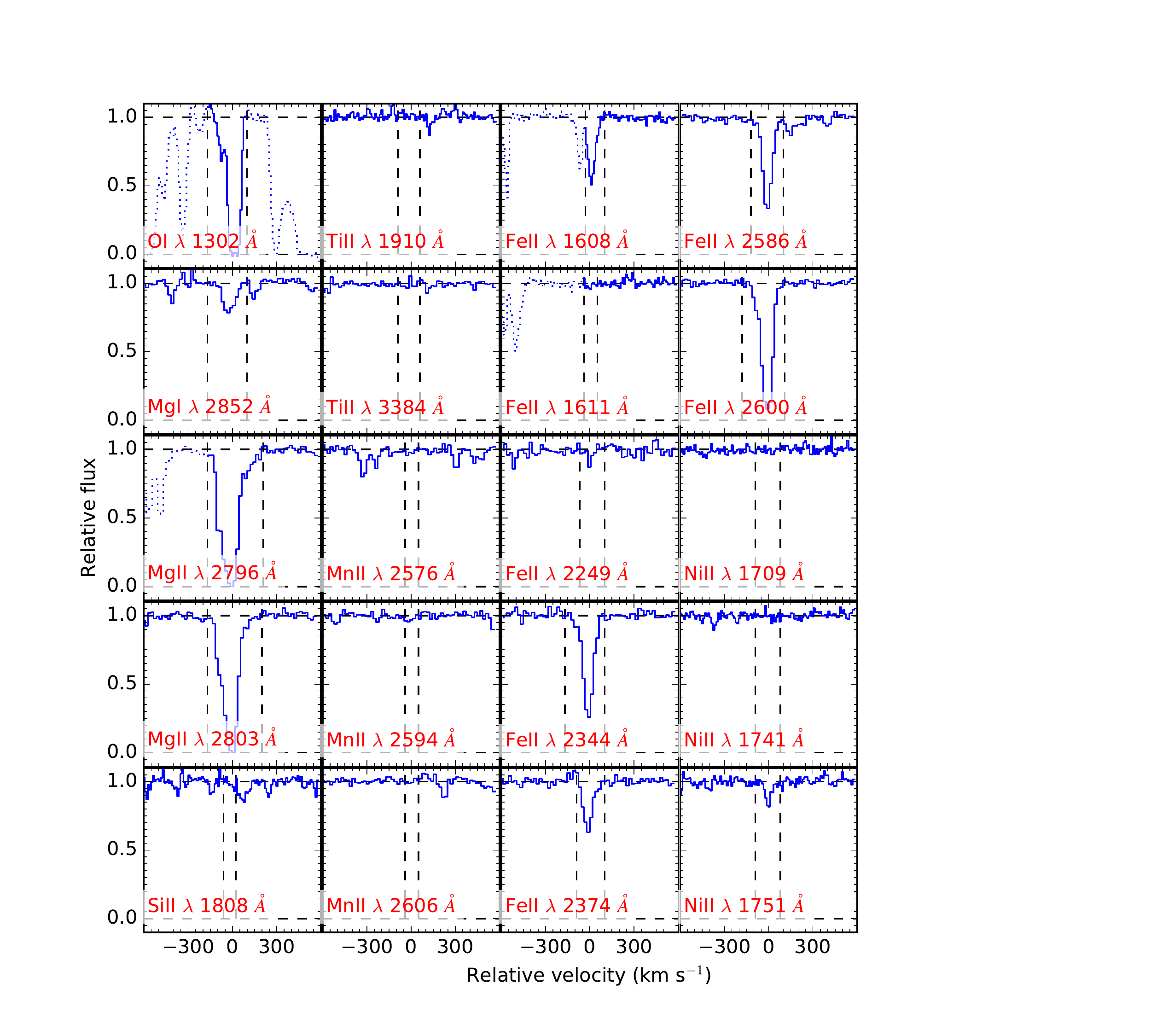}
\caption{Velocity profile of the XQ100 spectrum towards J0307-4945 (\zabs{}=3.591).}
\label{fig:J0307-4945,35910}
\end{center}
\end{figure*}

\begin{table*}
\begin{center}
\caption{X-Shooter metal column densities for J0307$-$4945 (\zabs{}=$4.466$)}
\label{tab:J0307-4945,44665}
\begin{tabular}{lcccccccc}
\hline
Ion& Line& $\lambda$& $f$& $v_{\rm min}$& $v_{\rm max}$& logN(X)& Included& logN$_{\rm adopt}$\\
 &  & \AA{}&  & \kms{}& \kms{}&  &  &  \\
\hline
C\textsc{ii}& 1334& 1334.532& 1.28E-01& -200& 140& $>15.01$& Y& --\\
C\textsc{ii}& & & & & & & & $>15.01$\\
\hline
O\textsc{i}& 1302& 1302.168& 4.89E-02& -200& 140& $>15.13$& Y& --\\
O\textsc{i}& & & & & & & & $>15.13$\\
\hline
Mg\textsc{i}& 2852& 2852.964& 1.81E+00& -160& 60& $11.81\pm0.10$& Y& --\\
Mg\textsc{i}& & & & & & & & $11.81\pm0.10$\\
\hline
Mg\textsc{ii}& 2796& 2796.352& 6.12E-01& -200& 140& $>14.00$& N& --\\
Mg\textsc{ii}& 2803& 2803.531& 3.05E-01& -200& 140& $>14.18$& Y& --\\
Mg\textsc{ii}& & & & & & & & $>14.18$\\
\hline
Si\textsc{ii}& 1304& 1304.370& 9.40E-02& -200& 140& $14.59\pm0.00$& Y& --\\
Si\textsc{ii}& 1526& 1526.707& 1.27E-01& -200& 140& $>14.46$& N& --\\
Si\textsc{ii}& & & & & & & & $14.59\pm0.05$\\
\hline
P\textsc{ii}& 1532& 1532.533& 7.61E-03& -160& 60& $<14.03$& Y& --\\
P\textsc{ii}& & & & & & & & $<14.03$\\
\hline
Ti\textsc{ii}& 1910& 1910.750& 2.02E-01& -160& 60& $<12.32$& N& --\\
Ti\textsc{ii}& 3073& 3073.877& 1.09E-01& -160& 60& $<12.50$& N& --\\
Ti\textsc{ii}& 3242& 3242.929& 2.32E-01& -160& 60& $<11.97$& Y& --\\
Ti\textsc{ii}& & & & & & & & $<11.97$\\
\hline
Fe\textsc{ii}& 1608& 1608.451& 5.80E-02& -200& 140& $14.12\pm0.01$& Y& --\\
Fe\textsc{ii}& 1611& 1611.200& 1.36E-03& -160& 60& $<14.53$& N& --\\
Fe\textsc{ii}& 2249& 2249.877& 1.82E-03& -160& 60& $<14.59$& N& --\\
Fe\textsc{ii}& 2344& 2344.214& 1.14E-01& -200& 140& $14.10\pm0.01$& Y& --\\
Fe\textsc{ii}& 2374& 2374.461& 3.13E-02& -210& 140& $14.17\pm0.03$& Y& --\\
Fe\textsc{ii}& & & & & & & & $14.11\pm0.05$\\
\hline
Ni\textsc{ii}& 1751& 1751.916& 2.77E-02& -30& 60& $<13.43$& Y& --\\
Ni\textsc{ii}& & & & & & & & $<13.43$\\
\hline
Zn\textsc{ii}& 2026& 2026.136& 4.89E-01& -160& 60& $<12.31$& Y& --\\
Zn\textsc{ii}& & & & & & & & $<12.31$\\
\hline
\end{tabular}
\end{center}
\end{table*} 

\begin{figure*}
\begin{center}
\includegraphics[width=1.1\textwidth]{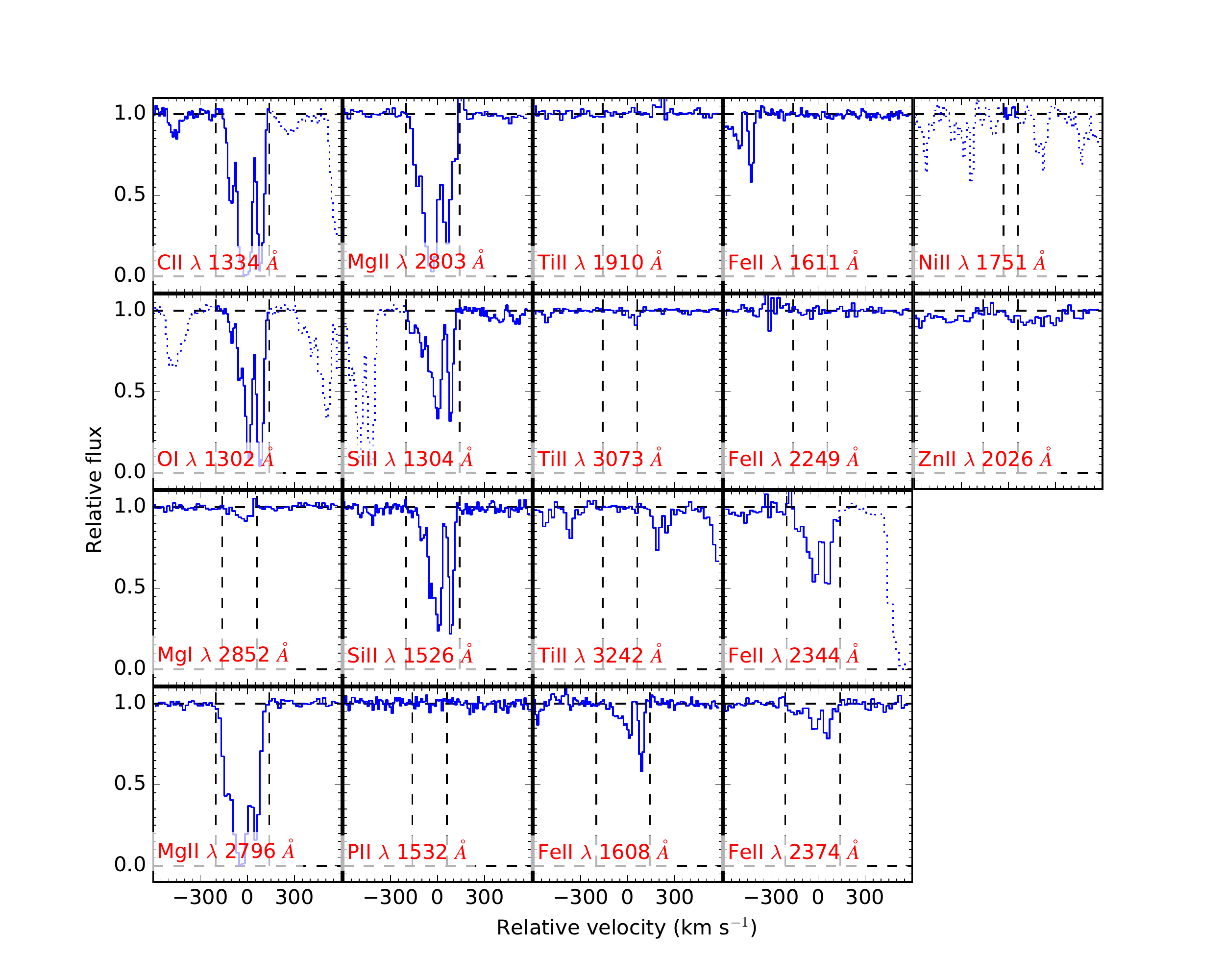}
\caption{Velocity profile of the XQ100 spectrum towards J0307-4945 (\zabs{}=4.466).}
\label{fig:J0307-4945,44665}
\end{center}
\end{figure*}

\clearpage

\begin{table*}
\begin{center}
\caption{X-Shooter metal column densities for J0415$-$4357 (\zabs{}=$3.808$)}
\label{tab:J0415-4357,38080}
\begin{tabular}{lcccccccc}
\hline
Ion& Line& $\lambda$& $f$& $v_{\rm min}$& $v_{\rm max}$& logN(X)& Included& logN$_{\rm adopt}$\\
 &  & \AA{}&  & \kms{}& \kms{}&  &  &  \\
\hline
C\textsc{ii}& 1334& 1334.532& 1.28E-01& -150& 130& $>14.98$& Y& --\\
C\textsc{ii}& & & & & & & & $>14.98$\\
\hline
O\textsc{i}& 1302& 1302.168& 4.89E-02& -210& 90& $>15.50$& Y& --\\
O\textsc{i}& & & & & & & & $>15.50$\\
\hline
Mg\textsc{ii}& 2796& 2796.352& 6.12E-01& -210& 110& $>13.95$& Y& --\\
Mg\textsc{ii}& & & & & & & & $>13.95$\\
\hline
\end{tabular}
\end{center}
\end{table*} 

\begin{figure*}
\begin{center}
\includegraphics[width=1.1\textwidth]{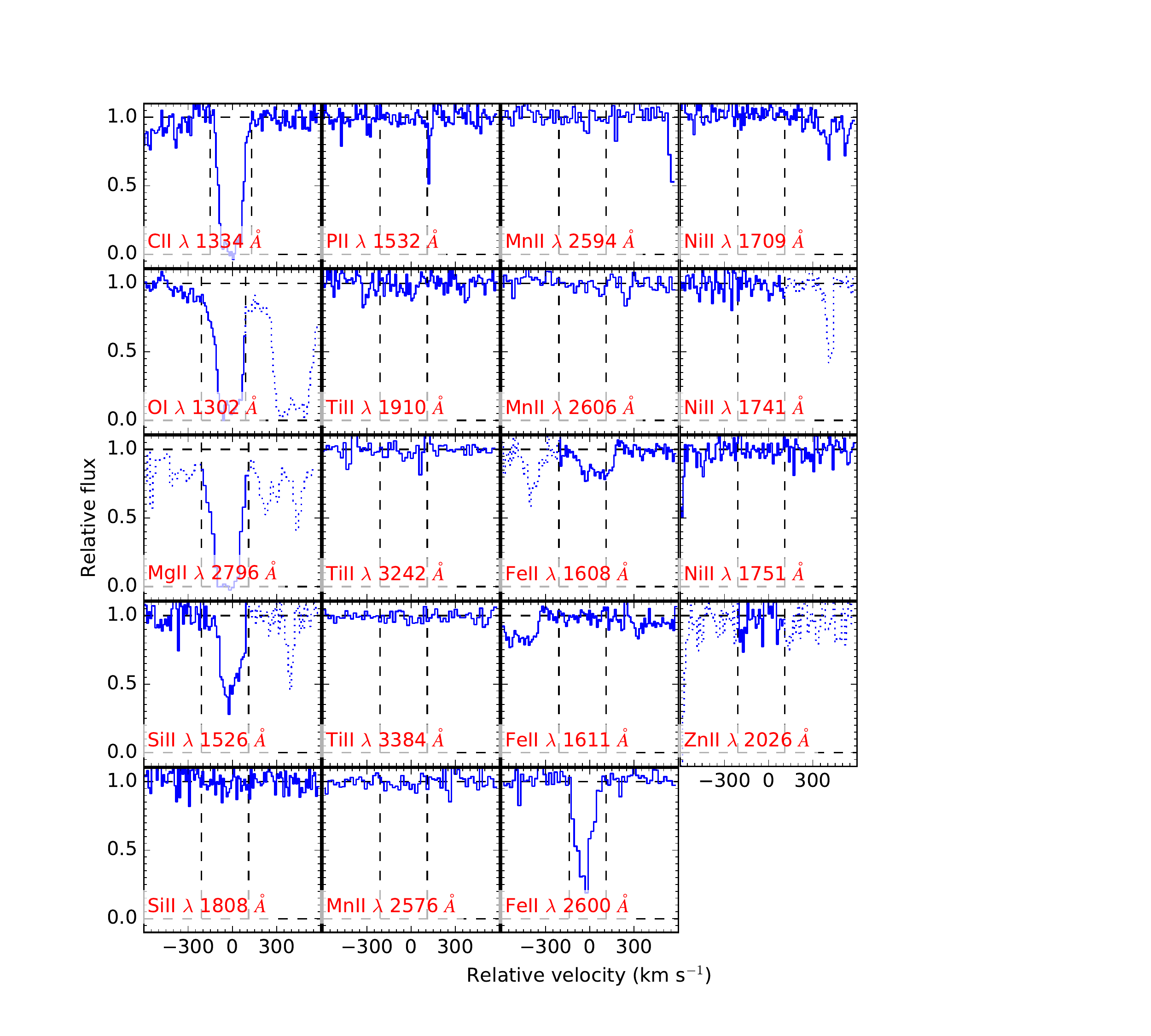}
\caption{Velocity profile of the XQ100 spectrum towards J0415-4357 (\zabs{}=3.808).}
\label{fig:J0415-4357,38080}
\end{center}
\end{figure*}

\begin{table*}
\begin{center}
\caption{X-Shooter metal column densities for J0424$-$2209 (\zabs{}=$2.982$)}
\label{tab:J0424-2209,29825}
\begin{tabular}{lcccccccc}
\hline
Ion& Line& $\lambda$& $f$& $v_{\rm min}$& $v_{\rm max}$& logN(X)& Included& logN$_{\rm adopt}$\\
 &  & \AA{}&  & \kms{}& \kms{}&  &  &  \\
\hline
Mg\textsc{ii}& 2796& 2796.352& 6.12E-01& -80& 100& $>13.62$& Y& --\\
Mg\textsc{ii}& & & & & & & & $>13.62$\\
\hline
Si\textsc{ii}& 1808& 1808.013& 2.19E-03& -80& 100& $<15.08$& Y& --\\
Si\textsc{ii}& & & & & & & & $14.44$--$15.08$\\
\hline
Ca\textsc{ii}& 3934& 3934.777& 6.50E-01& -80& 60& $<12.07$& Y& --\\
Ca\textsc{ii}& 3969& 3969.591& 3.22E-01& -80& 60& $<12.51$& N& --\\
\hline
\end{tabular}
\end{center}
\end{table*} 

\begin{figure*}
\begin{center}
\includegraphics[width=1.1\textwidth]{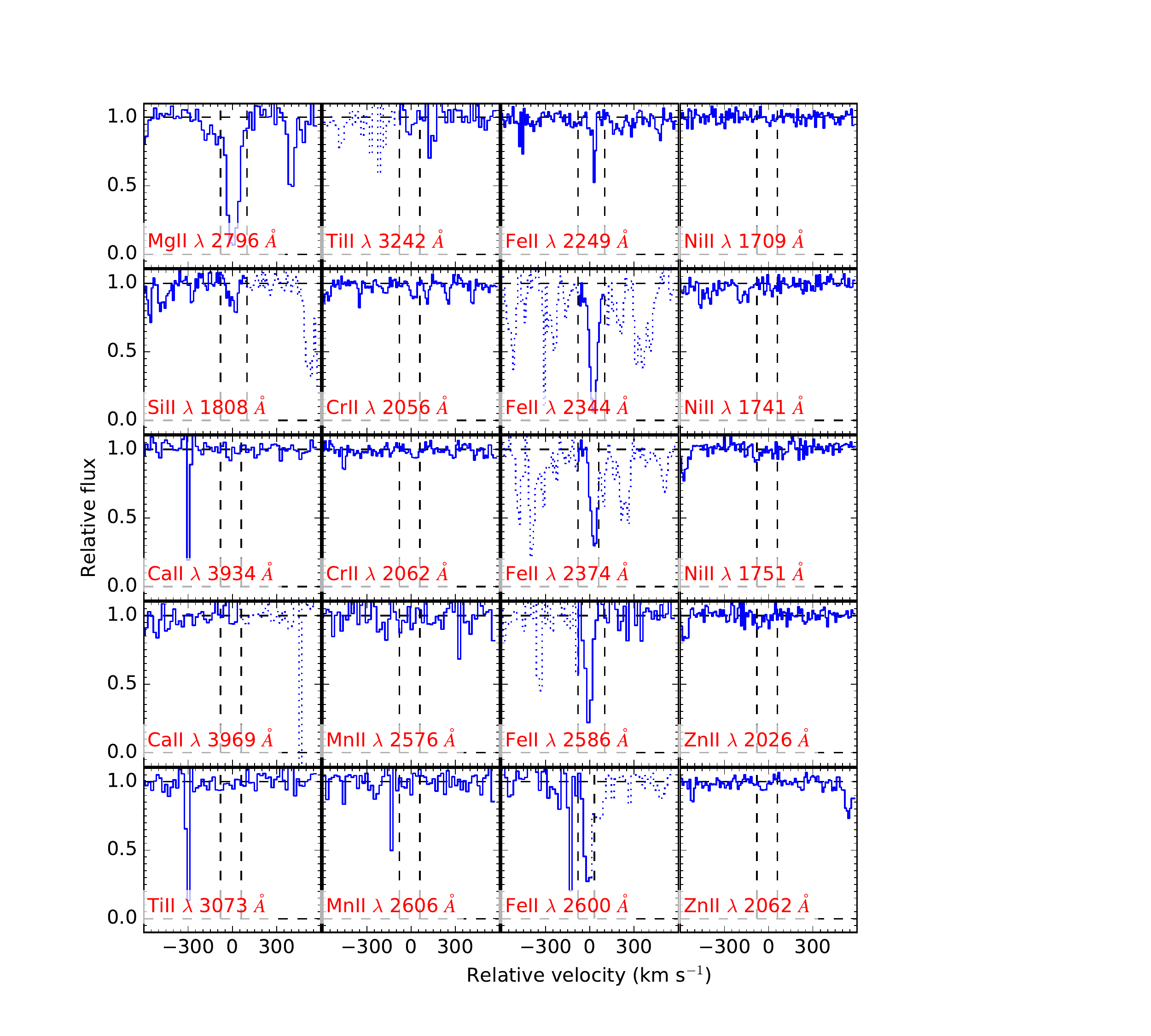}
\caption{Velocity profile of the XQ100 spectrum towards J0424-2209 (\zabs{}=2.982).}
\label{fig:J0424-2209,29825}
\end{center}
\end{figure*}

\clearpage
\begin{table*}
\begin{center}
\caption{X-Shooter metal column densities for J0529$-$3552 (\zabs{}=$3.684$)}
\label{tab:J0529-3552,36840}
\begin{tabular}{lcccccccc}
\hline
Ion& Line& $\lambda$& $f$& $v_{\rm min}$& $v_{\rm max}$& logN(X)& Included& logN$_{\rm adopt}$\\
 &  & \AA{}&  & \kms{}& \kms{}&  &  &  \\
\hline
Mg\textsc{ii}& 2796& 2796.352& 6.12E-01& -80& 50& $13.09\pm0.03$& Y& --\\
Mg\textsc{ii}& 2803& 2803.531& 3.05E-01& -80& 50& $13.08\pm0.07$& Y& --\\
Mg\textsc{ii}& & & & & & & & $13.08\pm0.08$\\
\hline
Si\textsc{ii}& 1526& 1526.707& 1.27E-01& -80& 50& $13.53\pm0.05$& Y& --\\
Si\textsc{ii}& & & & & & & & $13.53\pm0.05$\\
\hline
Fe\textsc{ii}& 1608& 1608.451& 5.80E-02& -80& 50& $<13.80$& Y& --\\
Fe\textsc{ii}& & & & & & & & $<13.80$\\
\hline
\end{tabular}
\end{center}
\end{table*} 

\begin{figure*}
\begin{center}
\includegraphics[width=1.1\textwidth]{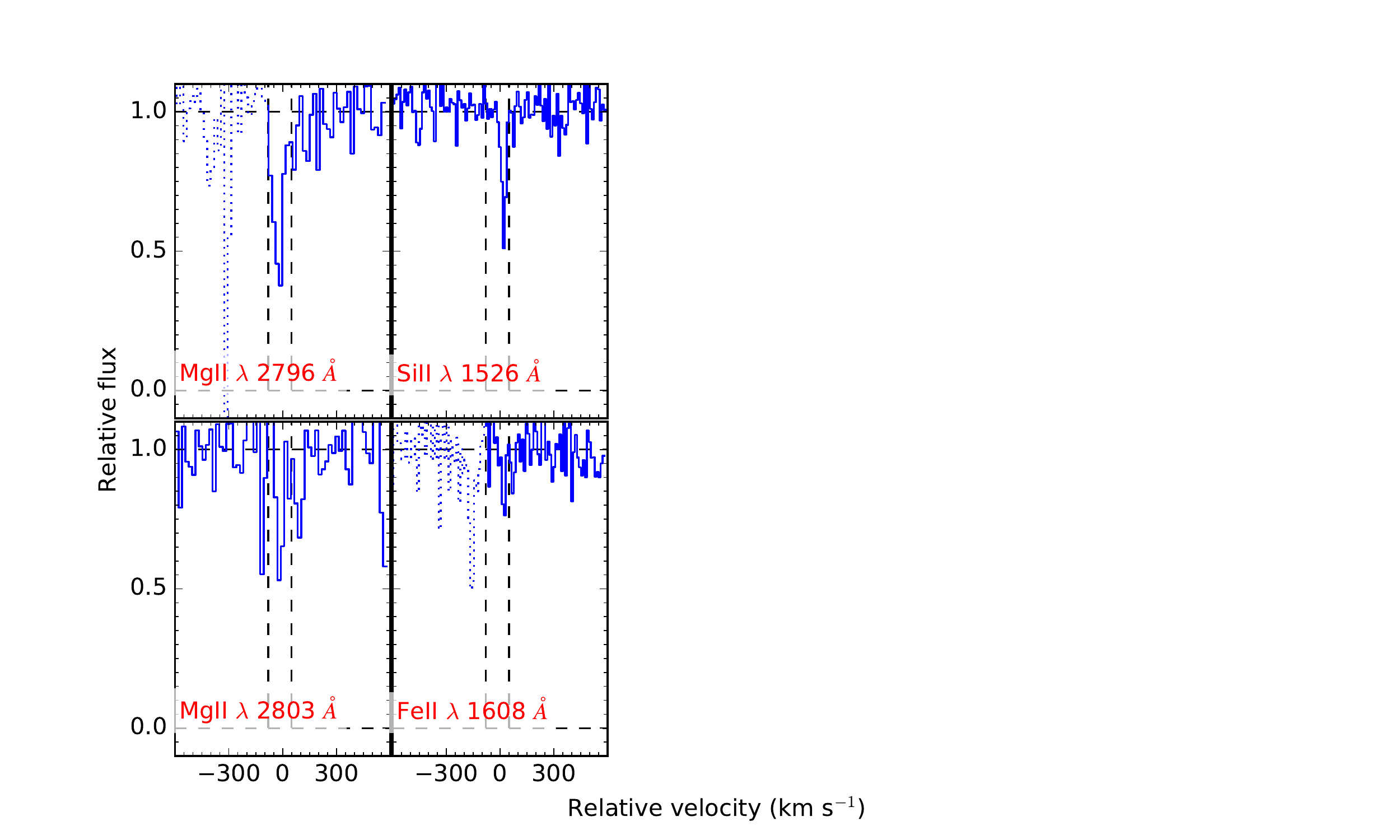}
\caption{Velocity profile of the XQ100 spectrum towards J0529-3552 (\zabs{}=3.684).}
\label{fig:J0529-3552,36840}
\end{center}
\end{figure*}

\begin{table*}
\begin{center}
\caption{X-Shooter metal column densities for J0747+2739 (\zabs{}=$3.424$)}
\label{tab:J0747+2739,34235}
\begin{tabular}{lcccccccc}
\hline
Ion& Line& $\lambda$& $f$& $v_{\rm min}$& $v_{\rm max}$& logN(X)& Included& logN$_{\rm adopt}$\\
 &  & \AA{}&  & \kms{}& \kms{}&  &  &  \\
\hline
Mg\textsc{i}& 2852& 2852.964& 1.81E+00& -200& 100& $<11.85$& Y& --\\
Mg\textsc{i}& & & & & & & & $<11.85$\\
\hline
Mg\textsc{ii}& 2796& 2796.352& 6.12E-01& -200& 100& $>13.99$& N& --\\
Mg\textsc{ii}& 2803& 2803.531& 3.05E-01& -200& 100& $>14.13$& Y& --\\
Mg\textsc{ii}& & & & & & & & $>14.13$\\
\hline
Si\textsc{ii}& 1526& 1526.707& 1.27E-01& -200& 100& $>14.58$& Y& --\\
Si\textsc{ii}& 1808& 1808.013& 2.19E-03& -200& 100& $<14.88$& Y& --\\
Si\textsc{ii}& & & & & & & & $14.58$--$14.88$\\
\hline
P\textsc{ii}& 1532& 1532.533& 7.61E-03& -180& 90& $<14.23$& Y& --\\
P\textsc{ii}& & & & & & & & $<14.23$\\
\hline
Ti\textsc{ii}& 1910& 1910.750& 2.02E-01& -180& 90& $<12.74$& Y& --\\
Ti\textsc{ii}& & & & & & & & $<12.74$\\
\hline
Fe\textsc{ii}& 1608& 1608.451& 5.80E-02& -200& 100& $14.43\pm0.01$& Y& --\\
Fe\textsc{ii}& 1611& 1611.200& 1.36E-03& -200& 100& $14.87\pm0.15$& N& --\\
Fe\textsc{ii}& 2344& 2344.214& 1.14E-01& -200& 100& $>14.29$& N& --\\
Fe\textsc{ii}& 2374& 2374.461& 3.13E-02& -200& 100& $14.47\pm0.00$& Y& --\\
Fe\textsc{ii}& 2586& 2586.650& 6.91E-02& -200& 100& $>14.82$& N& --\\
Fe\textsc{ii}& 2600& 2600.173& 2.39E-01& -200& 100& $>14.49$& N& --\\
Fe\textsc{ii}& & & & & & & & $14.47\pm0.05$\\
\hline
Ni\textsc{ii}& 1709& 1709.604& 3.24E-02& -180& 90& $<13.49$& N& --\\
Ni\textsc{ii}& 1741& 1741.553& 4.27E-02& -180& 90& $<13.46$& Y& --\\
Ni\textsc{ii}& 1751& 1751.916& 2.77E-02& -180& 90& $<13.55$& N& --\\
Ni\textsc{ii}& & & & & & & & $<13.46$\\
\hline
\end{tabular}
\end{center}
\end{table*} 

\begin{figure*}
\begin{center}
\includegraphics[width=1.1\textwidth]{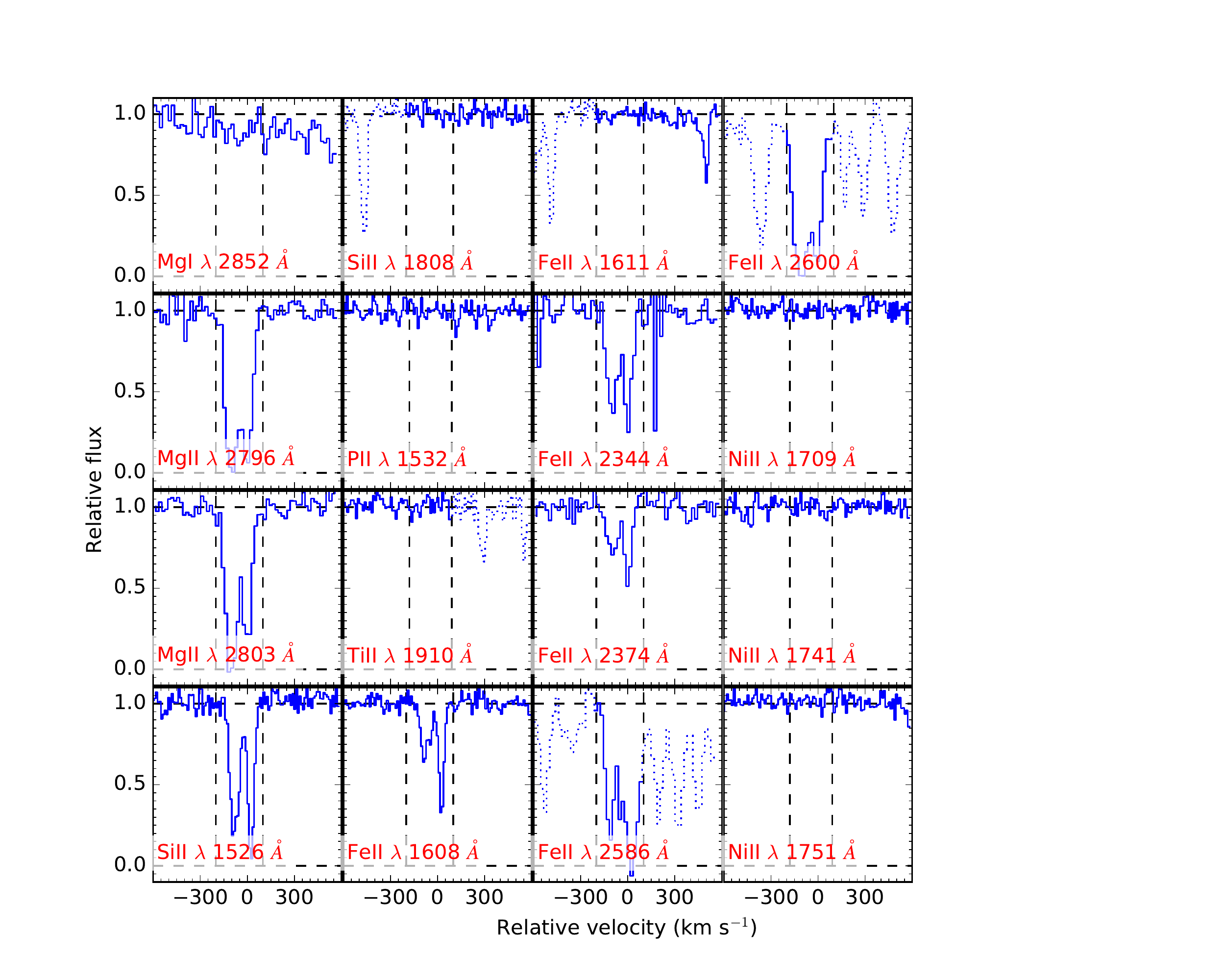}
\caption{Velocity profile of the XQ100 spectrum towards J0747+2739 (\zabs{}=3.424).}
\label{fig:J0747+2739,34235}
\end{center}
\end{figure*}

\begin{table*}
\begin{center}
\caption{X-Shooter metal column densities for J0747+2739 (\zabs{}=$3.901$)}
\label{tab:J0747+2739,39010}
\begin{tabular}{lcccccccc}
\hline
Ion& Line& $\lambda$& $f$& $v_{\rm min}$& $v_{\rm max}$& logN(X)& Included& logN$_{\rm adopt}$\\
 &  & \AA{}&  & \kms{}& \kms{}&  &  &  \\
\hline
C\textsc{ii}& 1036& 1036.337& 1.23E-01& -220& 50& $14.31\pm0.02$& Y& --\\
C\textsc{ii}& 1334& 1334.532& 1.28E-01& -220& 50& $>14.54$& N& --\\
C\textsc{ii}& & & & & & & & $14.31\pm0.05$\\
\hline
O\textsc{i}& 1302& 1302.168& 4.89E-02& -220& 50& $>14.90$& Y& --\\
O\textsc{i}& & & & & & & & $>14.90$\\
\hline
Si\textsc{ii}& 1304& 1304.370& 9.40E-02& -220& 50& $14.07\pm0.02$& Y& --\\
\hline
\end{tabular}
\end{center}
\end{table*} 

\begin{figure*}
\begin{center}
\includegraphics[width=1.1\textwidth]{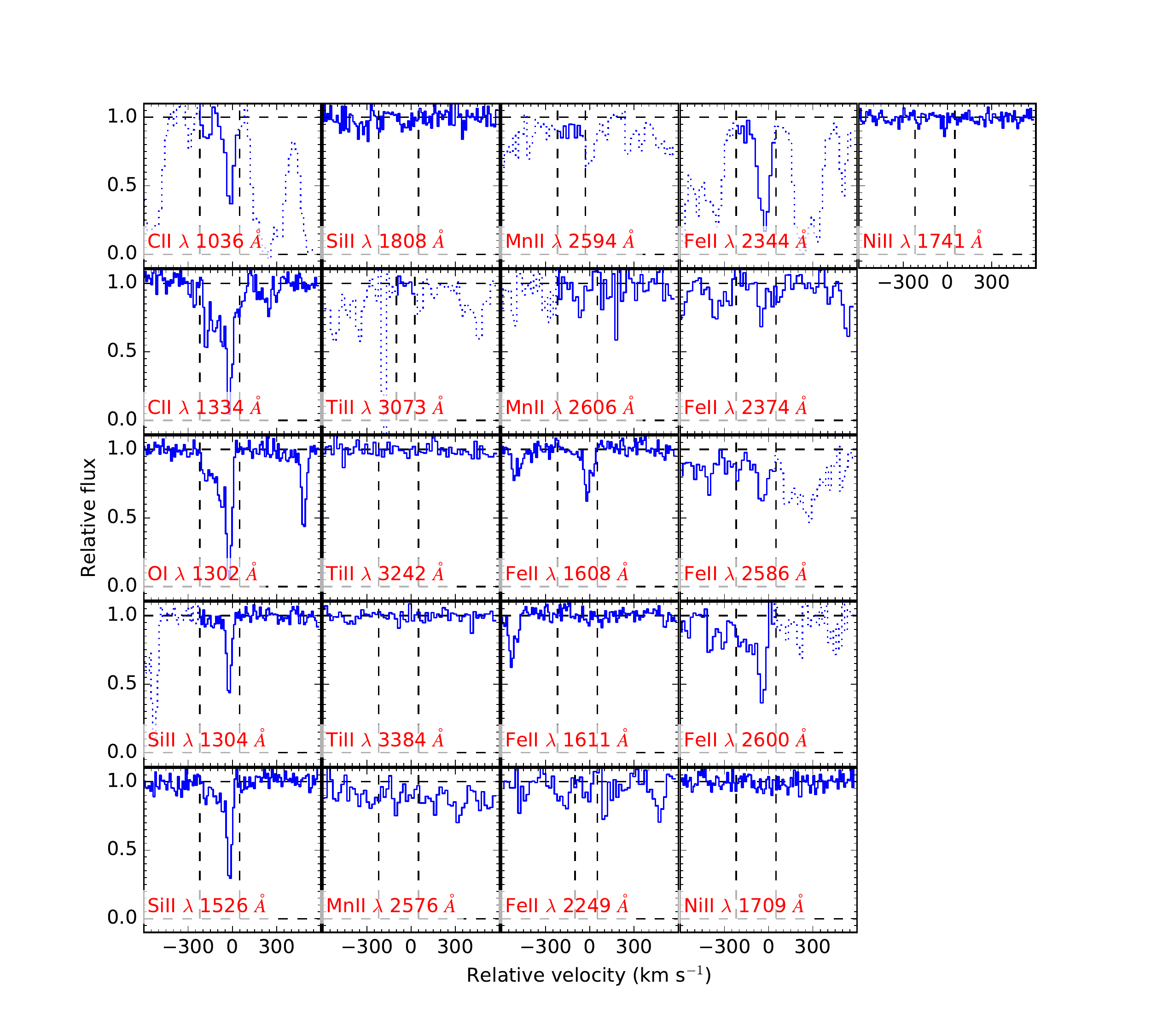}
\caption{Velocity profile of the XQ100 spectrum towards J0747+2739 (\zabs{}=3.901).}
\label{fig:J0747+2739,39010}
\end{center}
\end{figure*}

\clearpage
\begin{table*}
\begin{center}
\caption{X-Shooter metal column densities for J0800+1920 (\zabs{}=$3.946$)}
\label{tab:J0800+1920,39465}
\begin{tabular}{lcccccccc}
\hline
Ion& Line& $\lambda$& $f$& $v_{\rm min}$& $v_{\rm max}$& logN(X)& Included& logN$_{\rm adopt}$\\
 &  & \AA{}&  & \kms{}& \kms{}&  &  &  \\
\hline
C\textsc{ii}& 1334& 1334.532& 1.28E-01& -65& 60& $>14.58$& Y& --\\
C\textsc{ii}& & & & & & & & $>14.58$\\
\hline
O\textsc{i}& 1302& 1302.168& 4.89E-02& -65& 60& $>15.06$& Y& --\\
O\textsc{i}& & & & & & & & $>15.06$\\
\hline
Si\textsc{ii}& 1260& 1260.422& 1.01E+00& -165& 60& $>13.74$& N& --\\
Si\textsc{ii}& 1304& 1304.370& 9.40E-02& -165& 60& $14.04\pm0.01$& Y& --\\
\hline
\end{tabular}
\end{center}
\end{table*} 

\begin{figure*}
\begin{center}
\includegraphics[width=1.1\textwidth]{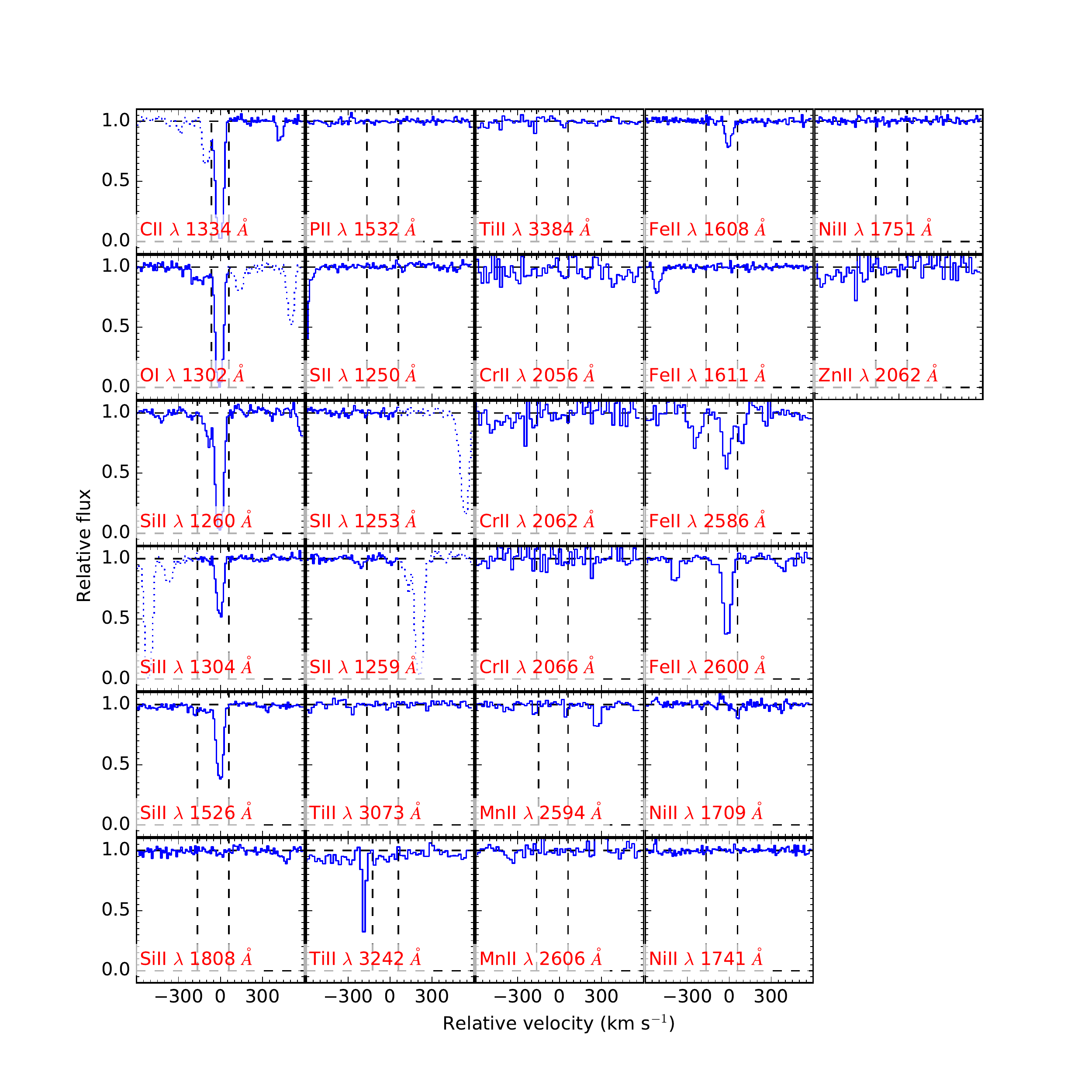}
\caption{Velocity profile of the XQ100 spectrum towards J0800+1920 (\zabs{}=3.946).}
\label{fig:J0800+1920,39465}
\end{center}
\end{figure*}

\begin{table*}
\begin{center}
\caption{X-Shooter metal column densities for J0818+0958 (\zabs{}=$3.306$)}
\label{tab:J0818+0958,33060}
\begin{tabular}{lcccccccc}
\hline
Ion& Line& $\lambda$& $f$& $v_{\rm min}$& $v_{\rm max}$& logN(X)& Included& logN$_{\rm adopt}$\\
 &  & \AA{}&  & \kms{}& \kms{}&  &  &  \\
\hline
C\textsc{ii}& 1334& 1334.532& 1.28E-01& -100& 300& $>14.88$& Y& --\\
C\textsc{ii}& & & & & & & & $>14.88$\\
\hline
Mg\textsc{ii}& 2796& 2796.352& 6.12E-01& -200& 300& $>13.97$& N& --\\
Mg\textsc{ii}& 2803& 2803.531& 3.05E-01& -200& 300& $>14.12$& Y& --\\
Mg\textsc{ii}& & & & & & & & $>14.12$\\
\hline
Si\textsc{ii}& 1526& 1526.707& 1.27E-01& -100& 135& $>14.59$& N& --\\
\hline
\end{tabular}
\end{center}
\end{table*} 

\begin{figure*}
\begin{center}
\includegraphics[width=1.1\textwidth]{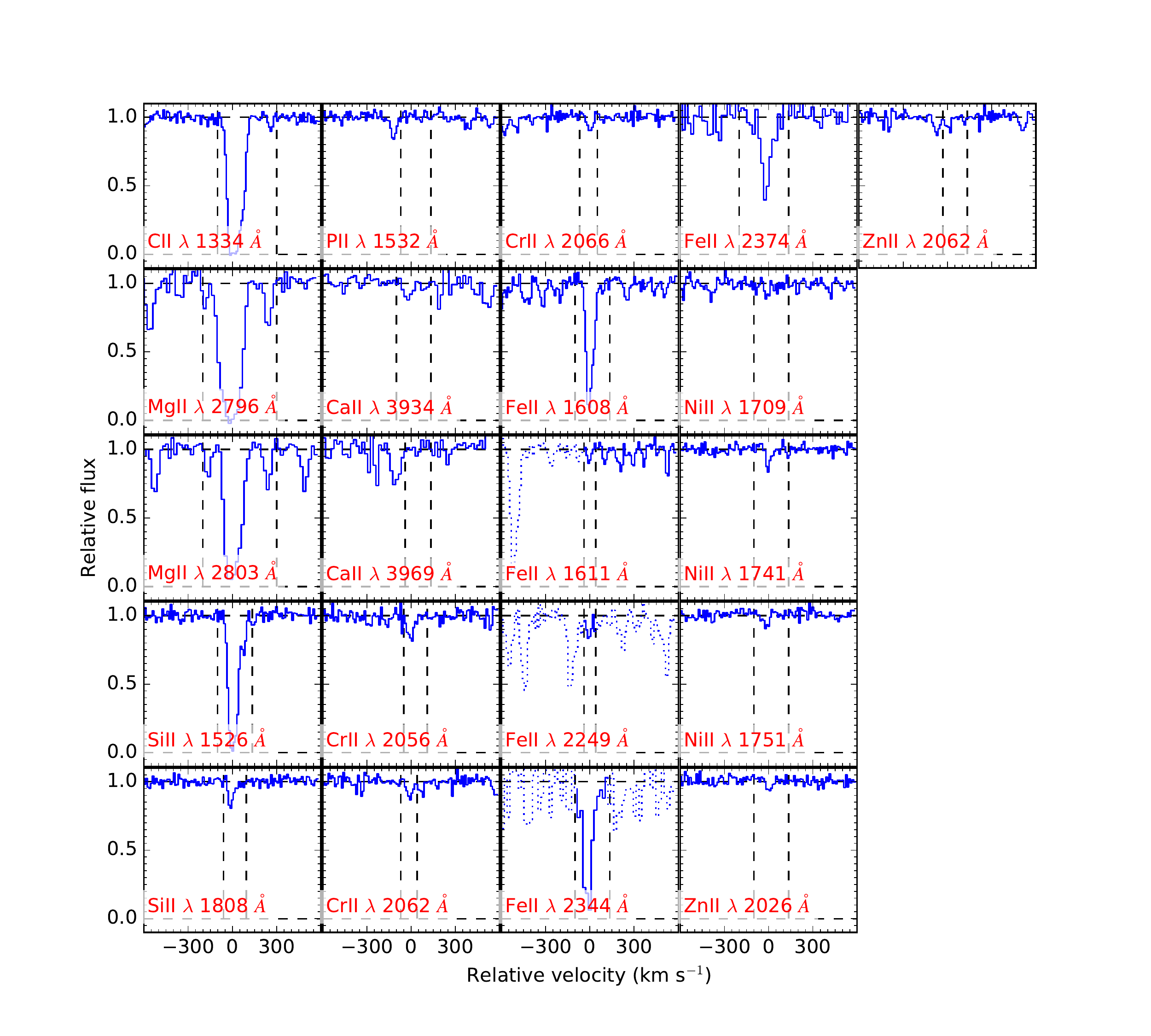}
\caption{Velocity profile of the XQ100 spectrum towards J0818+0958 (\zabs{}=3.306).}
\label{fig:J0818+0958,33060}
\end{center}
\end{figure*}

\begin{table*}
\begin{center}
\caption{X-Shooter metal column densities for J0835+0650 (\zabs{}=$3.955$)}
\label{tab:J0835+0650,39555}
\begin{tabular}{lcccccccc}
\hline
Ion& Line& $\lambda$& $f$& $v_{\rm min}$& $v_{\rm max}$& logN(X)& Included& logN$_{\rm adopt}$\\
 &  & \AA{}&  & \kms{}& \kms{}&  &  &  \\
\hline
C\textsc{ii}& 1334& 1334.532& 1.28E-01& -75& 75& $>14.62$& N& --\\
C\textsc{ii}& & & & & & & & $>14.68$\\
\hline
O\textsc{i}& 1302& 1302.168& 4.89E-02& -75& 90& $>15.02$& Y& --\\
O\textsc{i}& & & & & & & & $>15.02$\\
\hline
Si\textsc{ii}& 1260& 1260.422& 1.01E+00& -75& 90& $>13.72$& N& --\\
Si\textsc{ii}& 1304& 1304.370& 9.40E-02& -75& 90& $14.10\pm0.02$& Y& --\\
\hline
\end{tabular}
\end{center}
\end{table*} 

\begin{figure*}
\begin{center}
\includegraphics[width=1.1\textwidth]{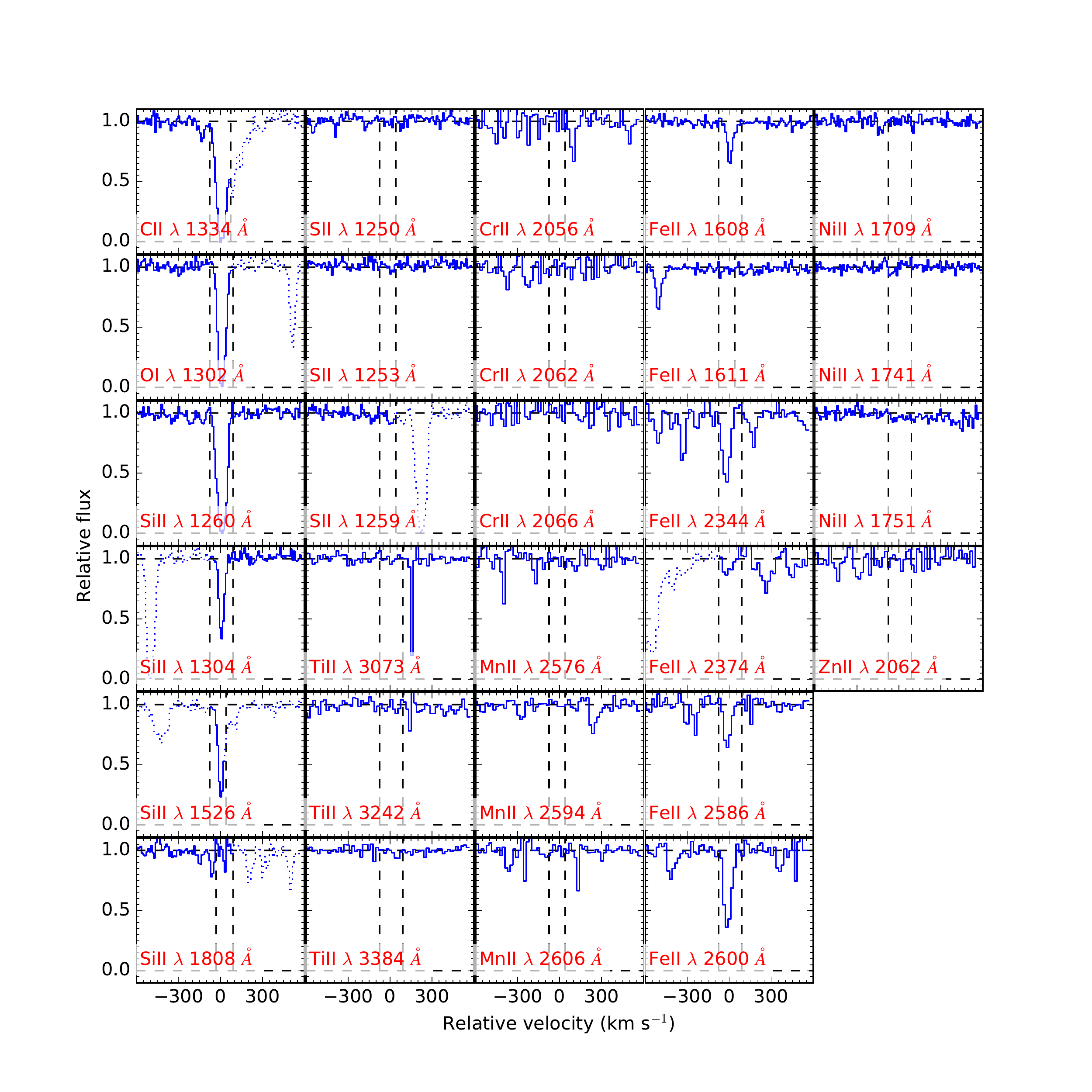}
\caption{Velocity profile of the XQ100 spectrum towards J0835+0650 (\zabs{}=3.955).}
\label{fig:J0835+0650,39555}
\end{center}
\end{figure*}

\begin{table*}
\begin{center}
\caption{X-Shooter metal column densities for J0920+0725 (\zabs{}=$2.238$)}
\label{tab:J0920+0725,22380}
\begin{tabular}{lcccccccc}
\hline
Ion& Line& $\lambda$& $f$& $v_{\rm min}$& $v_{\rm max}$& logN(X)& Included& logN$_{\rm adopt}$\\
 &  & \AA{}&  & \kms{}& \kms{}&  &  &  \\
\hline
Mg\textsc{ii}& 2796& 2796.352& 6.12E-01& -270& 0& $>13.93$& Y& --\\
Mg\textsc{ii}& 2803& 2803.531& 3.05E-01& -270& 20& $>14.27$& N& --\\
Mg\textsc{ii}& & & & & & & & $>13.93$\\
\hline
Si\textsc{ii}& 1526& 1526.707& 1.27E-01& -200& 20& $>14.60$& Y& --\\
Si\textsc{ii}& 1808& 1808.013& 2.19E-03& -200& 0& $<15.08$& N& --\\
Si\textsc{ii}& & & & & & & & $>14.60$\\
\hline
\end{tabular}
\end{center}
\end{table*} 

\begin{figure*}
\begin{center}
\includegraphics[width=1.1\textwidth]{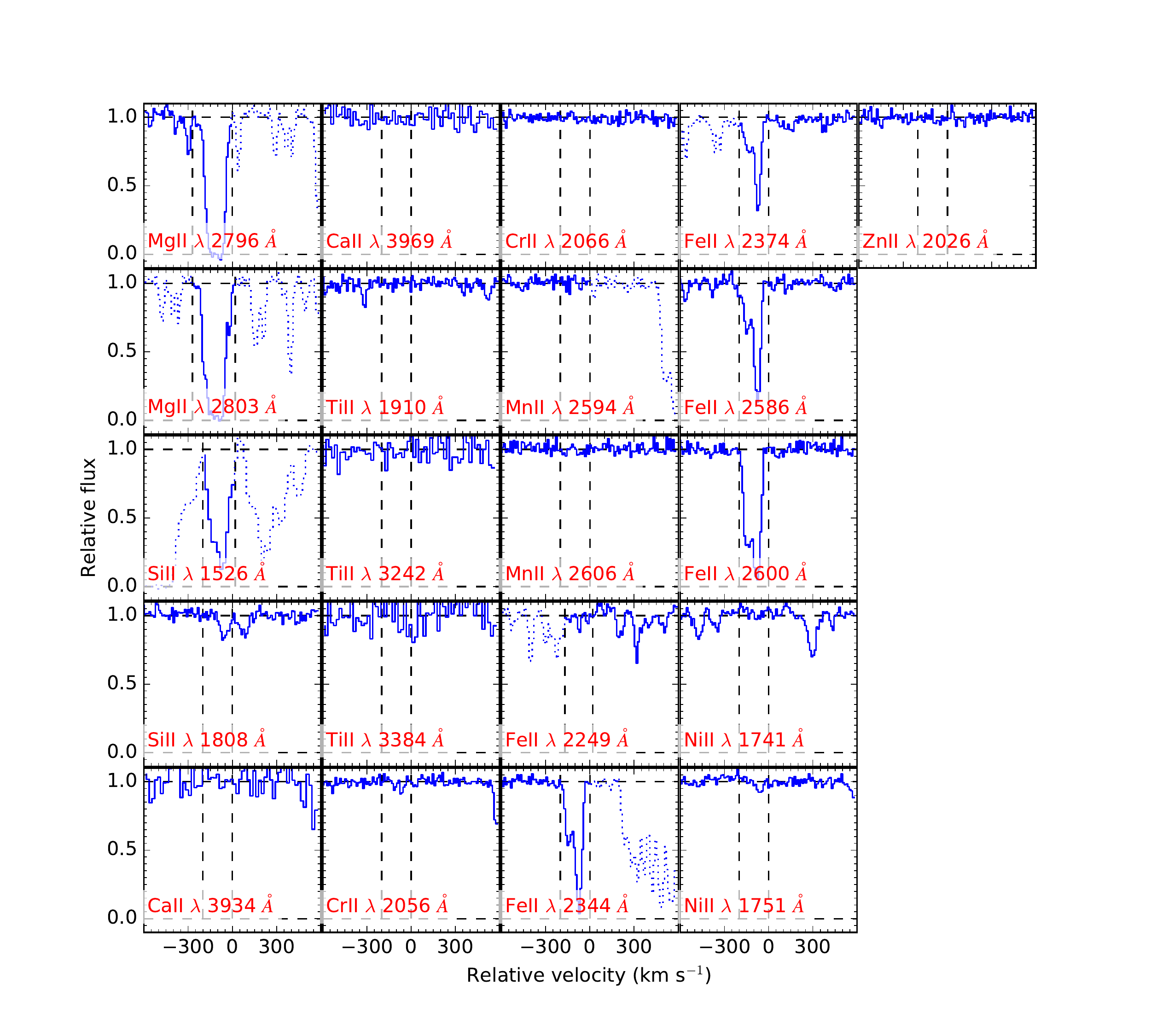}
\caption{Velocity profile of the XQ100 spectrum towards J0920+0725 (\zabs{}=2.238).}
\label{fig:J0920+0725,22380}
\end{center}
\end{figure*}

\clearpage
\begin{table*}
\begin{center}
\caption{X-Shooter metal column densities for J0955$-$0130 (\zabs{}=$4.024$)}
\label{tab:J0955-0130,40245}
\begin{tabular}{lcccccccc}
\hline
Ion& Line& $\lambda$& $f$& $v_{\rm min}$& $v_{\rm max}$& logN(X)& Included& logN$_{\rm adopt}$\\
 &  & \AA{}&  & \kms{}& \kms{}&  &  &  \\
\hline
C\textsc{ii}& 1334& 1334.532& 1.28E-01& -270& 190& $>15.18$& Y& --\\
C\textsc{ii}& & & & & & & & $>15.18$\\
\hline
Si\textsc{ii}& 1526& 1526.707& 1.27E-01& -270& 200& $>14.68$& Y& --\\
Si\textsc{ii}& & & & & & & & $>14.68$\\
\hline
P\textsc{ii}& 1532& 1532.533& 7.61E-03& -270& 200& $<14.16$& Y& --\\
P\textsc{ii}& & & & & & & & $<14.16$\\
\hline
Ti\textsc{ii}& 3073& 3073.877& 1.09E-01& -150& 200& $<12.73$& N& --\\
Ti\textsc{ii}& 3242& 3242.929& 2.32E-01& -270& 200& $<12.55$& N& --\\
Ti\textsc{ii}& 3384& 3384.740& 3.58E-01& -270& -5& $<12.24$& Y& --\\
Ti\textsc{ii}& & & & & & & & $<12.24$\\
\hline
Cr\textsc{ii}& 2066& 2066.161& 5.15E-02& -170& 200& $<13.87$& Y& --\\
Cr\textsc{ii}& & & & & & & & $<13.87$\\
\hline
Mn\textsc{ii}& 2576& 2576.877& 3.51E-01& -270& 200& $<12.66$& Y& --\\
Mn\textsc{ii}& & & & & & & & $<12.66$\\
\hline
Fe\textsc{ii}& 1608& 1608.451& 5.80E-02& -260& 200& $<14.61$& N& --\\
Fe\textsc{ii}& 2374& 2374.461& 3.13E-02& -270& 200& $14.31\pm0.03$& Y& --\\
Fe\textsc{ii}& 2600& 2600.173& 2.39E-01& -285& 200& $>14.05$& N& --\\
Fe\textsc{ii}& & & & & & & & $14.31\pm0.05$\\
\hline
Ni\textsc{ii}& 1709& 1709.604& 3.24E-02& -270& 200& $<13.41$& Y& --\\
Ni\textsc{ii}& 1741& 1741.553& 4.27E-02& -270& 200& $<13.44$& N& --\\
Ni\textsc{ii}& 1751& 1751.916& 2.77E-02& -270& 200& $<13.47$& N& --\\
Ni\textsc{ii}& & & & & & & & $<13.41$\\
\hline
Zn\textsc{ii}& 2026& 2026.136& 4.89E-01& -270& 200& $<12.96$& Y& --\\
Zn\textsc{ii}& & & & & & & & $<12.96$\\
\hline
\end{tabular}
\end{center}
\end{table*} 

\begin{figure*}
\begin{center}
\includegraphics[width=1.1\textwidth]{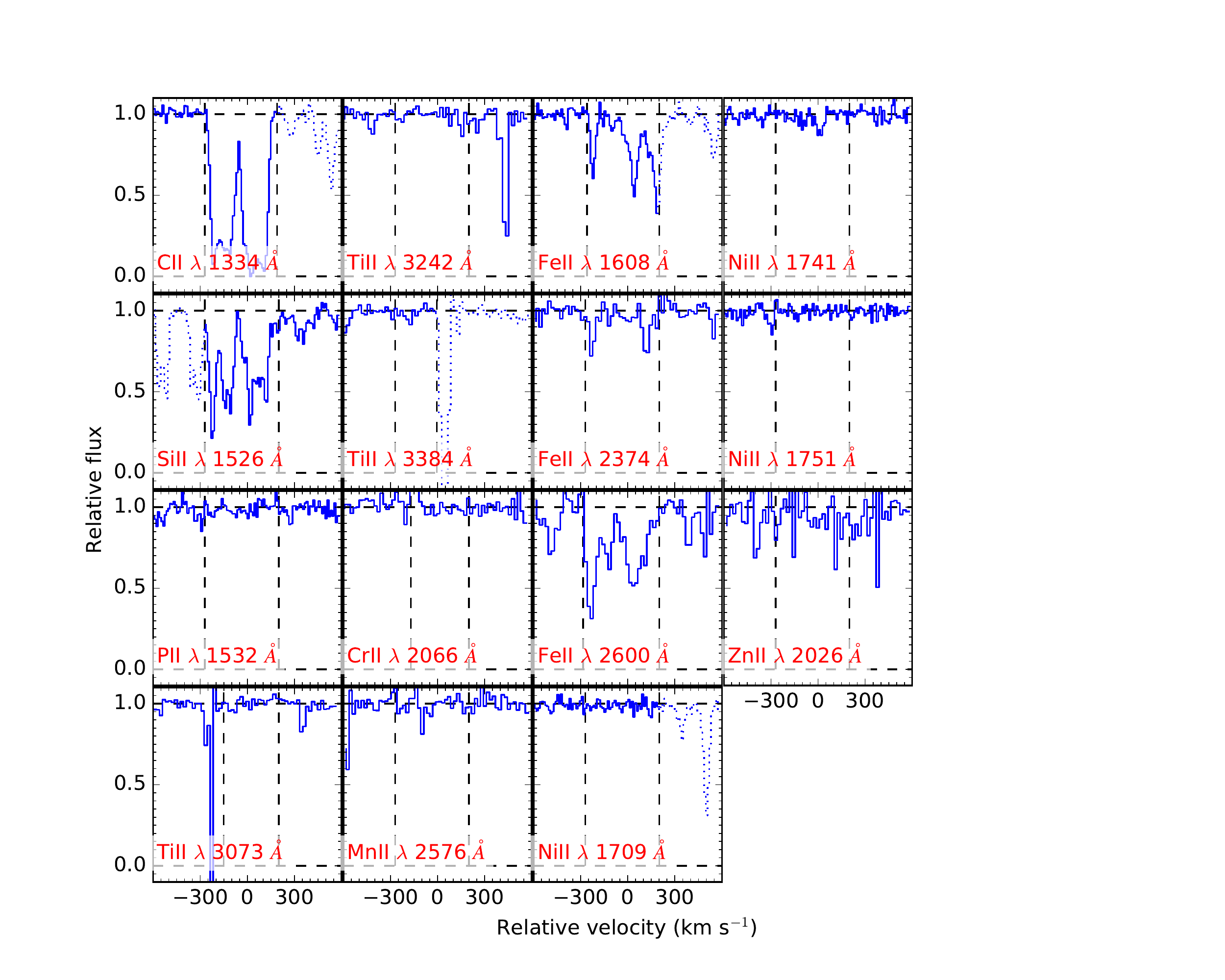}
\caption{Velocity profile of the XQ100 spectrum towards J0955-0130 (\zabs{}=4.024).}
\label{fig:J0955-0130,40245}
\end{center}
\end{figure*}

\begin{table*}
\begin{center}
\caption{X-Shooter metal column densities for J1020+0922 (\zabs{}=$2.592$)}
\label{tab:J1020+0922,25920}
\begin{tabular}{lcccccccc}
\hline
Ion& Line& $\lambda$& $f$& $v_{\rm min}$& $v_{\rm max}$& logN(X)& Included& logN$_{\rm adopt}$\\
 &  & \AA{}&  & \kms{}& \kms{}&  &  &  \\
\hline
O\textsc{i}& 1302& 1302.168& 4.89E-02& 0& 200& $>14.98$& Y& --\\
O\textsc{i}& & & & & & & & $>14.98$\\
\hline
Si\textsc{ii}& 1808& 1808.013& 2.19E-03& 0& 200& $15.26\pm0.04$& Y& --\\
Si\textsc{ii}& & & & & & & & $15.26\pm0.05$\\
\hline
Ti\textsc{ii}& 1910& 1910.750& 2.02E-01& 0& 200& $<12.97$& Y& --\\
Ti\textsc{ii}& 3073& 3073.877& 1.09E-01& 0& 200& $13.28\pm0.08$& N& --\\
\hline
\end{tabular}
\end{center}
\end{table*} 

\begin{figure*}
\begin{center}
\includegraphics[width=1.1\textwidth]{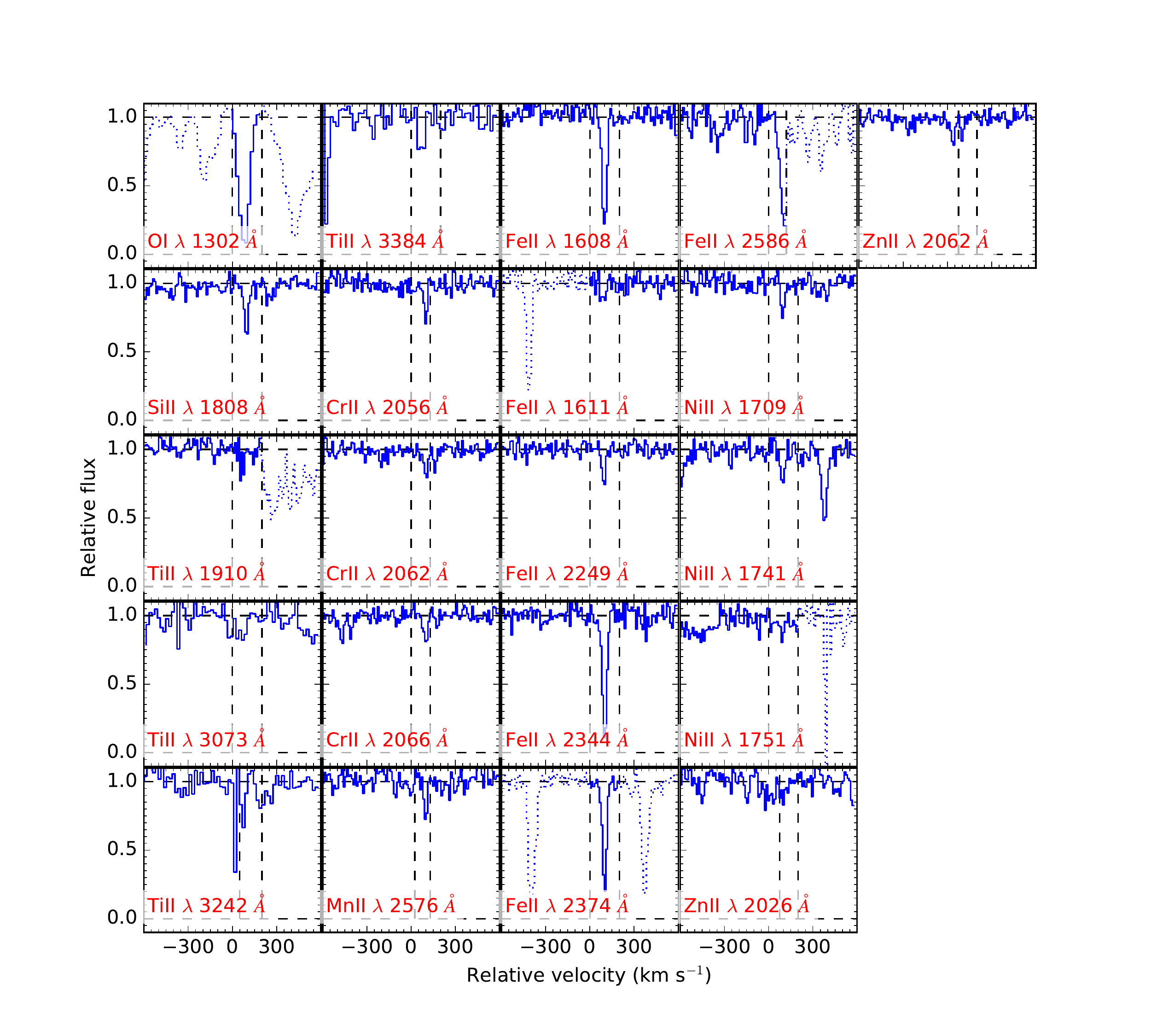}
\caption{Velocity profile of the XQ100 spectrum towards J1020+0922 (\zabs{}=2.592).}
\label{fig:J1020+0922,25920}
\end{center}
\end{figure*}

\begin{table*}
\begin{center}
\caption{X-Shooter metal column densities for J1024+1819 (\zabs{}=$2.298$)}
\label{tab:J1024+1819,22980}
\begin{tabular}{lcccccccc}
\hline
Ion& Line& $\lambda$& $f$& $v_{\rm min}$& $v_{\rm max}$& logN(X)& Included& logN$_{\rm adopt}$\\
 &  & \AA{}&  & \kms{}& \kms{}&  &  &  \\
\hline
Mg\textsc{ii}& 2796& 2796.352& 6.12E-01& -80& 90& $>13.71$& N& --\\
Mg\textsc{ii}& 2803& 2803.531& 3.05E-01& -80& 130& $>13.88$& Y& --\\
Mg\textsc{ii}& & & & & & & & $>13.88$\\
\hline
Si\textsc{ii}& 1260& 1260.422& 1.01E+00& -80& 130& $>13.93$& N& --\\
Si\textsc{ii}& 1526& 1526.707& 1.27E-01& -80& 130& $>14.51$& N& --\\
Si\textsc{ii}& 1808& 1808.013& 2.19E-03& -80& 100& $15.36\pm0.03$& Y& --\\
\hline
\end{tabular}
\end{center}
\end{table*} 

\begin{figure*}
\begin{center}
\includegraphics[width=1.1\textwidth]{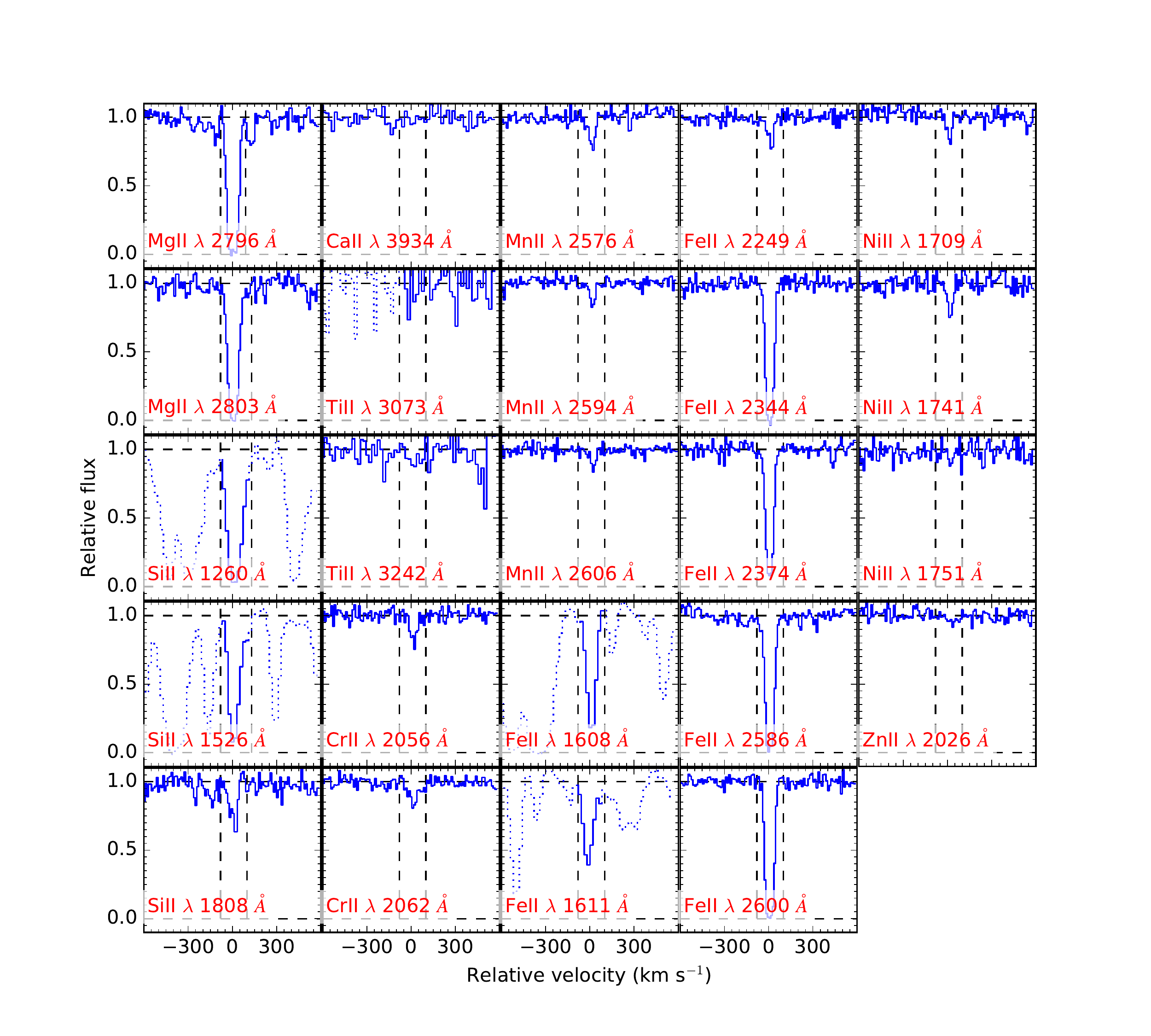}
\caption{Velocity profile of the XQ100 spectrum towards J1024+1819 (\zabs{}=2.298).}
\label{fig:J1024+1819,22980}
\end{center}
\end{figure*}

\clearpage
\begin{table*}
\begin{center}
\caption{X-Shooter metal column densities for J1057+1910 (\zabs{}=$3.373$)}
\label{tab:J1057+1910,33735}
\begin{tabular}{lcccccccc}
\hline
Ion& Line& $\lambda$& $f$& $v_{\rm min}$& $v_{\rm max}$& logN(X)& Included& logN$_{\rm adopt}$\\
 &  & \AA{}&  & \kms{}& \kms{}&  &  &  \\
\hline
Mg\textsc{ii}& 2796& 2796.352& 6.12E-01& -350& 100& $>13.86$& N& --\\
Mg\textsc{ii}& 2803& 2803.531& 3.05E-01& -320& 100& $>14.17$& Y& --\\
Mg\textsc{ii}& & & & & & & & $>14.17$\\
\hline
Si\textsc{ii}& 1304& 1304.370& 9.40E-02& -320& 60& $>14.63$& N& --\\
Si\textsc{ii}& 1526& 1526.707& 1.27E-01& -320& 90& $>14.51$& Y& --\\
Si\textsc{ii}& 1808& 1808.013& 2.19E-03& -320& 100& $<14.71$& Y& --\\
Si\textsc{ii}& & & & & & & & $14.51$--$14.71$\\
\hline
P\textsc{ii}& 1532& 1532.533& 7.61E-03& -320& 100& $<14.32$& Y& --\\
P\textsc{ii}& & & & & & & & $<14.32$\\
\hline
Ca\textsc{ii}& 3934& 3934.777& 6.50E-01& -320& -50& $<11.99$& N& --\\
Ca\textsc{ii}& 3969& 3969.591& 3.22E-01& -320& 100& $<13.06$& Y& --\\
Ca\textsc{ii}& & & & & & & & $<13.06$\\
\hline
Ti\textsc{ii}& 1910& 1910.750& 2.02E-01& -190& 100& $<12.90$& Y& --\\
Ti\textsc{ii}& & & & & & & & $<12.90$\\
\hline
Cr\textsc{ii}& 2066& 2066.161& 5.15E-02& -200& 100& $<13.43$& Y& --\\
Cr\textsc{ii}& & & & & & & & $<13.43$\\
\hline
Fe\textsc{ii}& 1608& 1608.451& 5.80E-02& -300& 100& $14.20\pm0.04$& Y& --\\
Fe\textsc{ii}& 1611& 1611.200& 1.36E-03& -320& 100& $15.20\pm0.11$& N& --\\
Fe\textsc{ii}& 2344& 2344.214& 1.14E-01& -320& 100& $>14.12$& Y& --\\
Fe\textsc{ii}& 2374& 2374.461& 3.13E-02& -300& 100& $14.30\pm0.07$& Y& --\\
Fe\textsc{ii}& 2586& 2586.650& 6.91E-02& -290& 100& $14.31\pm0.06$& Y& --\\
Fe\textsc{ii}& & & & & & & & $14.24\pm0.10$\\
\hline
Ni\textsc{ii}& 1709& 1709.604& 3.24E-02& -320& 100& $<13.73$& Y& --\\
Ni\textsc{ii}& & & & & & & & $<13.73$\\
\hline
Zn\textsc{ii}& 2026& 2026.136& 4.89E-01& -320& 100& $<12.56$& Y& --\\
Zn\textsc{ii}& & & & & & & & $<12.56$\\
\hline
\end{tabular}
\end{center}
\end{table*} 

\begin{figure*}
\begin{center}
\includegraphics[width=1.1\textwidth]{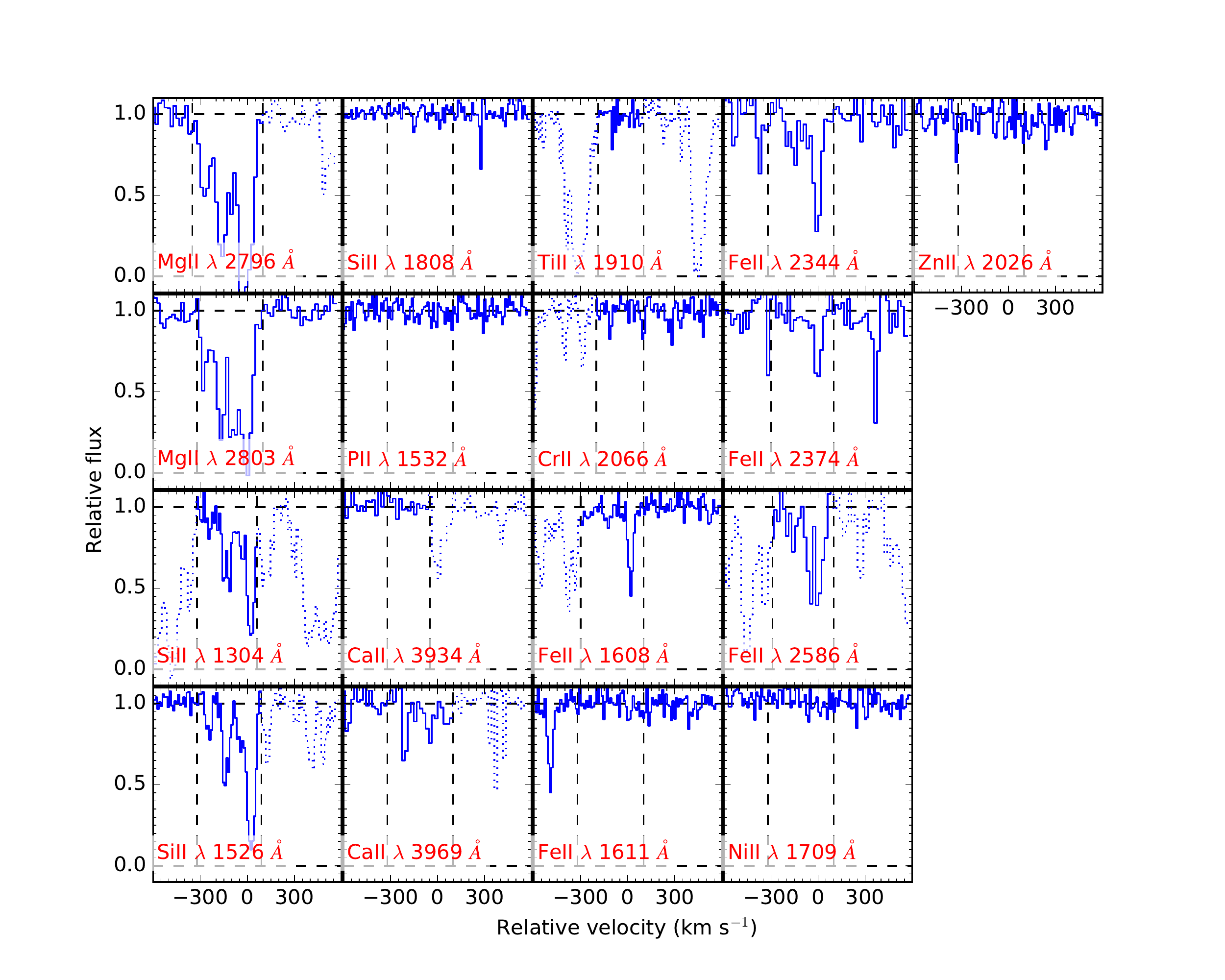}
\caption{Velocity profile of the XQ100 spectrum towards J1057+1910 (\zabs{}=3.373).}
\label{fig:J1057+1910,33735}
\end{center}
\end{figure*}

\begin{table*}
\begin{center}
\caption{X-Shooter metal column densities for J1058+1245 (\zabs{}=$3.432$)}
\label{tab:J1058+1245,34315}
\begin{tabular}{lcccccccc}
\hline
Ion& Line& $\lambda$& $f$& $v_{\rm min}$& $v_{\rm max}$& logN(X)& Included& logN$_{\rm adopt}$\\
 &  & \AA{}&  & \kms{}& \kms{}&  &  &  \\
\hline
Mg\textsc{ii}& 2796& 2796.352& 6.12E-01& -210& 70& $>13.93$& N& --\\
Mg\textsc{ii}& 2803& 2803.531& 3.05E-01& -210& 70& $>14.04$& Y& --\\
Mg\textsc{ii}& & & & & & & & $>14.04$\\
\hline
Si\textsc{ii}& 1526& 1526.707& 1.27E-01& -210& 70& $14.26\pm0.01$& Y& --\\
Si\textsc{ii}& 1808& 1808.013& 2.19E-03& -210& 70& $<15.02$& N& --\\
Si\textsc{ii}& & & & & & & & $14.26\pm0.05$\\
\hline
P\textsc{ii}& 1532& 1532.533& 7.61E-03& -210& 70& $<14.06$& Y& --\\
P\textsc{ii}& & & & & & & & $<14.06$\\
\hline
Ca\textsc{ii}& 3934& 3934.777& 6.50E-01& -210& 70& $<11.98$& Y& --\\
Ca\textsc{ii}& 3969& 3969.591& 3.22E-01& -210& 70& $<12.15$& Y& --\\
Ca\textsc{ii}& & & & & & & & $<11.98$\\
\hline
Ti\textsc{ii}& 1910& 1910.750& 2.02E-01& -210& 70& $<12.80$& Y& --\\
Ti\textsc{ii}& & & & & & & & $<12.80$\\
\hline
Fe\textsc{ii}& 1608& 1608.451& 5.80E-02& -210& 70& $14.16\pm0.02$& Y& --\\
Fe\textsc{ii}& 1611& 1611.200& 1.36E-03& -210& 70& $15.12\pm0.08$& N& --\\
Fe\textsc{ii}& 2344& 2344.214& 1.14E-01& -210& 70& $13.97\pm0.03$& Y& --\\
Fe\textsc{ii}& 2374& 2374.461& 3.13E-02& -210& 70& $<13.75$& Y& --\\
Fe\textsc{ii}& & & & & & & & $14.10\pm0.05$\\
\hline
Ni\textsc{ii}& 1709& 1709.604& 3.24E-02& -210& 70& $<13.58$& Y& --\\
Ni\textsc{ii}& 1741& 1741.553& 4.27E-02& -210& 70& $<13.80$& Y& --\\
Ni\textsc{ii}& 1751& 1751.916& 2.77E-02& -210& 70& $<13.79$& Y& --\\
Ni\textsc{ii}& & & & & & & & $<13.58$\\
\hline
\end{tabular}
\end{center}
\end{table*} 

\begin{figure*}
\begin{center}
\includegraphics[width=1.1\textwidth]{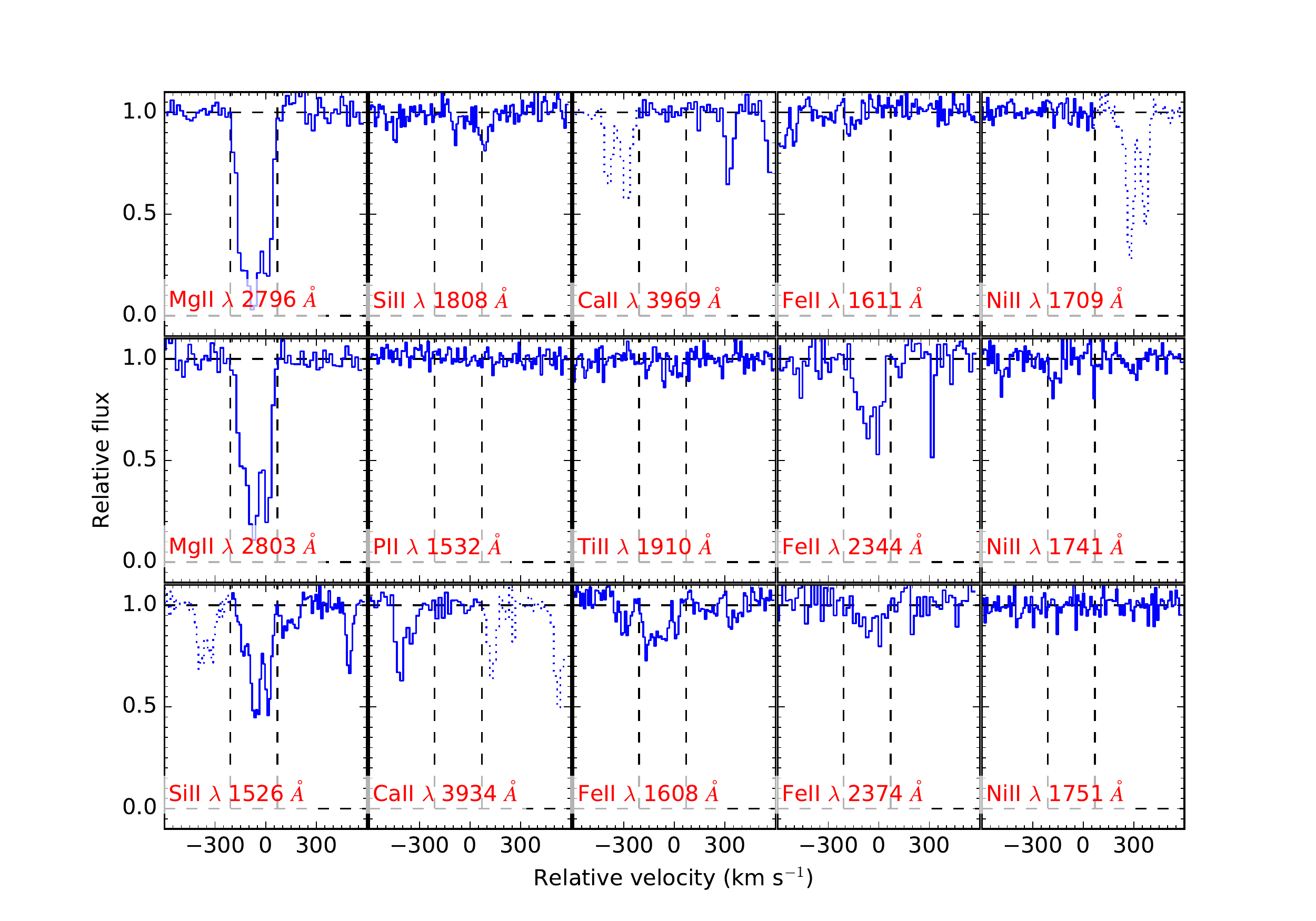}
\caption{Velocity profile of the XQ100 spectrum towards J1058+1245 (\zabs{}=3.432).}
\label{fig:J1058+1245,34315}
\end{center}
\end{figure*}

\begin{table*}
\begin{center}
\caption{X-Shooter metal column densities for J1108+1209 (\zabs{}=$3.397$)}
\label{tab:J1108+1209,33965}
\begin{tabular}{lcccccccc}
\hline
Ion& Line& $\lambda$& $f$& $v_{\rm min}$& $v_{\rm max}$& logN(X)& Included& logN$_{\rm adopt}$\\
 &  & \AA{}&  & \kms{}& \kms{}&  &  &  \\
\hline
C\textsc{ii}& 1334& 1334.532& 1.28E-01& -100& 70& $>13.98$& Y& --\\
C\textsc{ii}& & & & & & & & $>13.98$\\
\hline
O\textsc{i}& 1302& 1302.168& 4.89E-02& -100& 70& $14.64\pm0.01$& Y& --\\
O\textsc{i}& & & & & & & & $14.64\pm0.05$\\
\hline
Mg\textsc{i}& 2852& 2852.964& 1.81E+00& -50& 70& $11.94\pm0.05$& Y& --\\
Mg\textsc{i}& & & & & & & & $11.94\pm0.05$\\
\hline
\end{tabular}
\end{center}
\end{table*} 

\begin{figure*}
\begin{center}
\includegraphics[width=1.1\textwidth]{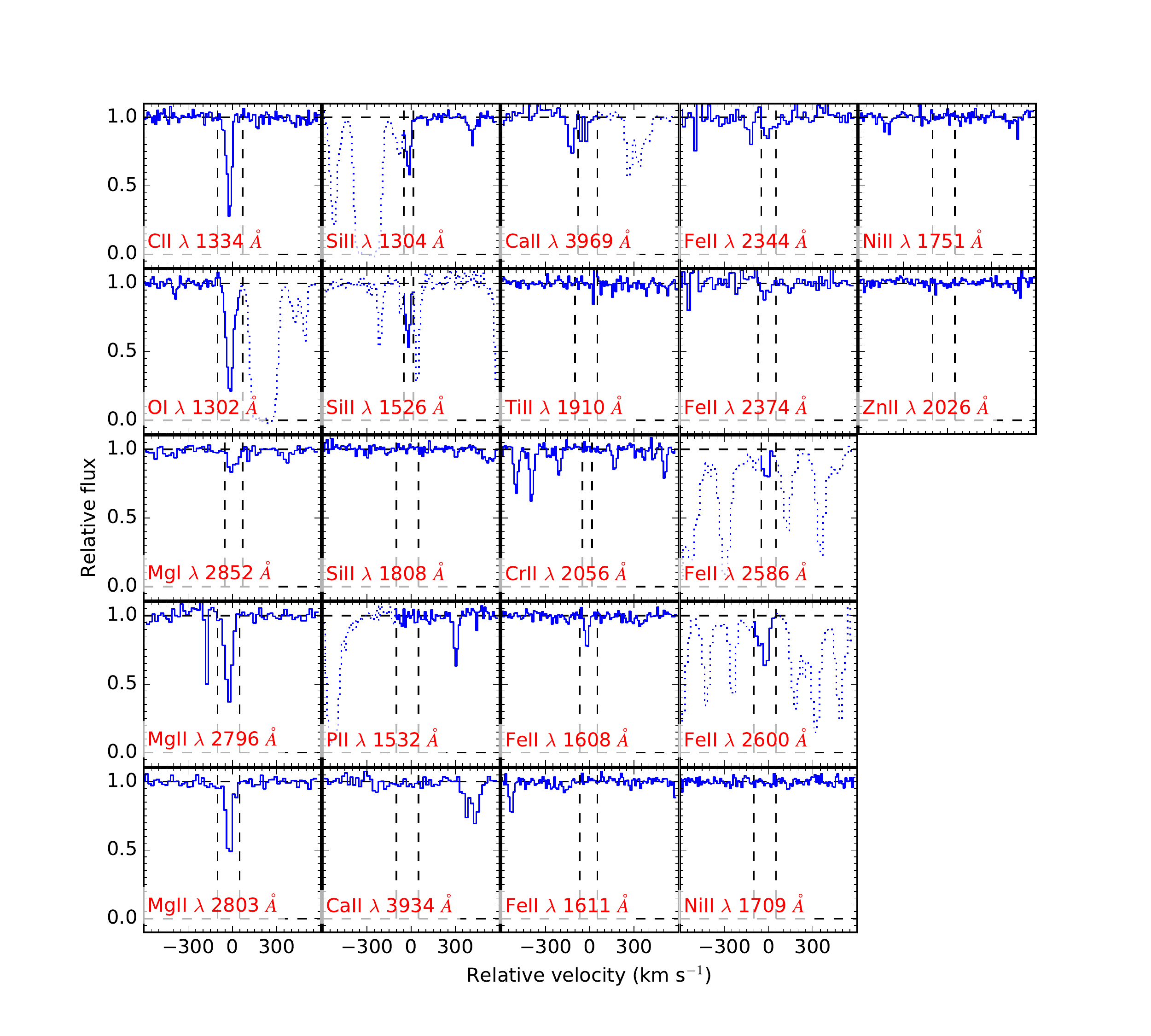}
\caption{Velocity profile of the XQ100 spectrum towards J1108+1209 (\zabs{}=3.397).}
\label{fig:J1108+1209,33965}
\end{center}
\end{figure*}

\begin{table*}
\begin{center}
\caption{X-Shooter metal column densities for J1108+1209 (\zabs{}=$3.546$)}
\label{tab:J1108+1209,35460}
\begin{tabular}{lcccccccc}
\hline
Ion& Line& $\lambda$& $f$& $v_{\rm min}$& $v_{\rm max}$& logN(X)& Included& logN$_{\rm adopt}$\\
 &  & \AA{}&  & \kms{}& \kms{}&  &  &  \\
\hline
C\textsc{ii}& 1334& 1334.532& 1.28E-01& -200& 110& $>15.06$& Y& --\\
C\textsc{ii}& & & & & & & & $>15.06$\\
\hline
O\textsc{i}& 1302& 1302.168& 4.89E-02& -200& 110& $>15.46$& Y& --\\
O\textsc{i}& & & & & & & & $>15.46$\\
\hline
Mg\textsc{i}& 2852& 2852.964& 1.81E+00& -200& 80& $12.28\pm0.02$& Y& --\\
Mg\textsc{i}& & & & & & & & $12.28\pm0.05$\\
\hline
\end{tabular}
\end{center}
\end{table*} 

\begin{figure*}
\begin{center}
\includegraphics[width=1.1\textwidth]{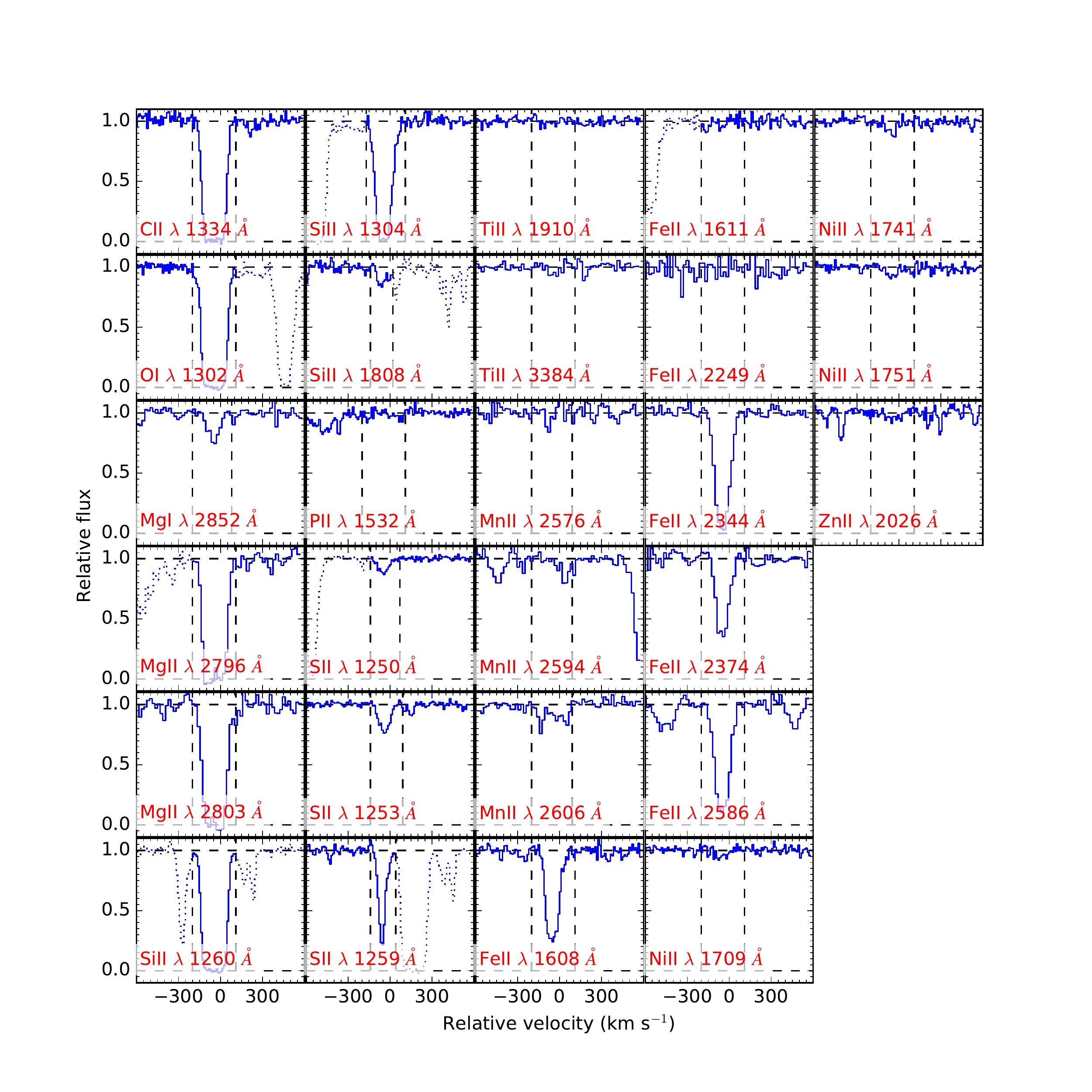}
\caption{Velocity profile of the XQ100 spectrum towards J1108+1209 (\zabs{}=3.546).}
\label{fig:J1108+1209,35460}
\end{center}
\end{figure*}

\clearpage

\begin{table*}
\begin{center}
\caption{X-Shooter metal column densities for J1312+0841 (\zabs{}=$2.660$)}
\label{tab:J1312+0841,26600}
\begin{tabular}{lcccccccc}
\hline
Ion& Line& $\lambda$& $f$& $v_{\rm min}$& $v_{\rm max}$& logN(X)& Included& logN$_{\rm adopt}$\\
 &  & \AA{}&  & \kms{}& \kms{}&  &  &  \\
\hline
Mg\textsc{i}& 2852& 2852.964& 1.81E+00& -160& 180& $<11.67$& Y& --\\
Mg\textsc{i}& & & & & & & & $<11.67$\\
\hline
Mg\textsc{ii}& 2796& 2796.352& 6.12E-01& -160& 180& $>13.74$& N& --\\
Mg\textsc{ii}& 2803& 2803.531& 3.05E-01& -160& 180& $>13.82$& Y& --\\
Mg\textsc{ii}& & & & & & & & $>13.82$\\
\hline
Si\textsc{ii}& 1260& 1260.422& 1.01E+00& -160& 180& $>13.94$& Y& --\\
\hline
\end{tabular}
\end{center}
\end{table*} 

\begin{figure*}
\begin{center}
\includegraphics[width=1.1\textwidth]{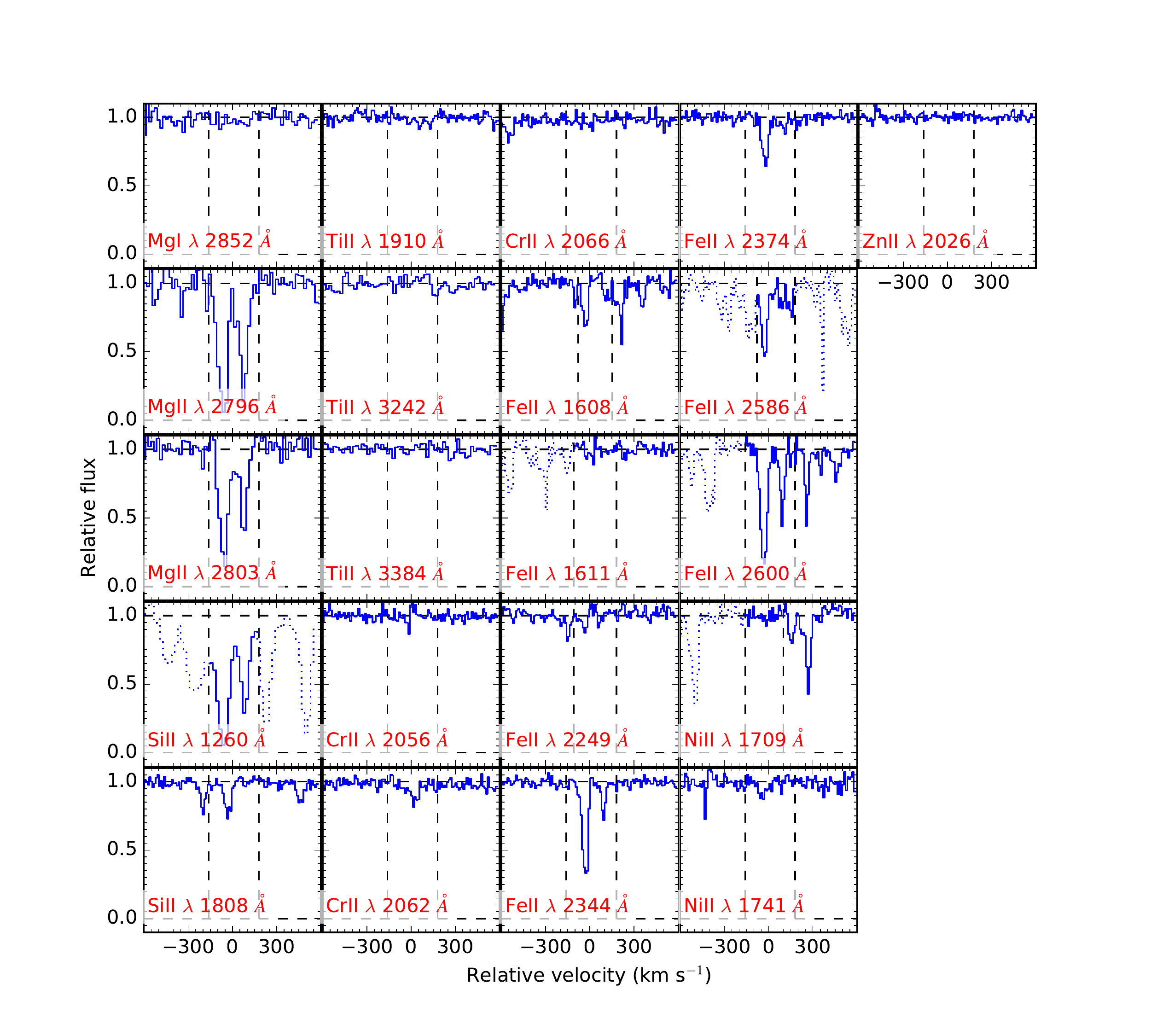}
\caption{Velocity profile of the XQ100 spectrum towards J1312+0841 (\zabs{}=2.660).}
\label{fig:J1312+0841,26600}
\end{center}
\end{figure*}

\begin{table*}
\begin{center}
\caption{X-Shooter metal column densities for J1421$-$0643 (\zabs{}=$3.449$)}
\label{tab:J1421-0643,34490}
\begin{tabular}{lcccccccc}
\hline
Ion& Line& $\lambda$& $f$& $v_{\rm min}$& $v_{\rm max}$& logN(X)& Included& logN$_{\rm adopt}$\\
 &  & \AA{}&  & \kms{}& \kms{}&  &  &  \\
\hline
C\textsc{ii}& 1334& 1334.532& 1.28E-01& -160& 200& $>14.68$& Y& --\\
C\textsc{ii}& & & & & & & & $>14.68$\\
\hline
O\textsc{i}& 1302& 1302.168& 4.89E-02& -150& 200& $>14.97$& Y& --\\
O\textsc{i}& & & & & & & & $>14.97$\\
\hline
Mg\textsc{ii}& 2796& 2796.352& 6.12E-01& -180& 200& $>13.58$& N& --\\
Mg\textsc{ii}& 2803& 2803.531& 3.05E-01& -180& 200& $>13.77$& Y& --\\
\hline
\end{tabular}
\end{center}
\end{table*} 

\begin{figure*}
\begin{center}
\includegraphics[width=1.1\textwidth]{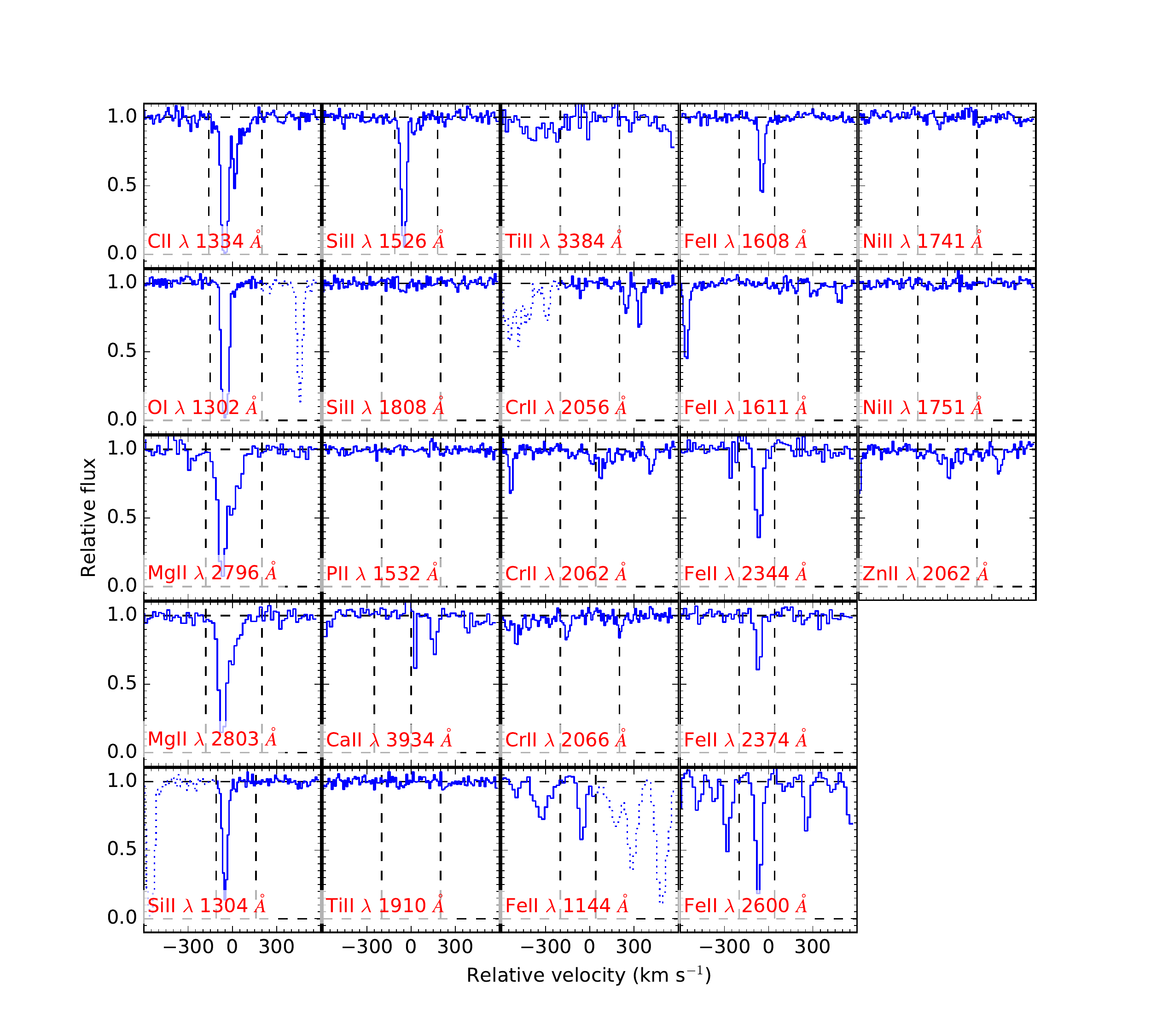}
\caption{Velocity profile of the XQ100 spectrum towards J1421-0643 (\zabs{}=3.449).}
\label{fig:J1421-0643,34490}
\end{center}
\end{figure*}

\clearpage
\begin{table*}
\begin{center}
\caption{X-Shooter metal column densities for J1517+0511 (\zabs{}=$2.688$)}
\label{tab:J1517+0511,26885}
\begin{tabular}{lcccccccc}
\hline
Ion& Line& $\lambda$& $f$& $v_{\rm min}$& $v_{\rm max}$& logN(X)& Included& logN$_{\rm adopt}$\\
 &  & \AA{}&  & \kms{}& \kms{}&  &  &  \\
\hline
Mg\textsc{i}& 2852& 2852.964& 1.81E+00& -200& 70& $<11.72$& Y& --\\
Mg\textsc{i}& & & & & & & & $<11.72$\\
\hline
Mg\textsc{ii}& 2796& 2796.352& 6.12E-01& -130& 70& $>13.63$& N& --\\
Mg\textsc{ii}& 2803& 2803.531& 3.05E-01& -150& 70& $>13.85$& Y& --\\
Mg\textsc{ii}& & & & & & & & $>13.85$\\
\hline
Si\textsc{ii}& 1526& 1526.707& 1.27E-01& -120& 20& $>14.43$& N& --\\
Si\textsc{ii}& 1808& 1808.013& 2.19E-03& -100& 30& $14.85\pm0.05$& Y& --\\
Si\textsc{ii}& & & & & & & & $14.85\pm0.05$\\
\hline
P\textsc{ii}& 1532& 1532.533& 7.61E-03& -120& 70& $<13.86$& Y& --\\
P\textsc{ii}& & & & & & & & $<13.86$\\
\hline
Ti\textsc{ii}& 1910& 1910.750& 2.02E-01& -120& 70& $<12.33$& Y& --\\
Ti\textsc{ii}& 3242& 3242.929& 2.32E-01& -90& 70& $<12.27$& Y& --\\
Ti\textsc{ii}& 3384& 3384.740& 3.58E-01& -50& 70& $<12.26$& Y& --\\
Ti\textsc{ii}& & & & & & & & $<12.26$\\
\hline
Cr\textsc{ii}& 2056& 2056.254& 1.05E-01& -70& 20& $12.91\pm0.05$& Y& --\\
Cr\textsc{ii}& & & & & & & & $12.91\pm0.05$\\
\hline
Fe\textsc{ii}& 1608& 1608.451& 5.80E-02& -130& 20& $14.38\pm0.01$& Y& --\\
Fe\textsc{ii}& 1611& 1611.200& 1.36E-03& -90& 20& $<15.26$& N& --\\
Fe\textsc{ii}& 2249& 2249.877& 1.82E-03& -50& 50& $<14.33$& N& --\\
Fe\textsc{ii}& 2344& 2344.214& 1.14E-01& -120& 70& $>14.21$& N& --\\
Fe\textsc{ii}& 2374& 2374.461& 3.13E-02& -120& 35& $14.39\pm0.01$& Y& --\\
Fe\textsc{ii}& & & & & & & & $14.39\pm0.05$\\
\hline
Ni\textsc{ii}& 1709& 1709.604& 3.24E-02& -120& 70& $<13.44$& N& --\\
Ni\textsc{ii}& 1741& 1741.553& 4.27E-02& -120& 70& $<13.21$& Y& --\\
Ni\textsc{ii}& 1751& 1751.916& 2.77E-02& -120& 70& $<13.51$& N& --\\
Ni\textsc{ii}& & & & & & & & $<13.21$\\
\hline
Zn\textsc{ii}& 2026& 2026.136& 4.89E-01& -120& 70& $<12.06$& Y& --\\
Zn\textsc{ii}& & & & & & & & $<12.06$\\
\hline
\end{tabular}
\end{center}
\end{table*} 

\begin{figure*}
\begin{center}
\includegraphics[width=1.1\textwidth]{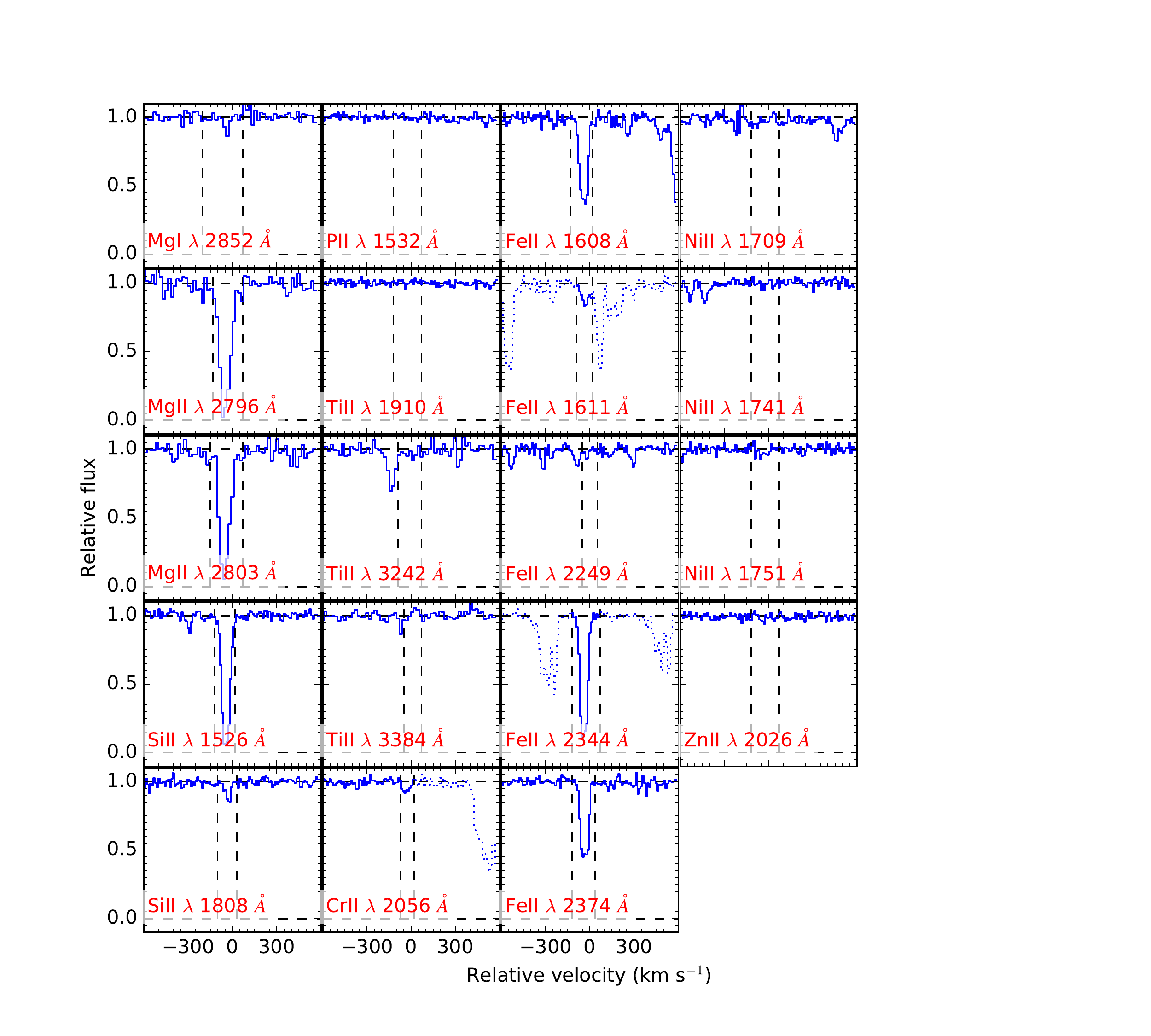}
\caption{Velocity profile of the XQ100 spectrum towards J1517+0511 (\zabs{}=2.688).}
\label{fig:J1517+0511,26885}
\end{center}
\end{figure*}

\begin{table*}
\begin{center}
\caption{X-Shooter metal column densities for J1552+1005 (\zabs{}=$3.601$)}
\label{tab:J1552+1005,36010}
\begin{tabular}{lcccccccc}
\hline
Ion& Line& $\lambda$& $f$& $v_{\rm min}$& $v_{\rm max}$& logN(X)& Included& logN$_{\rm adopt}$\\
 &  & \AA{}&  & \kms{}& \kms{}&  &  &  \\
\hline
C\textsc{ii}& 1334& 1334.532& 1.28E-01& -120& 80& $>14.77$& Y& --\\
C\textsc{ii}& & & & & & & & $>14.77$\\
\hline
O\textsc{i}& 1302& 1302.168& 4.89E-02& -120& 80& $>15.12$& Y& --\\
O\textsc{i}& & & & & & & & $>15.12$\\
\hline
Mg\textsc{ii}& 2796& 2796.352& 6.12E-01& -120& 80& $>13.66$& N& --\\
Mg\textsc{ii}& 2803& 2803.531& 3.05E-01& -150& 80& $>14.04$& Y& --\\
\hline
\end{tabular}
\end{center}
\end{table*} 

\begin{figure*}
\begin{center}
\includegraphics[width=1.1\textwidth]{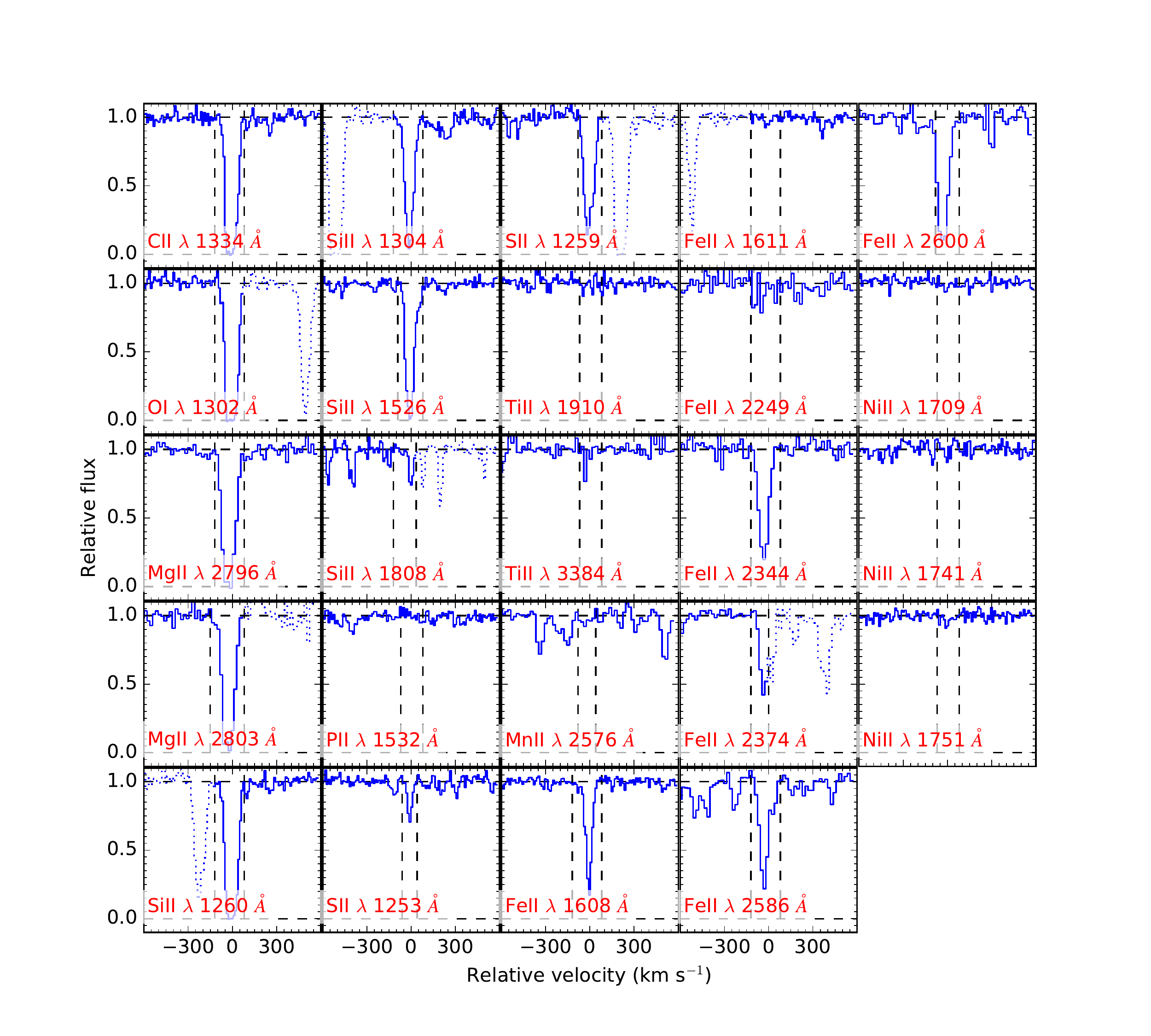}
\caption{Velocity profile of the XQ100 spectrum towards J1552+1005 (\zabs{}=3.601).}
\label{fig:J1552+1005,36010}
\end{center}
\end{figure*}

\begin{table*}
\begin{center}
\caption{X-Shooter metal column densities for J1552+1005 (\zabs{}=$3.667$)}
\label{tab:J1552+1005,36665}
\begin{tabular}{lcccccccc}
\hline
Ion& Line& $\lambda$& $f$& $v_{\rm min}$& $v_{\rm max}$& logN(X)& Included& logN$_{\rm adopt}$\\
 &  & \AA{}&  & \kms{}& \kms{}&  &  &  \\
\hline
C\textsc{ii}& 1334& 1334.532& 1.28E-01& -260& 200& $>14.87$& Y& --\\
C\textsc{ii}& & & & & & & & $>14.87$\\
\hline
O\textsc{i}& 1302& 1302.168& 4.89E-02& -260& 200& $>15.18$& Y& --\\
O\textsc{i}& & & & & & & & $>15.18$\\
\hline
Mg\textsc{ii}& 2796& 2796.352& 6.12E-01& -270& 200& $>13.86$& N& --\\
Mg\textsc{ii}& 2803& 2803.531& 3.05E-01& -270& 200& $>13.96$& Y& --\\
\hline
\end{tabular}
\end{center}
\end{table*} 

\begin{figure*}
\begin{center}
\includegraphics[width=1.1\textwidth]{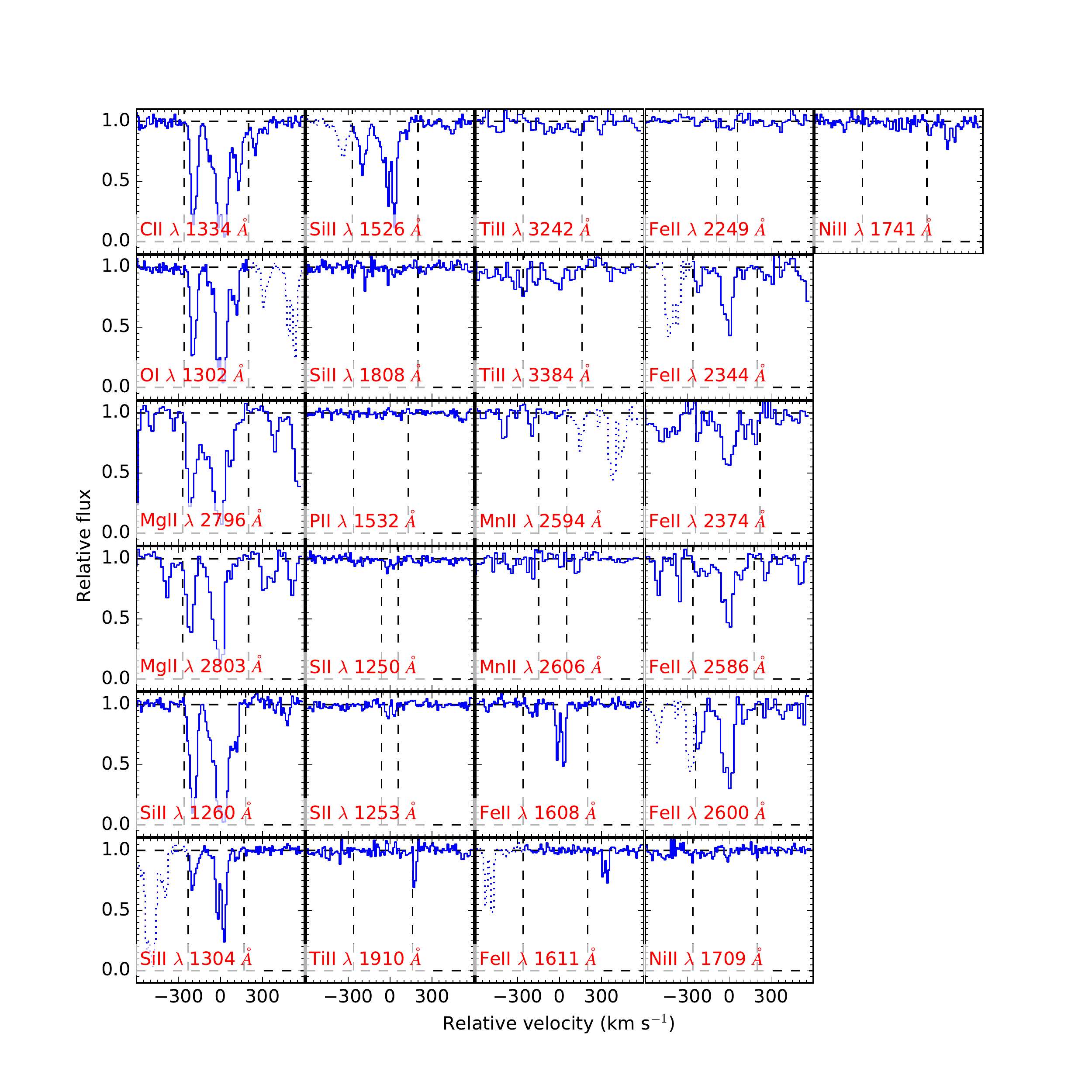}
\caption{Velocity profile of the XQ100 spectrum towards J1552+1005 (\zabs{}=3.667).}
\label{fig:J1552+1005,36665}
\end{center}
\end{figure*}

\clearpage
\begin{table*}
\begin{center}
\caption{X-Shooter metal column densities for J1633+1411 (\zabs{}=$2.882$)}
\label{tab:J1633+1411,28820}
\begin{tabular}{lcccccccc}
\hline
Ion& Line& $\lambda$& $f$& $v_{\rm min}$& $v_{\rm max}$& logN(X)& Included& logN$_{\rm adopt}$\\
 &  & \AA{}&  & \kms{}& \kms{}&  &  &  \\
\hline
Mg\textsc{i}& 2852& 2852.964& 1.81E+00& -200& -50& $<11.60$& Y& --\\
Mg\textsc{i}& & & & & & & & $<11.60$\\
\hline
Mg\textsc{ii}& 2796& 2796.352& 6.12E-01& -200& -50& $>13.09$& N& --\\
Mg\textsc{ii}& 2803& 2803.531& 3.05E-01& -200& -50& $>13.27$& Y& --\\
Mg\textsc{ii}& & & & & & & & $>13.27$\\
\hline
Si\textsc{ii}& 1808& 1808.013& 2.19E-03& -200& -120& $<14.45$& Y& --\\
\hline
\end{tabular}
\end{center}
\end{table*} 

\begin{figure*}
\begin{center}
\includegraphics[width=1.1\textwidth]{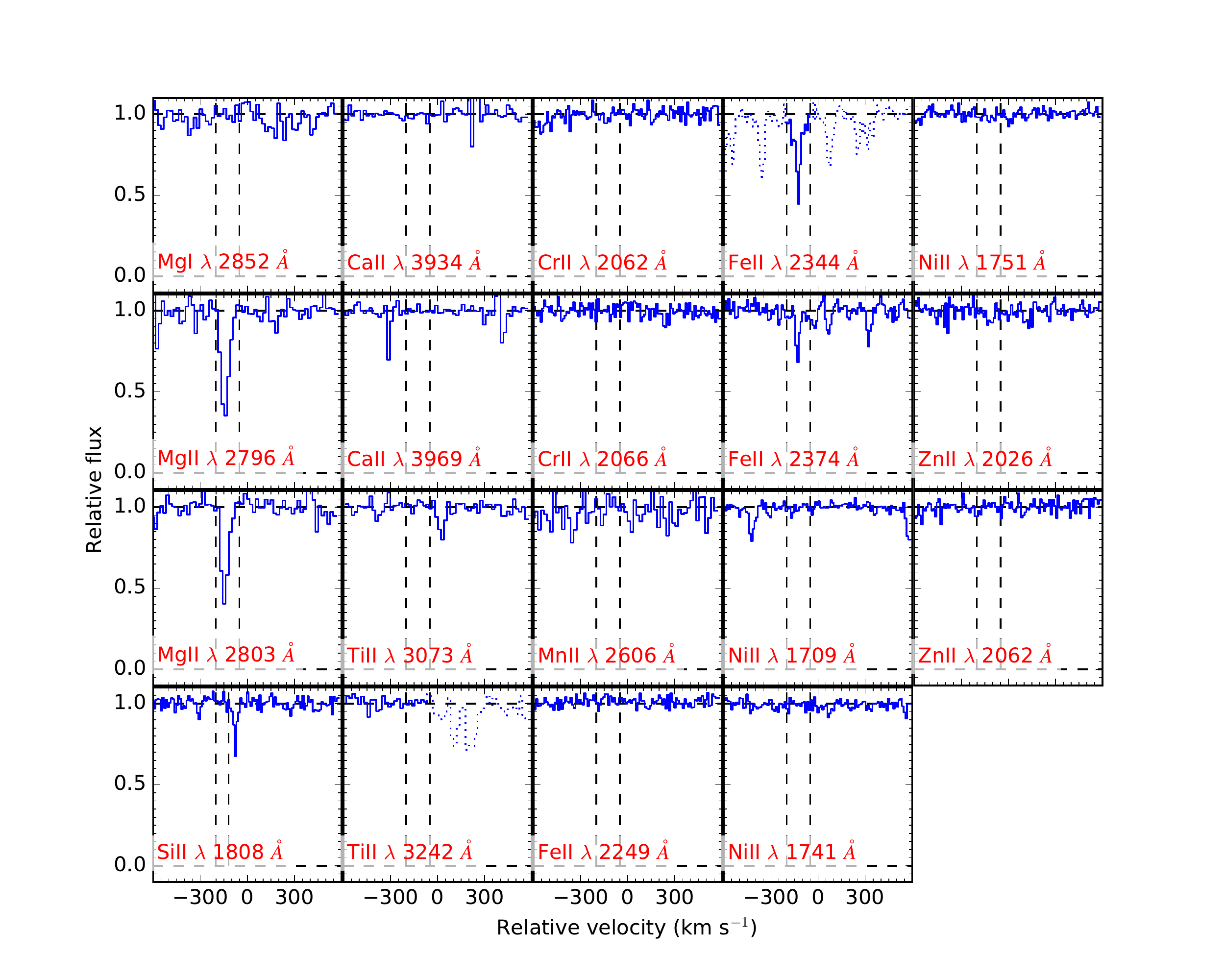}
\caption{Velocity profile of the XQ100 spectrum towards J1633+1411 (\zabs{}=2.882).}
\label{fig:J1633+1411,28820}
\end{center}
\end{figure*}

\begin{table*}
\begin{center}
\caption{X-Shooter metal column densities for J1723+2243 (\zabs{}=$3.698$)}
\label{tab:J1723+2243,36980}
\begin{tabular}{lcccccccc}
\hline
Ion& Line& $\lambda$& $f$& $v_{\rm min}$& $v_{\rm max}$& logN(X)& Included& logN$_{\rm adopt}$\\
 &  & \AA{}&  & \kms{}& \kms{}&  &  &  \\
\hline
Mg\textsc{ii}& 2796& 2796.352& 6.12E-01& -280& 300& $>14.41$& N& --\\
Mg\textsc{ii}& 2803& 2803.531& 3.05E-01& -280& 300& $>14.57$& Y& --\\
Mg\textsc{ii}& & & & & & & & $>14.57$\\
\hline
Si\textsc{ii}& 1526& 1526.707& 1.27E-01& -280& 260& $>14.89$& Y& --\\
Si\textsc{ii}& 1808& 1808.013& 2.19E-03& -210& 240& $<15.37$& Y& --\\
Si\textsc{ii}& & & & & & & & $14.89$--$15.37$\\
\hline
Ti\textsc{ii}& 3242& 3242.929& 2.32E-01& -230& 300& $<12.51$& N& --\\
Ti\textsc{ii}& 3384& 3384.740& 3.58E-01& -230& 300& $<12.10$& Y& --\\
Ti\textsc{ii}& & & & & & & & $<12.10$\\
\hline
Mn\textsc{ii}& 2576& 2576.877& 3.51E-01& -230& 240& $<12.55$& N& --\\
Mn\textsc{ii}& 2594& 2594.499& 2.71E-01& -240& 240& $<12.71$& N& --\\
Mn\textsc{ii}& 2606& 2606.462& 1.93E-01& -280& 300& $<12.46$& Y& --\\
Mn\textsc{ii}& & & & & & & & $<12.46$\\
\hline
Fe\textsc{ii}& 1608& 1608.451& 5.80E-02& -230& 300& $14.62\pm0.01$& Y& --\\
Fe\textsc{ii}& 1611& 1611.200& 1.36E-03& -230& 300& $15.09\pm0.09$& N& --\\
Fe\textsc{ii}& 2249& 2249.877& 1.82E-03& -230& 300& $14.72\pm0.00$& N& --\\
Fe\textsc{ii}& 2344& 2344.214& 1.14E-01& -280& 300& $>14.47$& N& --\\
Fe\textsc{ii}& 2586& 2586.650& 6.91E-02& -280& 300& $>14.55$& N& --\\
Fe\textsc{ii}& 2600& 2600.173& 2.39E-01& -280& 300& $>14.35$& N& --\\
Fe\textsc{ii}& & & & & & & & $14.62\pm0.05$\\
\hline
Ni\textsc{ii}& 1709& 1709.604& 3.24E-02& -230& 300& $<13.63$& Y& --\\
Ni\textsc{ii}& & & & & & & & $<13.63$\\
\hline
\end{tabular}
\end{center}
\end{table*} 

\begin{figure*}
\begin{center}
\includegraphics[width=1.1\textwidth]{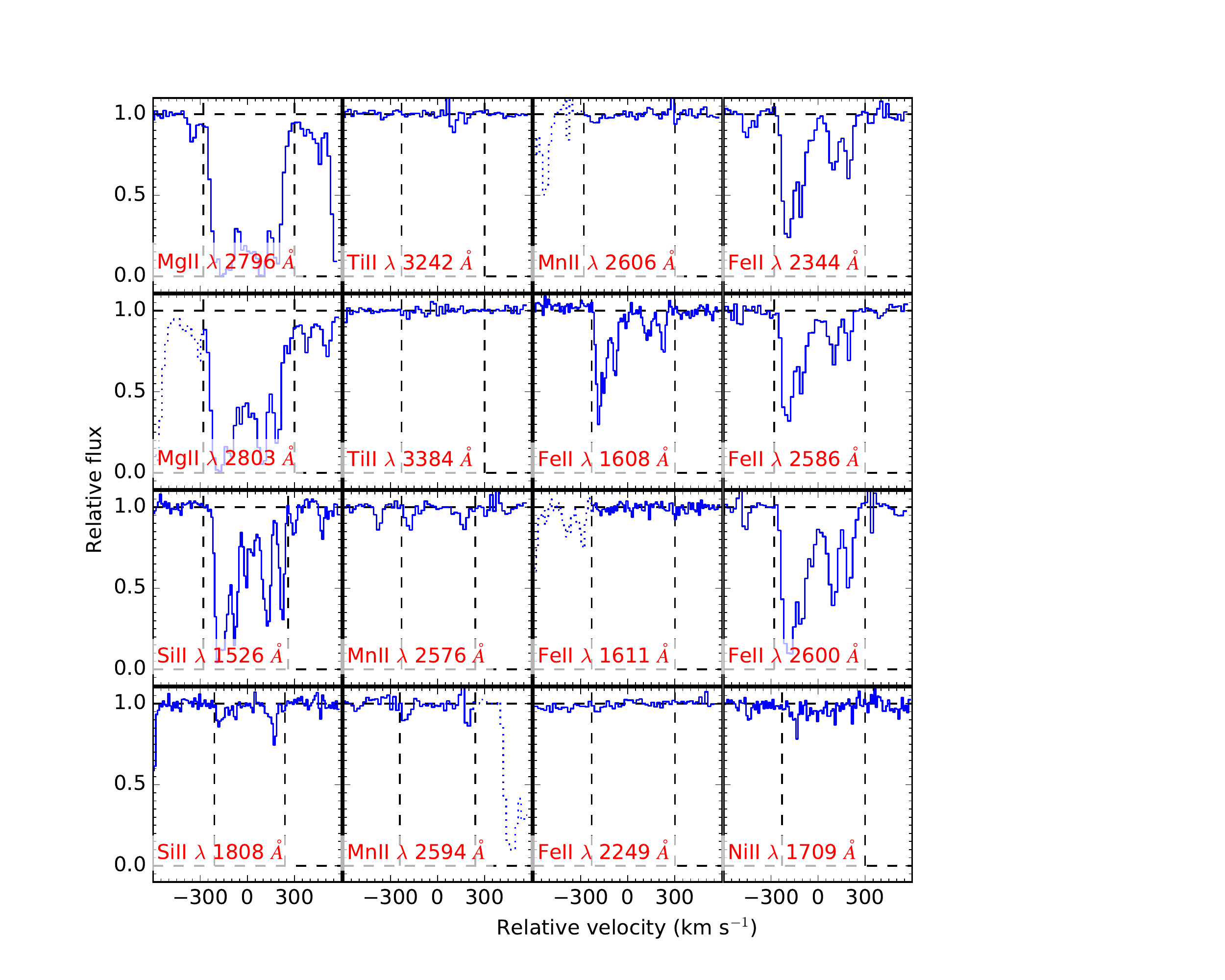}
\caption{Velocity profile of the XQ100 spectrum towards J1723+2243 (\zabs{}=3.698).}
\label{fig:J1723+2243,36980}
\end{center}
\end{figure*}

\input tb_J2239-0552,40805_adopt.tex 

\begin{figure*}
\begin{center}
\includegraphics[width=1.1\textwidth]{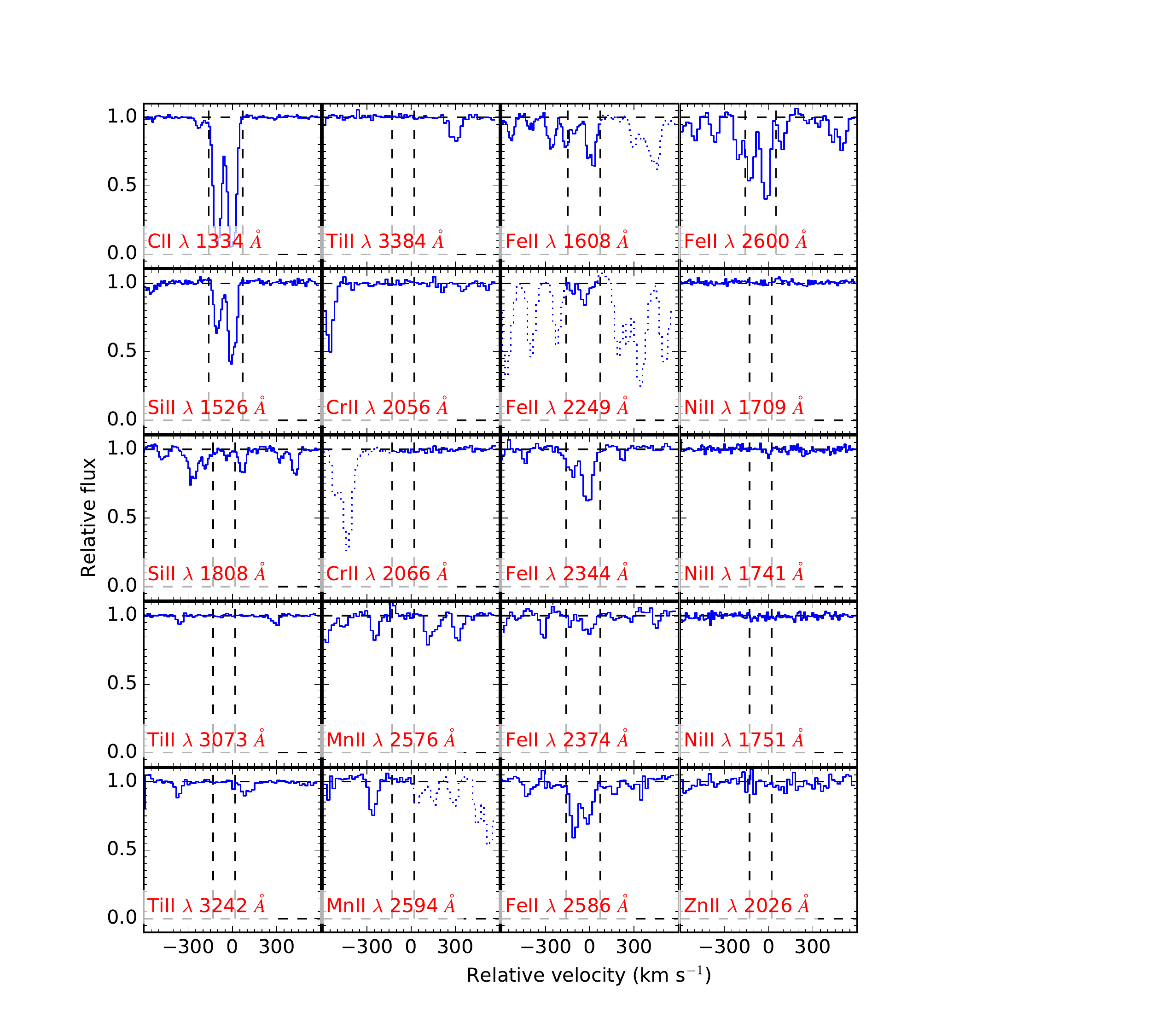}
\caption{Velocity profile of the XQ100 spectrum towards J2239-0552 (\zabs{}=4.080).}
\label{fig:J2239-0552,40805}
\end{center}
\end{figure*}

\clearpage
\begin{table*}
\begin{center}
\caption{X-Shooter metal column densities for J2344+0342 (\zabs{}=$3.220$)}
\label{tab:J2344+0342,32200}
\begin{tabular}{lcccccccc}
\hline
Ion& Line& $\lambda$& $f$& $v_{\rm min}$& $v_{\rm max}$& logN(X)& Included& logN$_{\rm adopt}$\\
 &  & \AA{}&  & \kms{}& \kms{}&  &  &  \\
\hline
Mg\textsc{i}& 2852& 2852.964& 1.81E+00& -70& 80& $12.56\pm0.01$& Y& --\\
Mg\textsc{i}& & & & & & & & $12.56\pm0.05$\\
\hline
Mg\textsc{ii}& 2796& 2796.352& 6.12E-01& -210& 205& $>13.74$& N& --\\
Mg\textsc{ii}& 2803& 2803.531& 3.05E-01& -170& 205& $>14.18$& Y& --\\
Mg\textsc{ii}& & & & & & & & $>14.18$\\
\hline
Ca\textsc{ii}& 3934& 3934.777& 6.50E-01& -90& 90& $<11.78$& Y& --\\
\hline
\end{tabular}
\end{center}
\end{table*} 

\begin{figure*}
\begin{center}
\includegraphics[width=1.1\textwidth]{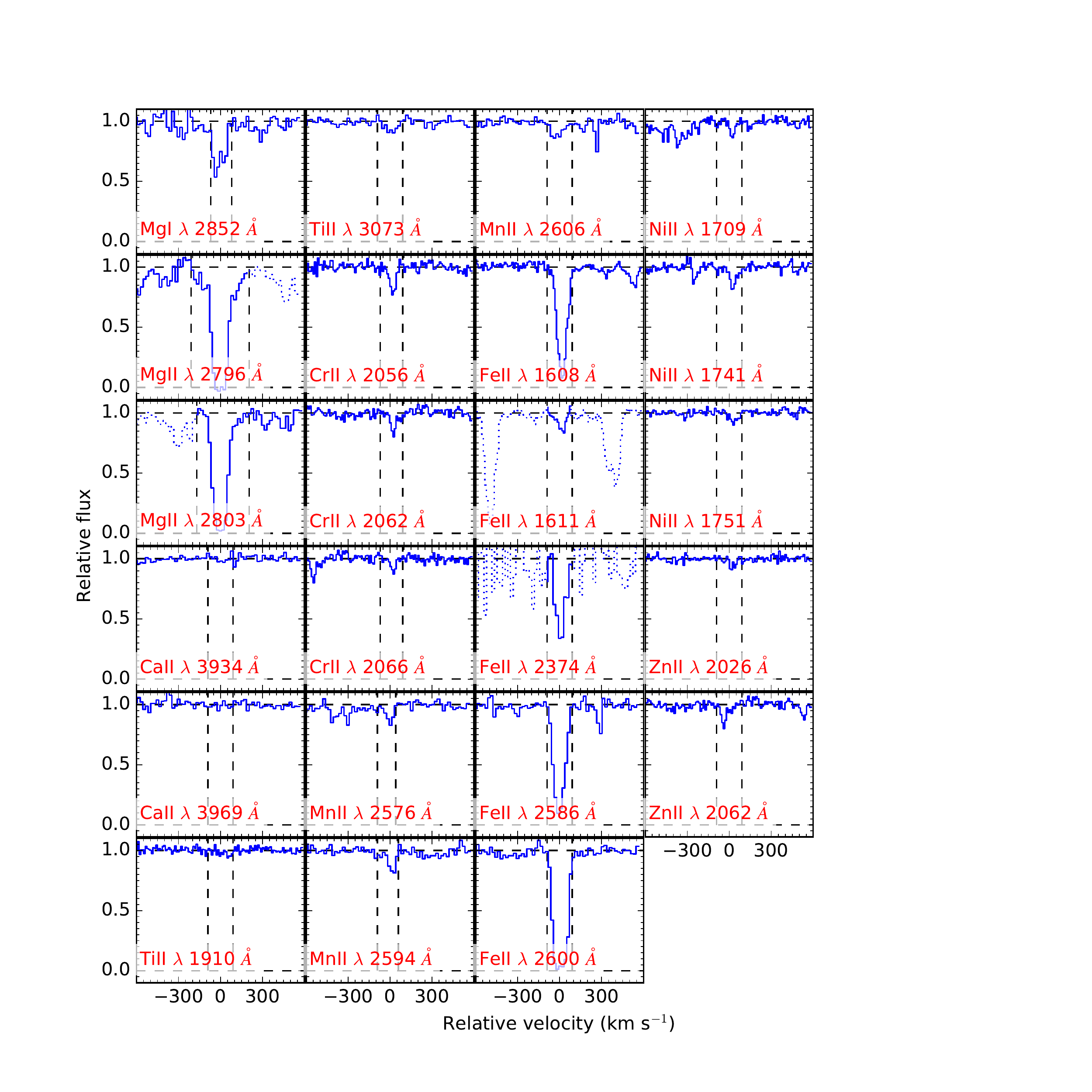}
\caption{Velocity profile of the XQ100 spectrum towards J2344+0342 (\zabs{}=3.220).}
\label{fig:J2344+0342,32200}
\end{center}
\end{figure*}

\clearpage
\subsection{UVES data J0034+1639}
\label{app:udata}
\input tb_J0034+1639,37525_uves_adopt.tex 

\begin{figure*}
\begin{center}
\includegraphics[width=1.1\textwidth]{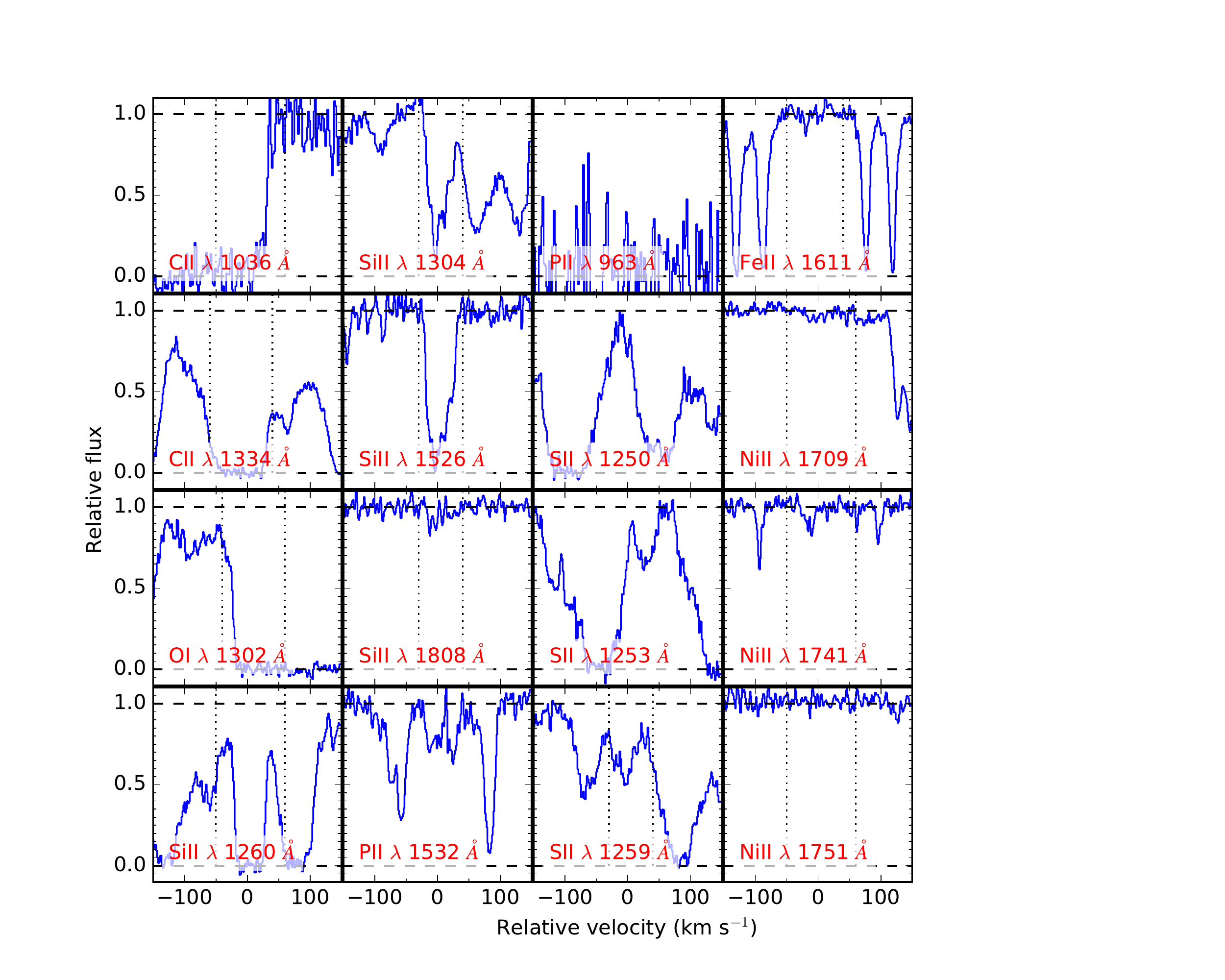}
\caption{Velocity profile of the UVES spectrum towards J0034+1639 (\zabs{}=3.752).}
\label{fig:uvesJ0034+1639,37525}
\end{center}
\end{figure*}

\input tb_J0034+1639,42828_uves_adopt.tex 

\begin{figure*}
\begin{center}
\includegraphics[width=1.1\textwidth]{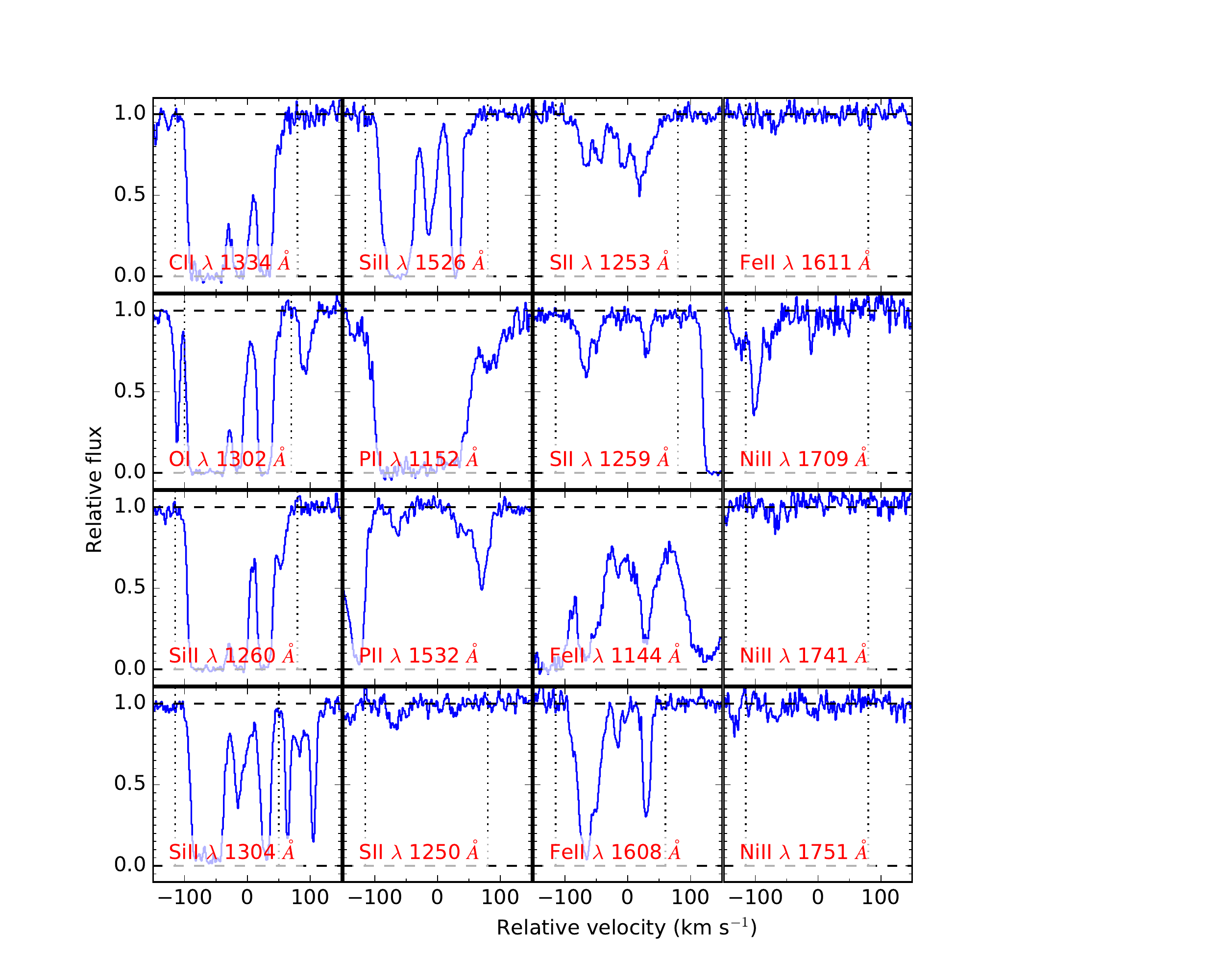}
\caption{Velocity profile of the UVES spectrum towards J0034+1639 (\zabs{}=4.282).}
\label{fig:uvesJ0034+1639,42828}
\end{center}
\end{figure*}

\section{Abundance discrepancies from literature}
\label{sec:AppXQ100}

\subsection*{J0003$-$2603 (\zabs{}$=3.39$)}
The Fe\sion{} literature column density was taken from \cite{DLAcat21}. The discrepancy with the XQ-100 derived column density of 0.12 dex is the result of a difference in the adopted oscillator strengths. \cite{DLAcat21} uses an Fe\sion{} 1611 \AA{} oscillator strength 33\% higher than the value adopted from \cite{Morton03}; matching the size of the discrepancy in column densities. We continue using the XQ-100 column density as the adopted value.

\subsection*{J0134+0400 (\zabs{}$=3.77$)}
The literature data is primarily taken from 
\cite{Prochaska03ApJS147}, and was obtained with ESI. Both the literature and XQ-100 spectra show Si\sion{} 1808 lines free from contamination, and provide abundances that are consistent with 
a saturated Si\sion{} 1526 lines. We suspect that our Si\sion{} 1808 measurement is more reliable as it reproduces the same velocity profile of other metal lines, 
whereas there appears to be an additional blue component ($\sim -75$\kms{}) in the \cite{Prochaska03ApJS147} profile.

The 0.14 dex discrepancy in the logN(Ni\sion{}) is likely due to inaccurate oscillator strengths of many Ni\sion{} lines \citep[e.g.][]{Jenkins06}. \cite{Prochaska03ApJS147} use the Ni\sion{} 1370\AA{} line, whereas our value uses a combination of Ni\sion{} 1741 and 1751 \AA{} lines. Given that adopted Ni\sion{} 1370\AA{} line from \cite{Prochaska03ApJS147} is also similarly higher ($\sim0.12$ dex) than from \citet[our source of oscillator strengths]{Morton03}, we believe accounting for this discrepancy in oscillator strengths would place the two column densities in agreement. We continue to adopt the XQ-100 derived column density.

\subsection*{J0255+0048 (\zabs{}$=3.25$ and 3.91)}
For both DLAs towards J0255+0048, we find discrepant Ni\sion{} from the values derived in \cite{DLAcat23}.  For the DLA at \zabs{}$=3.25$, \cite{DLAcat23} claim a detection of the Ni\sion{} 1709 line, whereas we do not see any significant absorption in the XQ-100 spectra for the Ni\sion{} 1709 line. As our 3$\sigma$ limit in column density is consistent with the detection by \cite{DLAcat23}, it is unclear whether the 
Ni\sion{} 1709 line is actually detected. We therefore conservatively adopt our logN(N\sion{}) limit.

The large discrepancy in the logN(Ni\sion{}) in the  DLA at \zabs{}$=3.91$ is likely due to inaccurate 
oscillator strengths of many Ni\sion{} lines. For the Ni\sion{} 1709, 1741, and 1751 lines; the oscillator strengths adopted in both works agree. However for Ni\sion{} 1317 and 1370, the oscillator strengths adopted in \cite{DLAcat23} are $\sim0.15$ dex higher than ours. Accounting for this discrepancy in oscillator strength, our derived column density would be consistent with that measured by \cite{DLAcat23}. \cite{Jenkins06} focused on independently measuring the oscillator strengths of Ni.  The oscillator strengths from \cite{Morton03} (which we adopt) are consistent with those measured by \cite{Jenkins06}. We therefore adopt the XQ-100 column densities.

\subsection*{J1421$-$0643 (\zabs{}$=3.45$)}
The Fe\sion{} abundance was taken from \cite{DLAcat50}, and is 0.09 dex higher than what we have measured. The discrepancy in column density is driven by the inclusion of the Fe\sion{} 1144\AA{} and 2374\AA{} lines in the XQ-100 value, as the lower Fe\sion{} 1608\AA{} column density agrees with the measured value of the 1608\AA{} line from the literature \citep{AkermanThesis,DLAcat50}. As multiple Fe\sion{} lines should provide a better estimate of the true column density, we continue to use the XQ-100 adopted column density.
\end{document}